\documentclass[final,5p,times,twocolumn,sort&compress]{elsarticle}
\usepackage[final]{changes}

\usepackage{amssymb}
\usepackage{amsmath}
\usepackage[version=4]{mhchem}
\usepackage[colorlinks=true, urlcolor=blue]{hyperref}
\usepackage[svgnames]{xcolor} 
\journal{Physica E: Low-dimensional Systems and Nanostructures}

\usepackage{booktabs}     
\usepackage{array}        
\usepackage{multirow}     
\usepackage{siunitx}  
\usepackage{textcomp} 
\usepackage{changes}
\usepackage{graphicx}
\usepackage{epstopdf}
\begin{document}

\begin{frontmatter}



\title{Correlated Quantum Phenomena in Confined Two-Dimensional Hexagonal Crystals}

\author[1]{Xiang Liu\fnref{fn1}}
\author[1]{Zheng Tao\fnref{fn1}}
\fntext[fn1]{These authors contributed equally to this work.}
\author[1]{Wenchen Luo\corref{cor1}}
\ead{luo.wenchen@csu.edu.cn}
\author[2]{Tapash Chakraborty\corref{cor1}}
\ead{Tapash.Chakraborty@umanitoba.ca}


\affiliation[1]{organization={School of Physics, Central South University},
	city={Changsha},
	postcode={410083}, 
	country={China}}
\affiliation[2]{organization={Department of Physics and Astronomy, University of Manitoba},
	city={Winnipeg},
	postcode={R3T 2N2}, 
	country={Canada}}
\cortext[cor1]{Corresponding authors}
\begin{abstract}
Low-energy fermionic excitations in two-dimensional materials deviate from the conventional Schrödinger description and are instead governed by Dirac equations. 
Such Dirac fermions give rise to a variety of unconventional quantum phenomena that have no direct analogues in traditional condensed matter systems. Among these materials, graphene and transition metal dichalcogenides (TMDs) represent two prototypical platforms, hosting massless and massive Dirac particles, respectively, and exhibiting rich electronic, optical, and valley dependent properties. Here we review the effect of the quantum confinement in these two-dimensional hexagonal materials that provides a powerful route to enhance Coulomb interactions and stabilizing correlated quantum states. In graphene- and TMD-based quantum dots and other nanostructures, externally imposed confinement leads to discrete levels or quantized subbands, where interaction effects are strongly amplified. In twisted van der Waals heterostructures, the moir\'e superlattices generate emergent confinement and induce nontrivial band topology, giving rise to a wealth of novel phenomena. 
More generally, reduced dimensionality and spatial localization in two-dimensional materials promote a diverse range of correlated states. Recent experimental and theoretical advances highlight the central role of confinement in shaping quantum behavior and reveal new opportunities for applications based on these states. In this review, we provide an overview of recent progress in confinement-induced correlated phenomena in two-dimensional materials from both theoretical and experimental perspectives.
\end{abstract}

%


\end{frontmatter}



\section{Introduction}
\label{sec1}

Symmetry and topology have become central organizing principles in modern condensed matter physics, particularly in the study of nanoscale quantum materials\cite{tkachov2022topological,chakraborty2002nano,vontapashqhe40,chakraborty2024encyclopedia,gruber2024interplay,Trambly2026reveiw}. Within this framework, atomically thin two-dimensional (2D) crystals provide a versatile platform for exploring symmetry-protected electronic states and topological phenomena. The immense potential of these systems for next-generation nanoelectronic and optoelectronic devices has stimulated intense interest in semiconducting 2D materials\cite{tapashl2010review,WehlingDirac,XuGraphene}. 

Over the past decade, a broad family of van der Waals layered materials has been identified, including insulating hexagonal boron nitride (hBN), anisotropic black phosphorus, Xenes, and related layered systems\cite{geim2013van,jung2026hbn,blackph2014,li2025Xenes,lu2017semimetals}. Within this expanding materials landscape, graphene and transition metal dichalcogenides (TMDs) have emerged as prominent platforms for both fundamental research and device-oriented investigations\cite{vontapashqhe40,graphenedetection2024,zhouyu2025TMDs,sunj2025TMDs}. Graphene hosts massless Dirac fermions with exceptionally high carrier mobility, whereas group-\textrm{VI} TMDs possess finite direct band gaps and strong excitonic effects, enabling a wide range of electronic and optical functionalities\cite{2Dmap,effectgraphnePR}.

\begin{figure}[htp]
	\centering
	\includegraphics[width=\linewidth]{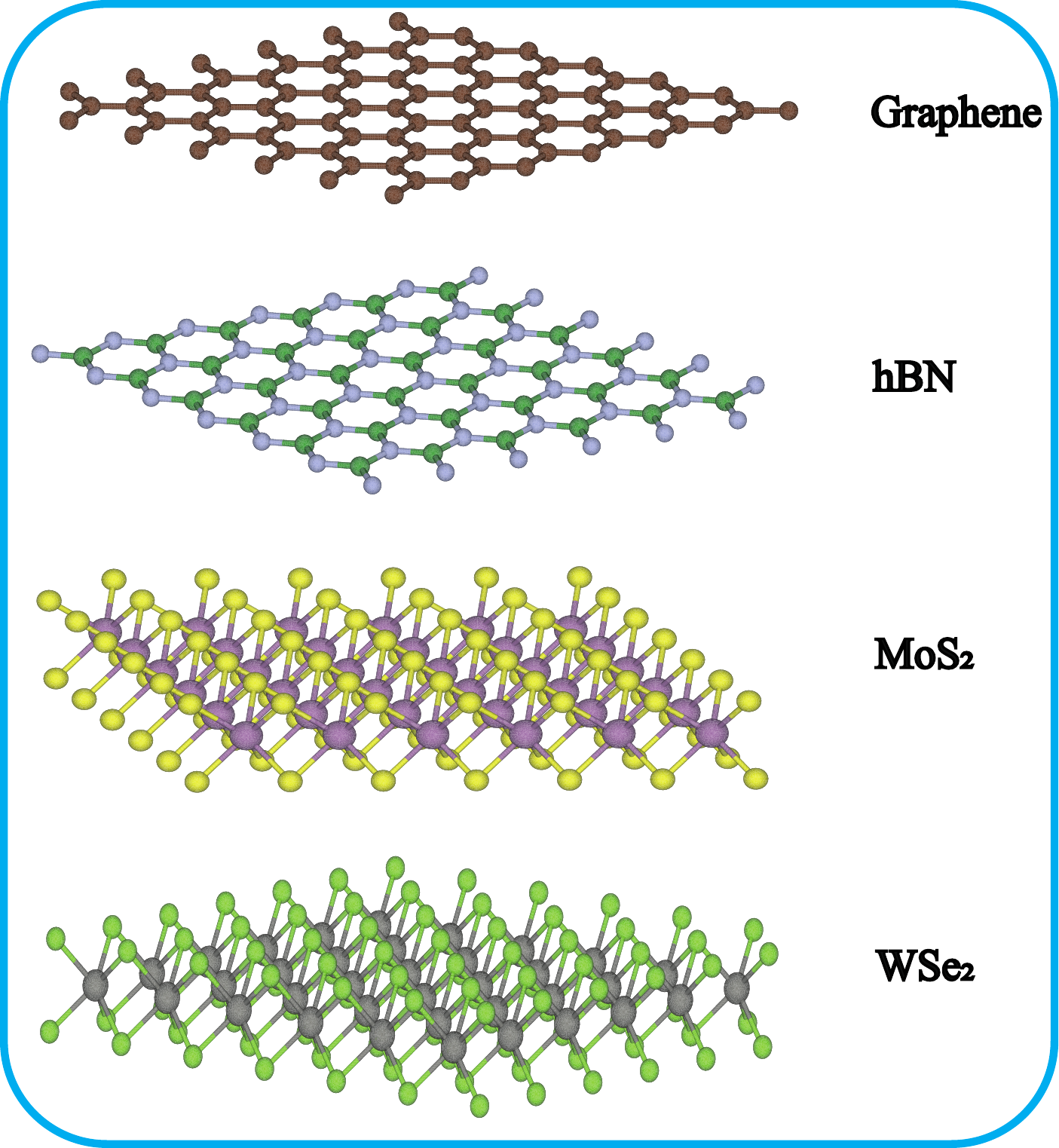}
	\caption{Crystal lattice structures of representative 2D materials including graphene, hBN, \ce{MoS2}, other transition metal dichalcogenides, and layered oxides.}
	\label{materials}
\end{figure}

Structurally, graphene forms a planar honeycomb lattice consisting of two equivalent carbon atoms occupying the A and B sublattices, which preserves both spatial inversion symmetry and time-reversal symmetry\cite{graphene2016pr}. In contrast, monolayer TMDs with chemical formula \ce{MX2} (where \ce{M} is a transition metal and \ce{X} is a chalcogen) adopt a trilayer \ce{X-M-X} structure in which inversion symmetry is intrinsically broken. This broken inversion symmetry, combined with strong spin-orbit coupling (SOC), lifts the A-B sublattice degeneracy and leads to coupled spin-valley physics that distinguishes TMDs from graphene-based systems\cite{yao2012tmds,2Dmater2015}. Representative crystal structures of several monolayer materials are shown in Fig.~\ref{materials}.

The low-energy electronic properties of these systems are commonly described using effective continuum models derived from $\mathbf{k}\cdot \mathbf{p}$ perturbation theory around high-symmetry points of the Brillouin zone \cite{2009electronicproperties,2010electronicproperties}. In graphene, this approach leads to the Dirac Hamiltonian describing massless Dirac fermions, which accounts for its characteristic optical absorption and ultrahigh carrier mobility\cite{2008opticalgra,2023highgraphene,zhao2024ultrahigh}. In contrast, TMDs host massive Dirac fermions with strong SOC, giving rise to coupled spin and valley degrees of freedom and enabling valley-selective optical excitation using circularly polarized light\cite{yaolight2015,mak2018light}.

\begin{figure}[t]
	\centering
	\includegraphics[width=0.95\linewidth]{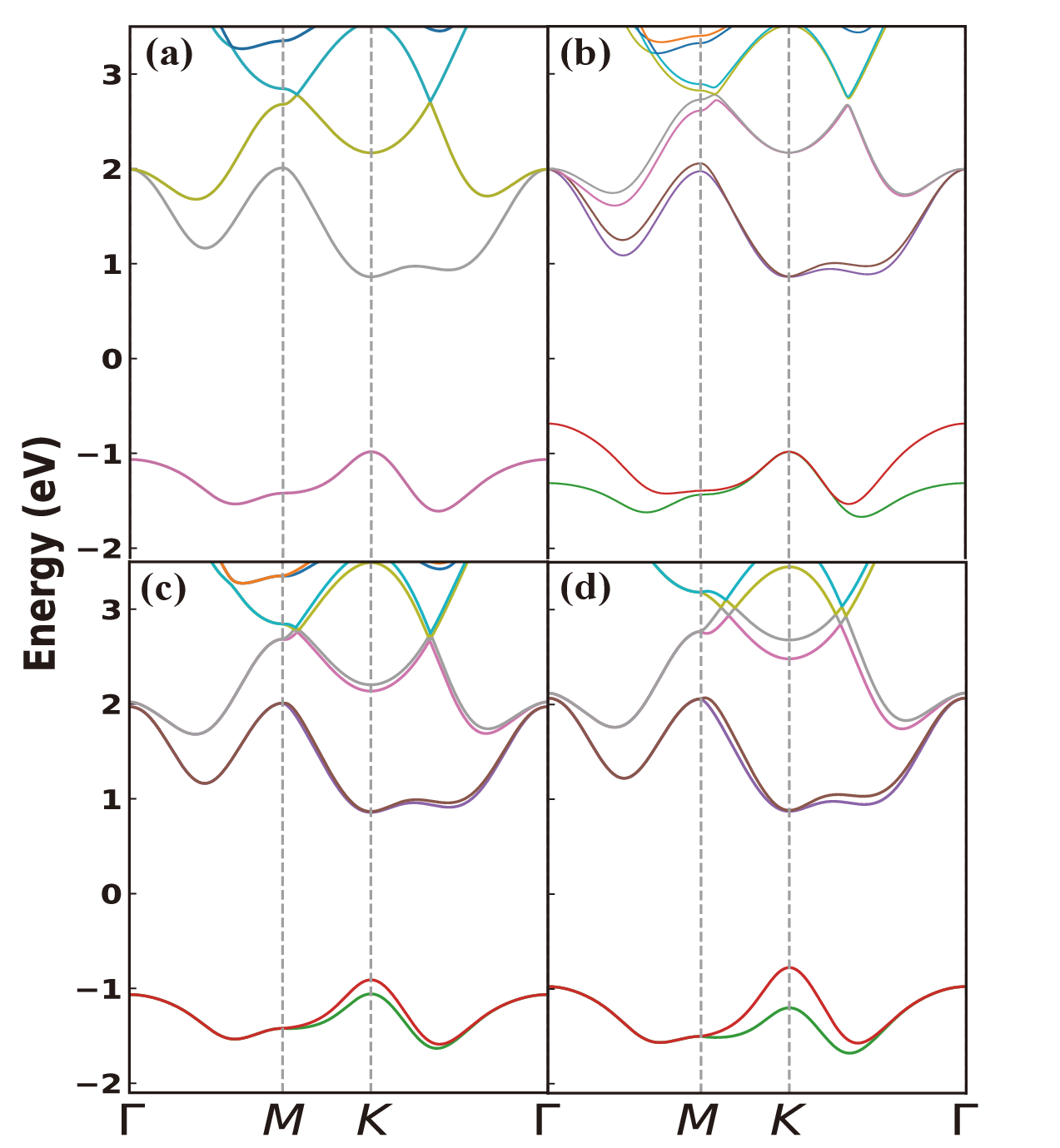}
	\caption{Electronic band structures of \ce{MoS2} and \ce{WS2} along the high-symmetry path $\Gamma$-$M$-$K$-$\Gamma$. (a) Monolayer \ce{MoS2} without spin-orbit coupling. (b) Bilayer \ce{MoS2} without spin-orbit coupling. (c) Monolayer \ce{MoS2} with spin-orbit coupling. (d) Monolayer \ce{WS2} with spin-orbit coupling. The band structures are obtained from tight-binding model calculations, with spin-orbit coupling included in panels (c) and (d) to capture the effects of spin-dependent interactions on the electronic properties of the materials.}
	\label{materialsband}
\end{figure}

As illustrated in Fig.~\ref{materialsband}, the band structures of TMDs exhibit significant differences between monolayer and bilayer configurations. Strong SOC leads to sizable spin splitting in both valence and conduction bands, accompanied by electron-hole asymmetry and trigonal warping effects near the valley extrema\cite{2013tw,2020tw}. In monolayer TMDs such as \ce{MoS2}, \ce{WS2}, \ce{MoSe2}, and \ce{WSe2}, the valence band spin splitting typically ranges from $70$ to $300~\mathrm{meV}$, while the conduction band splitting is on the order of $1$-$20~\mathrm{meV}$\cite{yao2012tmds,2Dmater2015}. These splittings determine the energetic ordering between bright and dark excitonic states and strongly influence optical emission efficiency\cite{2018darkexciton,xu2014spin}.

In addition to their intrinsic electronic structures, van der Waals stacking introduces new degrees of freedom associated with relative translation and rotation between layers. Such structural configurations can break inversion symmetry and generate spontaneous electric polarization. In particular, certain bilayer stacking registries give rise to sliding ferroelectricity in van der Waals materials, where polarization reversal is controlled by lateral interlayer displacement\cite{wu2021sliding}. When a small twist angle is introduced between two layers, the spatial variation of stacking across a moir\'e superlattice can further produce periodic ferroelectric domains, providing a platform for moir\'e ferroelectricity and electrically tunable correlated electronic states.

Beyond extended systems, additional quantum confinement can be realized by shaping 2D materials into zero-dimensional quantum dots (QDs) through top-down or bottom-up fabrication strategies\cite{ding2019defectqd,buqd2011,reviewqd2021}. When the confinement length becomes comparable to the carrier de Broglie wavelength, the continuous band spectrum evolves into discrete atom-like energy levels. These artificial atoms provide a powerful platform for exploring quantum confinement effects in solid-state systems\cite{tapsh1999129,tapashqdee}. \added{Graphene nanoribbons also provide a prototypical example: lateral confinement quantizes the spectrum into subbands, while zigzag edges support edge-localized states near the Fermi level \cite{brey2006graphene_nanoribbons}.} In particular, QDs based on graphene and TMDs have attracted considerable interest owing to the emergence of Dirac fermions under confinement and the presence of spin and valley degrees of freedom that enable controllable quantum states\cite{chen2007fock,kim2024confinement}. Their discrete optical transitions enable deterministic single-photon emission and provide promising building blocks for solid-state quantum devices and scalable qubit implementations\cite{2014TMDsqdqubit,2021TMDsqdqubit,2021TMDsqdqubitsBrooks}.

Strong Coulomb interactions and reduced dielectric screening in low-dimensional systems further promote the emergence of correlated electronic and excitonic phenomena. Early theoretical investigations of interaction effects in graphene and related Dirac systems clarified how Coulomb interactions and chirality govern electronic compressibility and collective properties\cite{2009Tapashchirality,2009TapashBilayerGraphene,tapashmoire,berashevich2011nature,CHAKRABORTY2013123,2014TapashHofstadterSystem,2012Tapashtrilayergraphene}. In extended graphene and TMD systems, many-body correlations can give rise to a wide range of collective behaviors, including correlated insulating states, excitonic complexes, and superconductivity in moir\'e superlattices\cite{regan2020mott,andrei2021marvels,ghiotto2021quantum,sheng2024terahertz}. The interplay between symmetry, topology, quantum confinement, and many-body interactions therefore establishes 2D materials as an exceptionally rich platform for exploring emergent quantum phases.

Given the rapid expansion of research on 2D materials, this review is not intended to be exhaustive. Instead, we adopt confinement in low-dimensional systems as a unifying physical theme, focusing on quantum dots and emergent confinement arising in moir\'e superlattices of twisted van der Waals heterostructures. Within this framework, we emphasize how spatial localization and reduced dimensionality enhance Coulomb interactions and promote a variety of correlated electronic phenomena.

This review is organized as follows. In Section~\ref{low_energy}, we introduce the theoretical framework of Dirac materials and summarize the essential low-energy electronic properties of graphene, TMDs, and Weyl systems. Section~\ref{QDs} discusses quantum confinement in graphene and TMD quantum dots. Section~\ref{correlated} reviews interaction-driven phenomena in 2D materials, including transport, collective excitations, moir\'e superlattices, and ferroelectricity. Finally, Section~\ref{applications} highlights emerging applications and technological prospects of 2D materials.

\section{Low-energy Properties of Dirac Materials Systems} \label{low_energy}
The low-energy electronic landscape of 2D Dirac materials is primarily governed by the underlying lattice symmetry, inversion symmetry breaking, and intrinsic SOC\cite{tapashl2010review,WehlingDirac,XuGraphene,geim2013van,yao2012tmds}. When inversion symmetry is broken, the resulting effective Hamiltonian typically describes massive Dirac quasiparticles endowed with valley-contrasting Berry curvature and coupled spin-valley degrees of freedom, which underpin phenomena in materials such as monolayer TMDs.

Monolayer graphene, as the first experimentally isolated 2D Dirac system, provides the foundational model from which many massive Dirac materials are understood as symmetry-broken derivatives. Its bipartite honeycomb lattice preserves inversion symmetry, enforcing a gapless spectrum described near the inequivalent $K$ and $K'$ valleys, and the low-energy excitations are described by the massless Dirac Hamiltonian
\begin{equation}
H_\mathrm{G}(\mathbf{k}) = \hbar v_F (\tau k_x \sigma_x + k_y \sigma_y),
\end{equation}
where $v_F$ is the Fermi velocity, $\tau = \pm 1$ denote the $K$ and $K'$ valleys, respectively, and $\sigma_{x(y)}$ acts on the sublattice pseudospin. This Hamiltonian gives rise to a quantized \(\pi\) Berry phase for closed paths around a Dirac point and locks the pseudospin orientation relative to the momentum, defining the chirality of the charge carriers \cite{2008berry,2019berry}. The gapless linear dispersion is topologically protected against symmetry-preserving perturbations. However, the continuous rotational symmetry is an emergent low-energy property. At higher carrier densities, the intrinsic trigonal warping from the $ C_3 $ lattice symmetry becomes dominant, breaking the isotropic approximation and inducing anisotropic transport behavior \cite{2020berry}.

In contrast to graphene, monolayer TMDs lack inversion symmetry, which opens a substantial bandgap $\Delta$ and confers a finite mass to the Dirac fermions. The presence of heavy metal atoms further introduces strong intrinsic SOC. The effective Hamiltonian governing these massive fermions is given by
\begin{equation}
	H_{\mathrm{TMD}}(\mathbf{k}) = a t (\tau k_x \sigma_x + k_y \sigma_y) + \frac{\Delta}{2} \sigma_z - \lambda \tau s_z \frac{\sigma_z - 1}{2},
\end{equation}
where \(a\) is the lattice constant, \(t\) denotes the effective hopping amplitude, and \(2\lambda\) represents the spin splitting of the valence band. The SOC term manifests as Ising-type coupling that pins the electron spin perpendicular to the atomic plane \cite{yao2012tmds,2011soczhu}. The sign reversal of this Ising SOC between valleys is dictated by time-reversal symmetry, leading to spin-valley locking. This locking originates from the combined effects of broken inversion symmetry and strong SOC, which lift the spin degeneracy in a valley dependent manner. Consequently, spin relaxation is strongly suppressed, since a spin flip necessarily requires a large momentum transfer between valleys. This robustness forms the physical basis of valleytronic applications \cite{Qi2014}, enabling optical initialization and manipulation of information encoded in the valley pseudospin \cite{wang2017valley,mak2018light,2021TMDsqdqubit}.

\begin{figure}[htp]
	\centering
	\includegraphics[width=\linewidth]{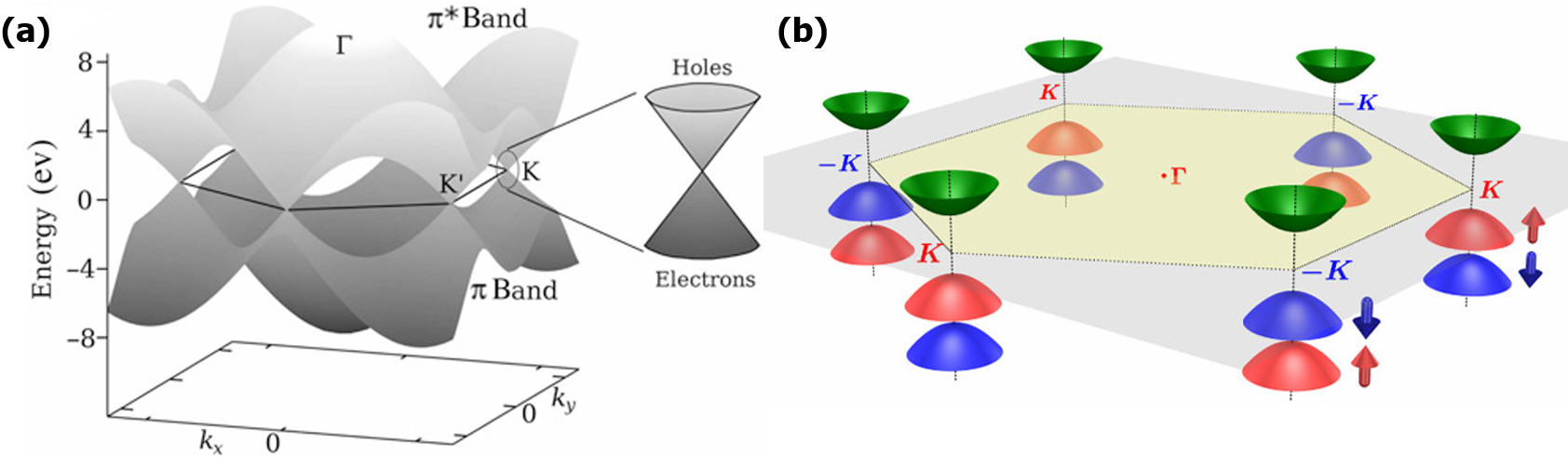}
	\caption{Schematic of graphene and TMDs drawing of
		the band structure at the band edges located at the $K$ and $K'$ points. Adapted from Ref.~\cite{tapashl2010review,yao2012tmds}}
	\label{energyband}
\end{figure}

The low-energy effective Hamiltonians of graphene and monolayer TMDs capture both the energy dispersion and the internal pseudospin and valley degrees of freedom of their Dirac quasiparticles. Figures~\ref{energyband}(a) and (b) show the band structures near the valley extrema at the $K$ and $K'$ points of graphene and TMDs, respectively. In monolayer TMDs, broken inversion symmetry together with strong SOC generates a finite, valley-contrasting Berry curvature $\Omega(\mathbf{k})$ concentrated near the $K$ and $K'$ valleys, giving rise to the valley Hall effect \cite{valleyqhe}. In pristine graphene, inversion symmetry enforces a vanishing Berry curvature throughout the Brillouin zone, with singularities confined only to the Dirac points, although nonzero Berry curvature can be induced through external symmetry breaking \cite{tapashl2010review,WehlingDirac,XuGraphene}.

While the continuum $\mathbf{k\cdot p}$ Hamiltonians discussed above capture the essential low-energy physics near the valley extrema, quantitative descriptions across the full Brillouin zone often require lattice-scale tight-binding models. For monolayer TMDs, a commonly used description is the 11-band tight-binding model constructed from the transition-metal $d$ orbitals $\{d_{z^2}, d_{x^2-y^2}, d_{xy}, d_{xz}, d_{yz}\}$ together with the $p$ orbitals of the top and bottom chalcogen atoms $\{p_x^{(t)}, p_y^{(t)}, p_z^{(t)}, p_x^{(b)}, p_y^{(b)}, p_z^{(b)}\}$, which faithfully reproduces the first-principles band structure over the entire Brillouin zone\cite{cappelluti2013tight}. \added{Moreover, if the spin-orbit coupling is important, a $22-$band spinful model is necessary.}

In three-dimensional Weyl semimetals (WSMs), conduction and valence bands touch at discrete Weyl nodes in the bulk BZ. Each node acts as a monopole of Berry curvature, endowing the electronic structure with a nontrivial Chern number on specific momentum-space planes \cite{lu2017semimetals,2018WeylRevModPhys,2025Vadym}. The low-energy Hamiltonian in the vicinity of a single Weyl node takes the form
\begin{equation}
	H_{\mathrm{WSM}}(\mathbf{k}) = \chi \hbar v_F (\sigma_x k_x + \sigma_y k_y + \sigma_z k_z),
\end{equation}
where \(\chi = \pm 1\) denotes the chirality and \(\sigma_{x,y,z}\) act on the relevant pseudospin space. A direct consequence of this topology is the presence of Fermi-arc surface states, which connect surface projections of Weyl nodes with opposite chirality and provide a clear manifestation of the bulk-boundary correspondence. When a WSM is confined along one spatial direction, its bulk spectrum quantizes into a series of 2D subbands. Crucially, these subbands inherit a nonzero Chern number from the parent Weyl nodes, thereby realizing chiral massive Dirac fermions. This outcome stands in sharp contrast to the massive Dirac fermions in monolayer TMDs or in symmetry-broken graphene, which possess a zero net Chern number. Consequently, the dimensional reduction of WSMs does not merely reproduce conventional 2D Dirac physics, but yields a distinct class of topologically nontrivial 2D systems with protected chiral edge states, offering a different pathway for engineering topological matter \cite{2014Weylhall}. It is interesting to note that a planar WSM can, under certain conditions, enhance the Coulomb interaction and behave as the incompressible Laughlin state~\cite{2025Vadym,lu2024realization,chakraborty2012fractional}.

\section{2D Material Quantum Dots}
\label{QDs}
In this section, we review the physics of quantum confinement in 2D materials, emphasizing its influence on electronic behavior and interaction driven phenomena. Starting from theoretical models of 2D QDs, we examine how confinement potentials modify single-particle spectra and enhance Coulomb interactions. Within these nanostructures, we discuss the impact of orbital hybridization and the emergence of Wigner molecules arising from strong Coulomb interactions. The review further considers the properties of confined excitons, the formation of nontrivial spin textures, and signatures of quantum chaos manifested through scarred states. We also address the role of light matter coupling and the formation of exciton polaritons. Finally, we highlight recent advances in engineered quantum architectures, including coupled dot arrays and superconducting hybrid structures, where tunneling induced correlations and Andreev bound states enable access to novel quantum phases.

\subsection{Models}
\label{sub_QDsmodel}

The theoretical modeling of QDs derived from 2D materials requires a framework that bridges the high accuracy of density functional theory with the computational efficiency of continuum $\mathbf{k \cdot p}$ methods\cite{yao2012tmds,2Dmater2015}. Within this hierarchy, first-principles calculations serve as rigorous benchmarks, providing the fidelity necessary to resolve subtle features of the band structure and to validate effective Hamiltonians against experimental measurements\cite{latini2017,emmanuele2020highly,lisi2021observation,riley2014direct}. A substantial body of work has demonstrated that such theoretical approaches accurately capture the electronic states of these materials. Nonetheless, a fundamental distinction emerges in the context of confined systems. While QDs based on two dimensional materials share the general feature of spatial confinement with conventional semiconductor QDs, they are distinctively defined by the relativistic nature of their charge carriers, often described as Dirac fermions\cite{2014TMDsqdqubit,ding2019defectqd,lxinterlayer2025,Vadym2024ultrafast}.

The primary challenge in confining Dirac fermions stems from Klein tunneling, where relativistic particles normally incident on a potential barrier transmit with unit probability regardless of barrier height \cite{tapashl2010review}. This obstacle is naturally absent in monolayer TMDs due to their substantial intrinsic bandgap, which suppresses tunneling and enables effective confinement via smooth electrostatic potentials \cite{WehlingDirac}. In contrast, for gapless graphene, theoretical descriptions often invoke the infinite mass boundary condition to enforce localization \cite{berrymassless1987}.

Circularly symmetric potentials are predominantly employed in modeling such confinement as they preserve the conservation of total angular momentum. Even when rotational symmetry is broken, the electronic states remain accessible through perturbation theory. While the detailed quantization characteristics depend on the geometry, lateral size, depth, and smoothness of the potential well, the infinite mass approximation provides a rigorous analytical baseline \cite{liu2014intervalley}. Within this model, the confinement potential $V(r)$ which vanishes within a disk of radius $R$ and diverges outside this region is defined as
\begin{equation} \label{hardwall}
	V(r) = 
	\begin{cases} 
		0 & \text{for } r < R, \\
		\infty & \text{for } r > R,
	\end{cases}
\end{equation}
where $R$ denotes the quantum dot radius. A key advantage of this framework is its exact analytical solvability. 
However, such a hard-wall potential is generally unsuitable for Dirac fermions, as imposing a vanishing wavefunction at the boundary can violate current conservation. The confinement of Dirac fermions in a QD can instead be modeled using the infinite-mass boundary condition, where the mass term diverges outside the dot and prevents particles from escaping. At the boundary ($r=R$), the spinor components satisfy
\begin{equation} \label{infinitymassbc}
\chi_2(R) = i\tau \chi_1(R),
\end{equation}
where $\chi_{1,2}$ are the spinor components and $\tau=\pm1$ is the valley index \cite{berrymassless1987,grujic2011electronic,paananen2011finite,paananen2011signatures,raca2017excitonic}. This condition ensures the vanishing of the normal probability current at the boundary and leads to a discrete energy spectrum in the QD.
Physically, the infinite mass term acts as a scalar potential that anticommutes with the kinetic energy operator. 

\begin{figure}[htp]
	\centering
	\includegraphics[width=\linewidth]{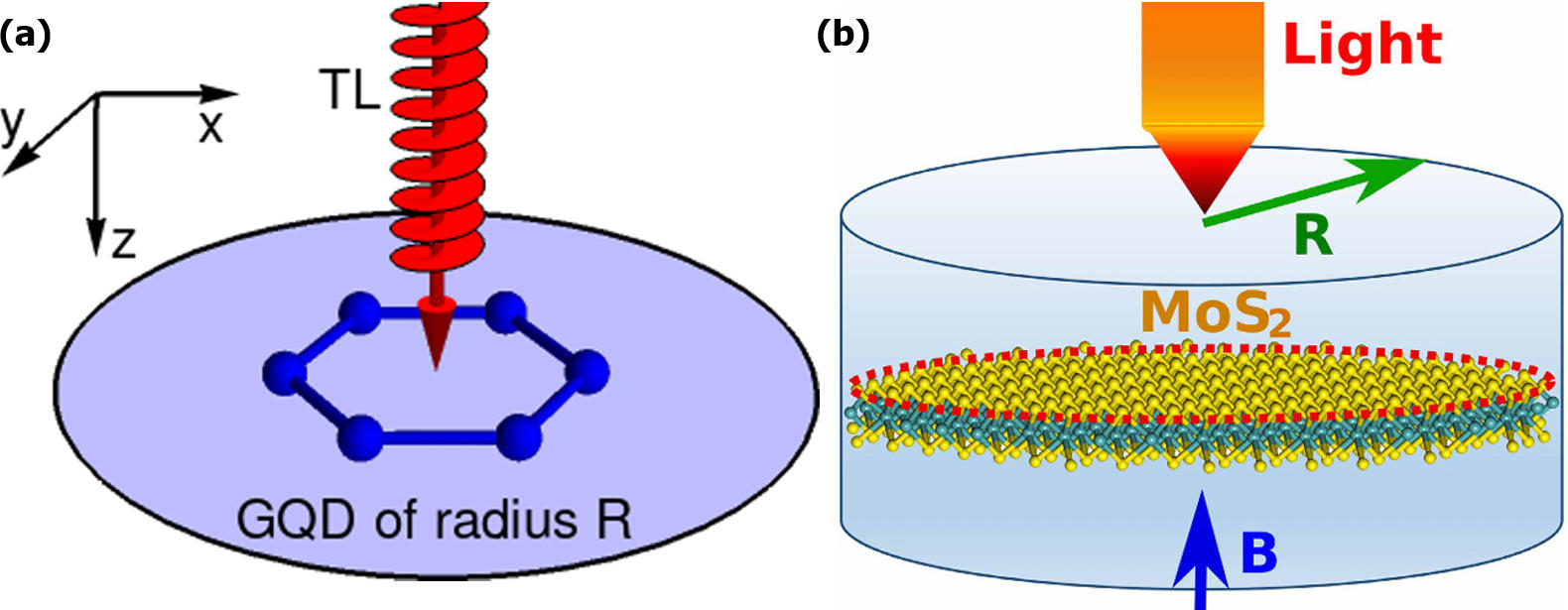}
	\caption{Schematic of different QDs structure. (a) A circular QD defined in monolayer graphene by a radial confinement potential of radius $R$. 
		(b)Monolayer \ce{MoS2} circular QD of radius $R$ (indicated by the red circle) and a perpendicular magnetic field $B$ is applied to the \ce{MoS2} plane.
		Adapted from Ref.~\cite{grapheneqds,qu2017tunable}}
	\label{syQD}
\end{figure}

The single-particle dynamics near the $K$ ($K'$) valley, indexed by $\tau = \pm 1$, are governed by an effective Dirac Hamiltonian of the form
\begin{equation}
	H = H_D + V(r) \sigma_z,
\end{equation}
where $H_D$ is the low-energy effective Dirac Hamiltonian of the specific 2D material and $\sigma_z$ is the Pauli matrix acting on the sublattice pseudospin space. The term $V(r)\sigma_z$ introduces a local staggered potential that breaks sublattice symmetry, effectively opening a spatially divergent band gap outside $r = R$ and thereby ensuring strict confinement of the Dirac fermions.
The 2D materials QDs system is displayed in Fig.~\ref{syQD}. 
The confinement discretizes the single-particle energy spectrum, converting the continuous bulk band structure of the extended material into the quantized level structure of a QD. The resulting eigenstates are labeled by a conserved total angular momentum quantum number and a radial node index, reflecting the cylindrical symmetry of the confinement. This quantization defines the electronic shell structure within the dot and establishes the theoretical foundation for analyzing excitonic and many-body correlations in these confined systems.

QDs based on monolayer, bilayer, trilayer, and even multilayer 2D materials have attracted substantial research interest due to their tunable optoelectronic properties and established experimental feasibility \cite{2017trilayerqd,hecker2023coherent,mirzakhani2020circular,tiutiunnyk2019opto}. In van der Waals heterostructures, interlayer coupling introduces an additional degree of freedom, enabling modulation of both the electronic structure and the confinement potential landscape in ways inaccessible in monolayer systems. Consequently, understanding the interplay between interlayer coupling and external confinement is crucial for accurately describing the electronic properties of multilayer QDs. The spatially electrostatic potential defining the confinement is determined by the Poisson equation. Finite element methods, often implemented in multiphysics platforms such as COMSOL, are widely employed for their ability to handle complex device geometries \cite{tapash2013ee,2016trilayerPeeters}. Alternatively, real-space discretization on a 2D grid offers an efficient approach for systems with regular lattice structures \cite{wang2025spin}. The layer-resolved potential profiles derived from these calculations are subsequently incorporated into effective Hamiltonian, enabling accurate determination of quantized energy levels and associated wavefunctions.

A comprehensive theoretical description of QDs in 2D materials must extend the single-particle picture to include many-body interactions, which are particularly strong in low dimensional nanostructures. The resulting Hamiltonians generally exceed the scope of analytical or perturbative methods, requiring advanced numerical techniques such as configuration interaction or exact diagonalization (ED). The ED scheme for QD was first described in detail in Ref.~\cite{Tapash2006}, and has since become a standard tool in many-body calculations~\cite{lxinterlayer2025,zeng2022strong,yannouleas2023quantum}.These methods provide a fully nonperturbative treatment of Coulomb correlations and are essential for accurately resolving the internal structure of few electron states.

In bulk semiconductors, electrostatic interactions are efficiently screened by the surrounding dielectric medium, whereas in two-dimensional systems spatial confinement greatly enhances interaction effects. Reduced dielectric screening in atomically thin materials fundamentally alters the nature of Coulomb interactions which can reach energy scales comparable to single-particle band energies, giving rise to strongly bound excitons, trions, and other correlated states. 
In van der Waals heterostructures, charges located in different layers interact through both intralayer and interlayer Coulomb potentials, with the relative strength controlled by the dielectric environment and the interlayer separation $d$. The interlayer Coulomb interaction can be written as
\begin{equation}
	V_{\mathrm{inter}}(\mathbf r_1,\mathbf r_2)
	= -\frac{e^2}{\varepsilon \sqrt{\left|\mathbf r_1-\mathbf r_2\right|^2 + d^2}},
\end{equation}
whereas for carriers residing within the same layer, the intralayer Coulomb interaction takes the form
\begin{equation}
	V_{\mathrm{intra}}(\mathbf r_1,\mathbf r_2)
	= \frac{e^2}{\varepsilon \left|\mathbf r_1-\mathbf r_2\right|}.
\end{equation}
This enhancement supports the formation of tightly bound excitons and long-lived interlayer excitonic states. When carriers are further confined within a QDs, the strong spatial localization leads to exchange interactions between electrons residing in different valleys. These exchange interactions become crucial in determining the fine structure of few particle states such as multiexciton complexes and trions \cite{lxinterlayer2025}, giving rise to distinct energy splittings and optical signatures. The resulting spectrum, which can be directly measured through Coulomb blockade spectroscopy, provides key information for understanding correlated states in 2D materials and their nanostructures \cite{brotons2019coulomb}.

These interactions significantly affect the optical response of the system. In QDs, optical absorption is governed by dipole transitions between discrete energy levels.Far-infrared (FIR) magneto-optical absorption spectroscopy is widely used to probe the optical properties of materials. In QDs, the optical selection rules are primarily governed by spatial confinement and the geometric characteristics of the system. The confinement converts the energy bands into discrete energy levels similar to atomic systems, making QDs artificial atoms~\cite{tapashqdee,Halonen_1996}. This strong confinement enhances the role of angular momentum $J_{\mathrm{tot}}$ and leads to unique light-matter interaction properties\cite{qu2017tunable,dias2016robust,mitra2024ultrafast_valley,mitra2024ultrafast_hhg}.

The radius of QDs considered here is typically of the order of tens of nanometers, which is much smaller than the wavelength of far-infrared radiation corresponding to the relevant energy scales. Therefore, the optical absorption can be described within the dipole approximation\cite{avetisyan2012strong,avetisyan2013superintense,chakraborty2018controllable}. The incident light is assumed to be perpendicular to the plane of TMDs. The dipole transition matrix element between states $|k\rangle$ and $|i\rangle$ is defined as $D_{ik}=\langle i|\boldsymbol{\varepsilon}\cdot\mathbf{r}|k\rangle$ where $\boldsymbol{\varepsilon}$ is the polarization vector lying in the $xOy$ plane.

For unpolarized light, the single-particle dipole transition matrix element is
\begin{equation}
	D_{ik}=\pi(\Omega_x+i\Omega_y)
	\int r^2 dr(\chi_{1i}^*\chi_{1k}+\chi_{2i}^*\chi_{2k})
\end{equation}
where $\Omega_{x,y}=\delta_{m_k+1,m_i}\pm\delta_{m_k-1,m_i}$, $m$ is the orbital angular momentum quantum number, and $\chi_{1,2}$ are the spinor components corresponding to two sublattices. According to the selection rule, dipole-allowed transitions satisfy $\Delta m = m_i - m_k = \pm1$. This rule applies to intra-exciton transitions between exciton states. The transition between an exciton state and the vacuum, such as exciton recombination, is not considered here. The dipole transition amplitude between the exciton states is
\begin{equation}
	A_{k\rightarrow i}=
	\sum_{i,k} C_i^* C_k
	\langle i_e,i_h|r_h-r_e|k_e,k_h\rangle
	\label{lightabs}
\end{equation}
and the absorption intensity is proportional to $|A_{k\rightarrow i}|^2$, where $C_i$ and $C_k$ are the expansion coefficients of the corresponding many-body states in the noninteracting basis. Only transitions from the ground state to excited states are considered.

In twisted bilayer TMDs, a small twist angle or lattice mismatch produces a long-wavelength moir\'e superlattice~\cite{berashevich2011nature,tapashmoire,angeli2021gamma}. Within each moir\'e unit cell, the local atomic registry varies smoothly, modulating the interlayer separation and stacking order. These spatial modulations act as an effective periodic potential $\Delta(\mathbf r)$ for carriers and excitons, giving rise to minibands, band-edge modulation, and shifts of excitonic energies~\cite{2023fuliang}. The single-particle dynamics in such a landscape can be captured by the effective Hamiltonian
This emergent periodic potential provides a pristine, highly tunable platform for exploring spatially quantized electronic states, enhanced many-body interactions, and correlated quantum phenomena in 2D materials \cite{2023fuliang}. The single-particle dynamics in such a landscape can be captured by the effective Hamiltonian
\begin{equation}
	H_{\text{moir\'e}} = \frac{\mathbf{p}^2}{2m^*} + \Delta(\mathbf{r}),
\end{equation}
where $m^*$ denotes the effective mass of the charge carrier. In the limit of a sufficiently deep and parabolic potential minimum, the quadratic term dominates, and the low-energy physics is well approximated by a 2D harmonic oscillator
\begin{equation}
	H_s \approx \frac{\mathbf{p}^2}{2m^*} + \frac{1}{2} \kappa r^2.
\end{equation}
Here, $\kappa$ represents the effective curvature of the moir\'e potential at its minimum.

This harmonic approximation reveals that the moir\'e superlattice inherently hosts a periodic array of nearly identical QDs. Within each QD, the low-energy spectrum consists of approximately equidistant levels, $E_n = \hbar \omega (n + 1)$. Under an applied perpendicular magnetic field, the confinement problem maps onto the canonical Fock-Darwin problem \cite{reviewChakraborty1992}. The magnetic field introduces a competing length scale set by the cyclotron frequency, which reorganizes the single-particle spectrum. 
The resulting eigenstates, known as Fock-Darwin states and labeled by radial and angular momentum quantum numbers, were first systematically illustrated in Ref.~\cite{reviewChakraborty1992}, and exhibit a characteristic evolution with the magnetic field~\cite{tapsh1999129}. Their level structure smoothly interpolates between discrete harmonic oscillator shells and degenerate spectra resembling Landau levels in the strong magnetic field regime. Consequently, the magnetic field provides a powerful tuning knob for controlling shell splitting, level crossings, and the intricate interplay among spin, valley, and many-body interaction effects in these confined moir\'e systems \cite{luo2023artificial,sanchez2025chiral,luo2025solving}.

\added{This QD-array picture can be recast as an effective lattice model, where intersite tunneling sets the hopping amplitude and local Coulomb or dipolar repulsion sets the local interaction scale. Twist angle, displacement field, dielectric environment, and carrier density then tune the filling, interaction strength, and lattice symmetry. In this form, moir\'e materials provide suitable platforms for fermionic carriers and bosonic excitons, enabling Mott-like localization, Wigner crystallization, magnetic ordering, and excitonic Hubbard physics~\cite{park2023dipole_ladders,pasek2023magnetic_moire_qd_arrays}.}

\added{Confinement in 2D Dirac materials is not restricted to quantum-dot geometries. One-dimensional confinement in nanoribbons provides a closely related and experimentally relevant platform. Graphene nanoribbons offer a direct example of boundary-controlled confinement: armchair ribbons exhibit width-dependent gaps, whereas zigzag edges support localized states near the Fermi level~\cite{nakada1996edge,brey2006graphene_nanoribbons}. A staggered sublattice potential acts as a Dirac mass and can localize carriers when it varies spatially. If the mass term changes sign across an interface, valley-dependent boundary modes may appear, linking real-space confinement to Berry curvature, valley-Hall physics, and topological edge states~\cite{yao2009edge_states,martin2008topological_confinement}.}

\added{Nanoribbon confinement has also been theoretically explored in TMDs, where edge termination and electron interactions generate physics distinct from graphene. Tight-binding and many-body calculations show that zigzag TMD nanoribbons can host interaction-sensitive metallic edge channels and filling-dependent edge magnetism\cite{Brito2022}, whereas geometry-engineered TMD nanoribbon heterojunctions can combine metallic zigzag edges with semiconducting armchair segments to realize tunneling and resonant-tunneling transport\cite{jiang2025}. These results indicate that TMD nanoribbons provide a one-dimensional confined platform in which edge states, correlations, and device functionality can be jointly controlled.}

\added{These idealized boundary and mass-domain pictures are further modified in real nanostructures, where electrostatic gating, interfacial potentials, local strain, edge roughness, dielectric inhomogeneity, and proximity effects can generate smooth or emergent confinement landscapes. For instance, interfacial QDs have been created and tuned in graphene/TMD heterostructures by STM-tip-induced patterning of interfacial TMD nanostructures, which locally introduce electrostatic confinement potentials~\cite{ren2024interfacial_qds}. Such nonideal confinement can modify level spacings, lift valley degeneracies, enhance intervalley scattering, and reshape the Coulomb matrix elements.}

\subsection{Single-Particle Energy Level Structure in QDs}

Determining the single-particle energy spectrum is a fundamental step toward understanding the optical, transport, and quantum information properties of 2D nanostructures\cite{lxinterlayer2025,Vadym2024ultrafast}. The single electron spectrum in QDs of 2D materials is governed by the interplay between Dirac like band dispersion and quantum confinement. In graphene QDs, confinement affects massless Dirac fermions, producing spectra that are strongly influenced by boundary conditions\cite{chen2007fock}. By contrast, TMD QDs host carriers with a finite band gap and valley dependent effective masses. This qualitative distinction significantly modifies the structure of the eigenstates and the arrangement of discrete energy levels, making TMD QDs more similar to conventional semiconductor dots while still preserving essential relativistic and valley dependent features.

\begin{figure}[htp]
	\centering
	\includegraphics[width=\linewidth]{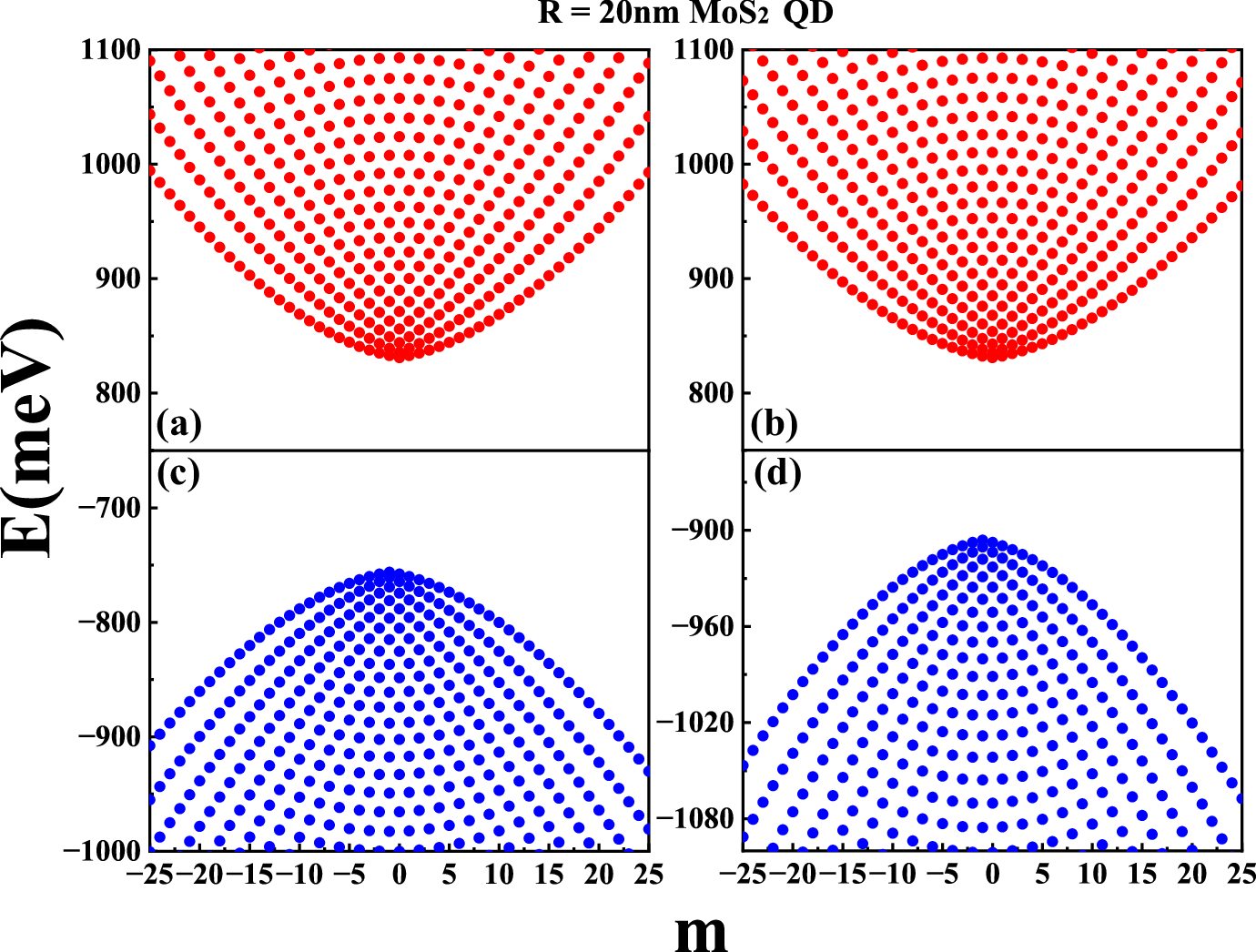}
	\caption{
		Low-lying energy spectra of a \text{MoS2} QD with radius $R = 20\text{nm}$ at zero magnetic field ($B = 0\text{T}$). The red and blue dots represent the electron and hole states, respectively. The left panels correspond to spin-up ($s = +1$) states, while the right panels correspond to spin-down ($s = -1$) states.
	}
	\label{energylevelzero}
\end{figure}

\begin{figure}[htp]
	\centering
	\includegraphics[width=\linewidth]{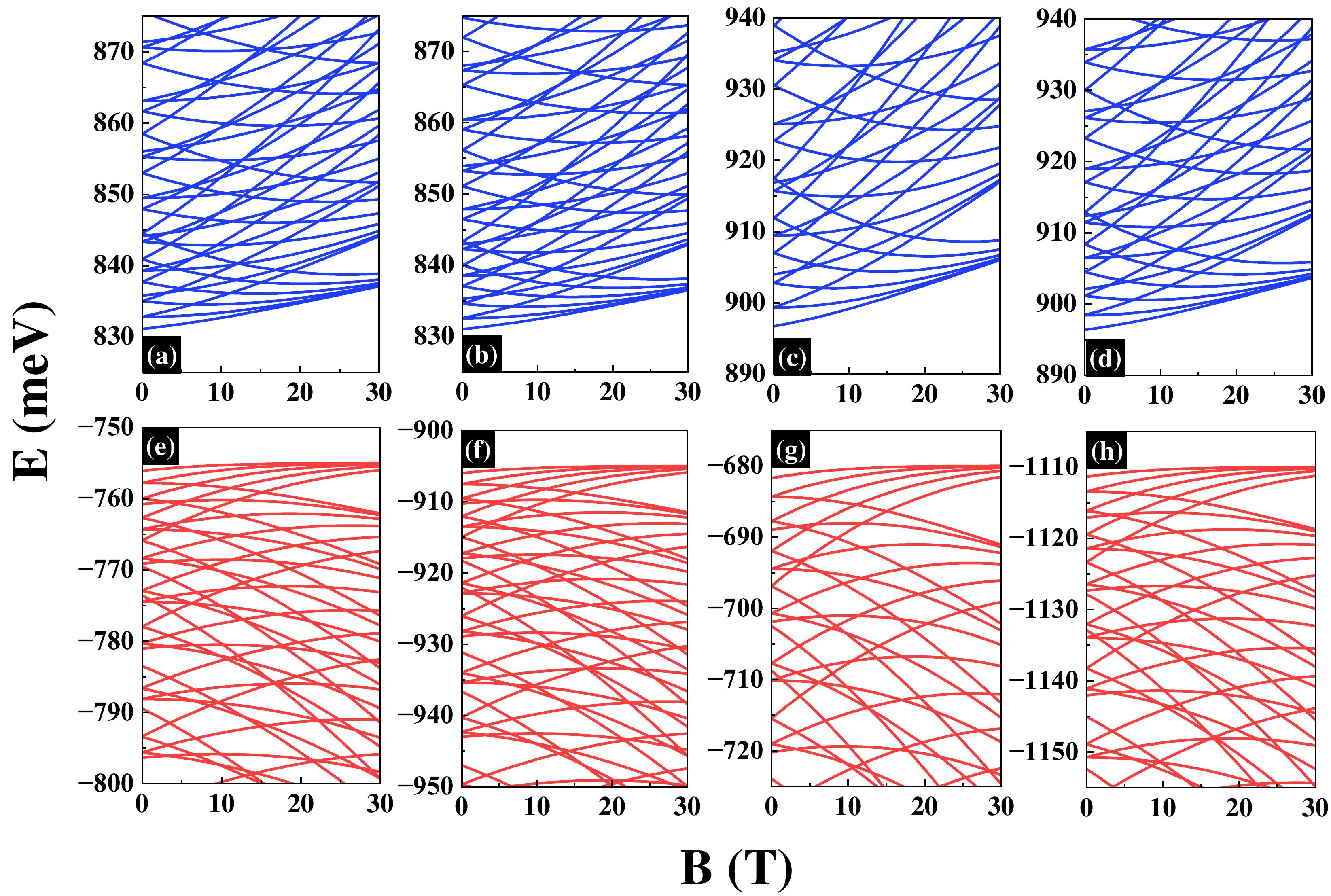}
	\caption{
		The low-lying energy spectra of a single electron and a single hole in TMD QDs with radius $R = 20$ nm vary with perpendicular magnetic fields. For simplicity, only the first five levels of a principal quantum number are displayed. 
		The energy spectra are provided for different TMDs with different spins: (a) \ce{MoS2}, the conduction band with spin $ s = 1 $; (b) \ce{MoS2}, conduction band with $ s = -1 $; (c) \ce{WS2} conduction band with $ s = 1 $; (d) \ce{WS2} conduction band with $ s = -1 $; (e) \ce{MoS2}, valence band with $ s = 1$ (f) \ce{MoS2}, valence band with $ s = -1 $ which is much  {farther} away from the Fermi surface; (g) \ce{WS2}, valence band with $ s = 1 $; (h) \ce{WS2}, valence band with $ s = -1 $. The electron and hole states are respectively indicated in blue and red.
	}
	\label{energy}
\end{figure}

In the absence of an external magnetic field, the circular symmetry of an ideal TMD QD guarantees the conservation of angular momentum. The single electron Hamiltonian commutes with the total angular momentum operator $J_z = L_z + \frac{\tau}{2} \sigma_z$, where $L_z$ is the orbital angular momentum, $\sigma_z$ acts on the sublattice pseudospin space, and $\tau = \pm 1$ denotes the valley index. Since the spin projection $s_z$ is also conserved, the eigenstates are fully characterized by the set of quantum numbers $(n, j, \tau, s_z)$, with $n$ being the radial quantum number and $j$ the total angular momentum. The low-lying energy levels of single-particle states in
The energy spectra of TMD QDs in the absence and presence of a magnetic field are shown in Fig.~\ref{energylevelzero} and Fig.~\ref{energy}, respectively. This classification naturally describes the electronic shell structure of QDs and the symmetry imposed degeneracies of their energy spectrum. When a homogeneous perpendicular magnetic field is applied, the rotational symmetry of the confinement is preserved, but the kinetic momentum couples to the vector potential, introducing a new magnetic length scale. Consequently, quantization from both spatial confinement and the magnetic field coexists, yielding a generalized Fock-Darwin spectrum for massive Dirac fermions. The magnetic field lifts residual degeneracies between states of opposite angular momentum and systematically reorganizes the energy levels. In the weak field limit, the spectrum is dominated by confinement effects, with magnetic corrections appearing as small energy shifts. At stronger fields, the discrete QD levels progressively evolve towards a Landau level structure \cite{keren2020quantum}.

The dependence of the energy spectrum on magnetic field in TMDs QDs reveals their distinct character as systems of massive Dirac fermions. At high fields, the spectrum approaches a set of gapped Landau levels, a direct consequence of the finite band gap and strong SOC that break valley degeneracy. This behavior contrasts sharply with the gapless, $\sqrt{B}$ scaling Landau levels of massless Dirac fermions in graphene \cite{chen2007fock}. The resulting single-particle spectrum underpins a variety of correlated phenomena. The shell structure and quantum number labeling govern optical selection rules and valley dependent transitions, while the magnetic field tunable level crossings and degeneracies provide essential control for investigating few-electron states and excitonic complexes.

\subsection{Correlated Electronic States in Quantum Dots}

A comprehensive understanding of QDs in 2D materials begins with the determination of their single-particle energy spectrum. The organization of these single-particle states, particularly the orbital hybridization between them, serves as the essential framework upon which the correlated many-body physics is built. At the atomic scale, isolated orbitals such as $s$ and $p$ possess well-defined angular momentum under isotropic potentials \cite{2Dmater2015,yao2012tmds}. The formation of a 2D crystal breaks this spherical symmetry through the periodic lattice potential, inducing atomic scale orbital hybridization into Bloch states adapted to the crystal symmetry. This process, analogous to $sp^2$ hybridization in graphene, defines the low-energy band structure and the effective quasiparticles of the extended material. Confinement into a QD introduces mesoscale orbital hybridization, governed by a distinct symmetry breaking mechanism. Here, the geometry of the confinement potential, rather than the atomic lattice, couples bound states of different orbital angular momenta such as $s$ and $d$ states, forming hybridized linear combinations of the original QD orbitals \cite{mao2025orbital,zhou2025recent}. These geometry tunable hybrid orbitals establish the single-particle level structure that ultimately dictates the arrangement of electrons, excitons, and correlated states within the dot.

\begin{figure*}[htp]
	\centering
	\includegraphics[width=\linewidth]{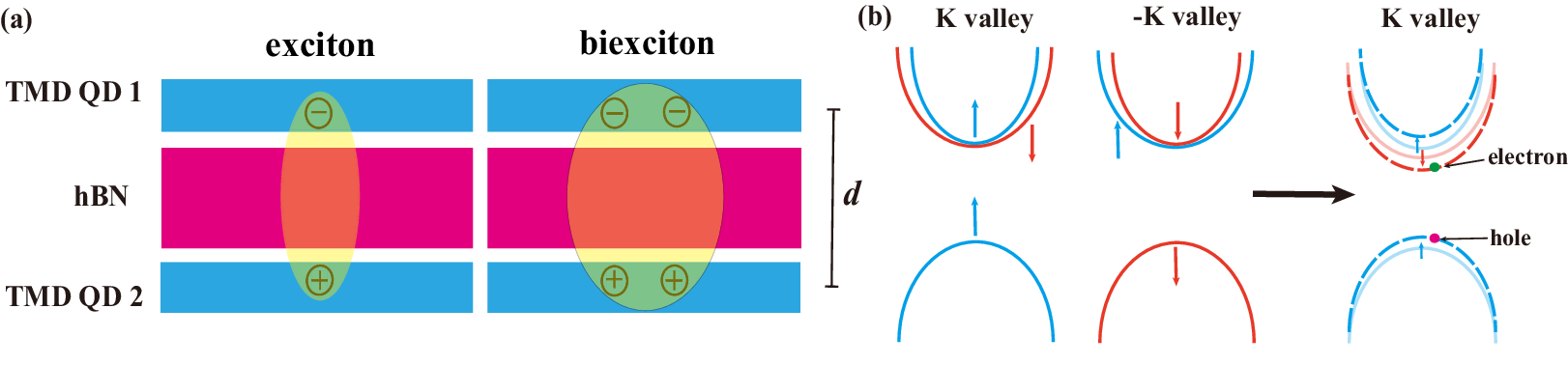}
	\caption{The schematic double layer system and bands structure of TMDs near Fermi surface. (a) Electron confined in a TMD layer, hole in another TMD layer. (b) The lower bands in valence band are neglected since they are far away from the Fermi surface. Only one valley, say valley $K$, is considered. The right panel indicates the bands shift by Zeeman coupling in a magnetic field. Electrons and holes are marked in different bands, meanwhile in different layers. }
	\label{band}
\end{figure*}

In QDs formed in 2D systems, spatial confinement fundamentally reshapes the electronic spectrum by converting extended states into a discrete set of quantized levels. This discretization provides a natural basis for constructing few-particle eigenstates, in which the electron-electron and electron-hole interactions play an essential role. As a consequence, the low-energy spectrum is no longer described by independent particles, but instead by a hierarchy of correlated excitations, including neutral excitons ($X^0$), charged trions ($X^\pm$), and biexcitons ($XX$)\cite{zeng2022strong,latini2017,lxinterlayer2025,kim2024confinement}. The double-layer structure and band selection scenario is
schematically illustrated in Fig. \ref{band}. The valley is supposed to
be polarized and electrons and holes are from different layers. 

To describe these correlated states in a systematic way, it is convenient to adopt the framework of second quantization. The electronic degrees of freedom in each layer are represented by field operators expanded in a suitable single-particle basis,
\begin{equation}
	\hat{\Phi}(\mathbf r)
	=
	\sum_{\alpha}
	\phi_{\alpha}(\mathbf r)\,
	\hat{c}_{\alpha},
\end{equation}
where the composite index $\alpha$ stands for the relevant orbital, spin, valley, and layer quantum numbers. Within this representation, Coulomb interactions are expressed in terms of the density operators constructed from the field operators, leading to a general many-body Hamiltonian of the form 
\begin{equation} 
	\hat{H}_{\mathrm{int}}
	=
	\frac{1}{2}
	\sum_{\alpha\beta\gamma\delta}
	V_{\alpha\beta\gamma\delta}
	\,
	\hat{c}^\dagger_\alpha
	\hat{c}^\dagger_\beta
	\hat{c}_\gamma
	\hat{c}_\delta,
\end{equation}
where $\alpha,\beta,\gamma,\delta$ label the discrete single-particle quantum numbers, and the Coulomb interaction matrix elements $V_{\alpha\beta\gamma\delta}$ are nonzero only when the corresponding quantum numbers satisfy the required conservation laws. In multilayer systems, these matrix elements naturally separate into intralayer and interlayer contributions, reflecting the spatial structure of the heterostructure. This formalism provides the microscopic foundation for understanding the 
emergence of excitons, trions, and multiexciton complexes in confined 2D systems.

The combined effects of SOC and electron-hole exchange produce a fine structure within the neutral exciton manifold, generating distinct bright and dark exciton states \cite{jiang2021interlayer}. Bright excitons arise from spin-allowed configurations that couple efficiently to light. In contrast, dark excitons originate from either spin-forbidden configurations or from possessing a finite momentum of the center of mass outside the light cone, rendering them optically inactive within the electric dipole approximation. In tungsten based 2D materials, both theory and experiment consistently place the dark exciton below the bright exciton in energy, leading to strongly suppressed photoluminescence at low temperatures. Although dark excitons do not emit light directly, their presence can be revealed through mixing induced by a magnetic field, recombination assisted by phonons, or hybridization due to quantum confinement \cite{robert2020measurement,kapuscinski2021rydberg}. Owing to their strongly suppressed radiative decay, dark excitons exhibit markedly extended lifetimes, establishing them as a key component of long lived excitonic dynamics in 2D quantum confined systems.

A distinct class of excitonic states emerges in van der Waals heterostructures formed by vertically stacked 2D materials. Due to the type-\textrm{II} band alignment, the lowest-energy excitation involves an electron and a hole localized in different layers, forming an interlayer exciton\cite{jiang2021interlayer}. The spatial separation between the charge carriers strongly suppresses their wavefunction overlap, resulting in radiative lifetimes that are orders of magnitude longer than those of intralayer excitons, as confirmed by time-resolved optical spectroscopy. In addition, interlayer excitons possess a permanent out-of-plane electric dipole moment, which makes their emission energy highly sensitive to vertical electric fields through the quantum confined Stark effect\cite{rivera2015observation}. Recent experiments have demonstrated that the emission energy of interlayer excitons can be widely tuned by applying an external electric field, providing a direct probe of the exciton dipole moment\cite{li2020dipolar,feng2024highly}. The formation of moir\'e superlattices in such heterostructures creates a periodic potential that traps interlayer excitons at specific stacking sites. This yields ordered arrays of localized quantum emitters which exhibit narrow optical linewidths and widely tunable dipolar interactions.

When such interlayer excitons are further confined into QDs, the confinement localizes the electron and hole at discrete sites, forming atom like excitonic states. Confinement enhances the localization of charge carriers, modifies the wavefunction overlap, and can increase the exciton binding energy. The resulting interlayer excitons exhibit tunable dipolar interactions, and extended radiative lifetimes, combining the long lived and electric field sensitive character of extended interlayer excitons with the discrete energy spectrum imposed by quantum confinement\cite{lxinterlayer2025}. These features make QDs with interlayer excitons a unique platform for exploring strongly localized excitonic physics in 2D materials.

In multielectron QDs based in 2D materials, electron-electron interactions dominate over single-particle confinement, leading to a profound restructuring of the many-body wavefunction. When the interaction energy becomes comparable to or larger than the spacing between confined levels, electrons localize in space, forming strongly correlated configurations known as Wigner molecules. In these states, electrons arrange themselves to maximize mutual separation, producing spatially ordered patterns that depart from the conventional delocalized orbital picture. ED and configuration interaction calculations have shown that as the strength of correlations increases, the charge density evolves from shell-like distributions to clearly localized arrangements, providing a microscopic description of the internal structure of these artificial molecules\cite{li2024wigner}.

\begin{figure}[ht]
	\centering
	\includegraphics[width=\linewidth]{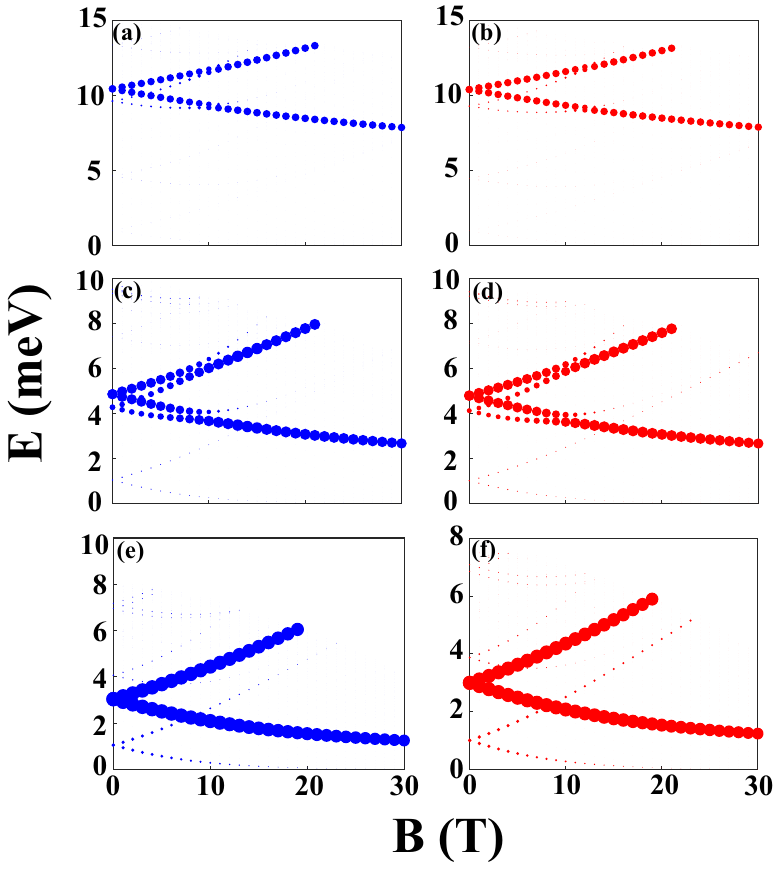}
	\caption{Dipole-allowed optical absorption spectra of \ce{MoS2}/\ce{WS2} double-layer QDs for different interlayer distances. The left panels correspond to spin-up electrons, while the right panels correspond to spin-down electrons. Panels (a) and (b) show the results for an interlayer distance $d=5\mathrm{nm}$, (c) and (d) for $d=10\mathrm{nm}$, and (e) and (f) for $d=15\mathrm{nm}$. The radii of the electron and hole layers are $R_e=30\mathrm{nm}$ and $R_h=20\mathrm{nm}$, respectively. The size of the symbols is proportional to the absorption intensity.}
	\label{lightabsfig}
\end{figure}

The optical absorption of interlayer excitons confined in QDs can be evaluated using Eq.~(\ref{lightabs}). Figure~\ref{lightabsfig} presents the dipole-allowed absorption spectra for different interlayer distances under varying magnetic fields. Coulomb interactions modify the optical transitions. While several modes satisfy the dipole selection rules in the noninteracting limit, electron-hole interactions suppress some of these transitions. Systems with smaller interlayer separations tend to favor higher-energy modes, whereas the transition intensity generally increases with increasing interlayer distance. 

Wigner molecules exhibit crystallized geometric arrangements, such as polygonal rings or nested shells, that reflect the symmetry of the external confinement\cite{yannouleas2023quantum}. Although the total wavefunction retains rotational invariance in the laboratory frame, the internal charge distribution behaves like a rigid molecule, giving rise to collective rotational and vibrational excitations distinct from single-particle Fock-Darwin transitions. Recent experiments in high quality 2D systems and moir\'e superlattices have observed signatures consistent with Wigner localization, including discrete low-energy modes and spatially ordered electron distributions, demonstrating that QDs in 2D materials provide a versatile platform for exploring strongly correlated electron physics\cite{goldberg2024electronic,yannouleas2024crystal}.

\subsection{Pseudspin topology and Quantum Chaos} 

\begin{figure}[htb]
	\centering
	\includegraphics[width=8.40cm]{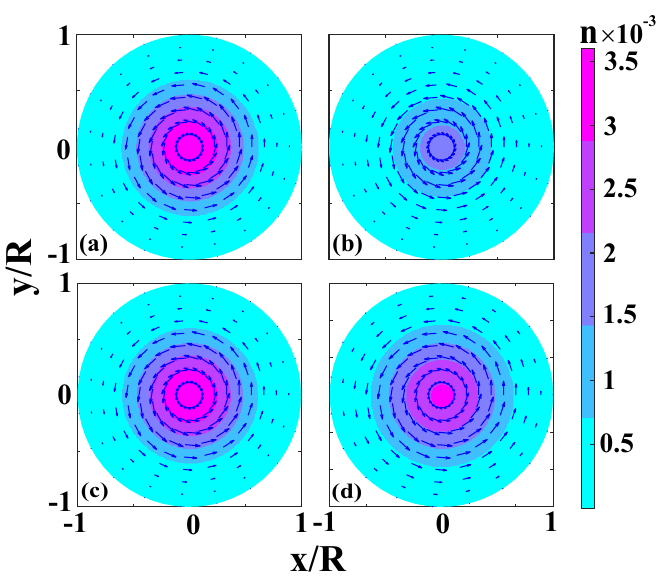}
	\caption{Ground state pseudo-spin fields $(\sigma^{\alpha}_{x},\sigma^{\alpha}_{y}) $ of interlayer excitons with electron and hole spin-up at magnetic field $ B = 1$ T. (a) and (c) The pseudospin textures of the electron and the hole in interlayer exciton in \ce{MoS2}/\ce{MoS2} double-layer QDs with radii $ R_{e} = R_{h}= 20\text{nm}$, respectively. The distance between the two layers is $5\text{nm}$. (b) and (d) The pseudospin textures of the electron and the hole in interlayer exciton in \ce{MoS2}/\ce{WS2} double-layer QDs with $ R_{h} = 20\text{nm} $ and $R_{e} = 30\text{nm} $. The distance between the two layers is $10\text{nm}$.}
	\label{spinture}
\end{figure}

The strong SOC in 2D material QDs generates rich, spatially varying spin textures, where the orientation of the spin continuously evolves across the electronic wavefunction\cite{luo2019unique,luo2019tuning,lxinterlayer2025}. In Fig. \ref{spinture}, the pseudospins of the electron and hole are textured with nontrivial topological charge $ Q=1 $ in interlayer exciton. The winding number $Q$ characterizes the topological structure of the in-plane spin texture in real space. It is defined as the total change of the azimuthal angle $\phi$ along a closed contour $C$
\begin{equation}
	Q=\frac{1}{2\pi}\oint_C d\phi
	=\frac{1}{2\pi}\oint_C
	\frac{\sigma_x(\mathbf r)\, d\sigma_y(\mathbf r)
		-\sigma_y(\mathbf r)\, d\sigma_x(\mathbf r)}
	{\sigma_x^2(\mathbf r)+\sigma_y^2(\mathbf r)},
\end{equation}
where $\sigma_x(\mathbf r)$ and $\sigma_y(\mathbf r)$ denote the in-plane components of the spin field. This noncollinear configuration emerges from the interplay between the valley dependent band structure, quantum confinement, and the local crystalline environment, producing spin patterns that are not only spatially modulated but also topologically nontrivial. Such textures can be characterized by local winding numbers or Berry curvature distributions, giving rise to topologically protected features in the energy spectrum and in the optical selection rules. The spatial variation of spin orientation across the dot can induce effective spin mixing, modify valley dependent optical transitions, and stabilize long lived spin valley excitations\cite{liu2024unconventional}. These topological spin textures provide a promising platform for applications in quantum technologies, including robust spin valley qubits, optically addressable topological states, and the manipulation of quantum information via light matter interactions in nanoscale 2D systems\cite{xia2023universal}.

Bridging these topologically nontrivial spin textures with nanophotonics, recent studies have shown that the spin and orbital angular momentum of light can interact with electronic states in a manner that mirrors their topological structure\cite{yang2021non,wan2023optical}. Illuminating QDs with circularly polarized light enables deterministic all-optical control through the inverse Faraday effect, where an oscillating electromagnetic field generates an effective, noncontact magnetic field that couples directly to the carrier spins. Unlike conventional thermal or magnetic switching, this optical approach allows ultrafast and coherent manipulation of the electronic spin textures, providing a means to imprint the topological properties of photons onto the QD states on femtosecond timescales. This optomagnetic control offers a versatile platform to explore spin valley physics, topological excitations, and light matter interactions in nanoscale 2D systems\cite{bliokh2015spin}.

\begin{figure}[htp]
	\centering
	\includegraphics[width=\linewidth]{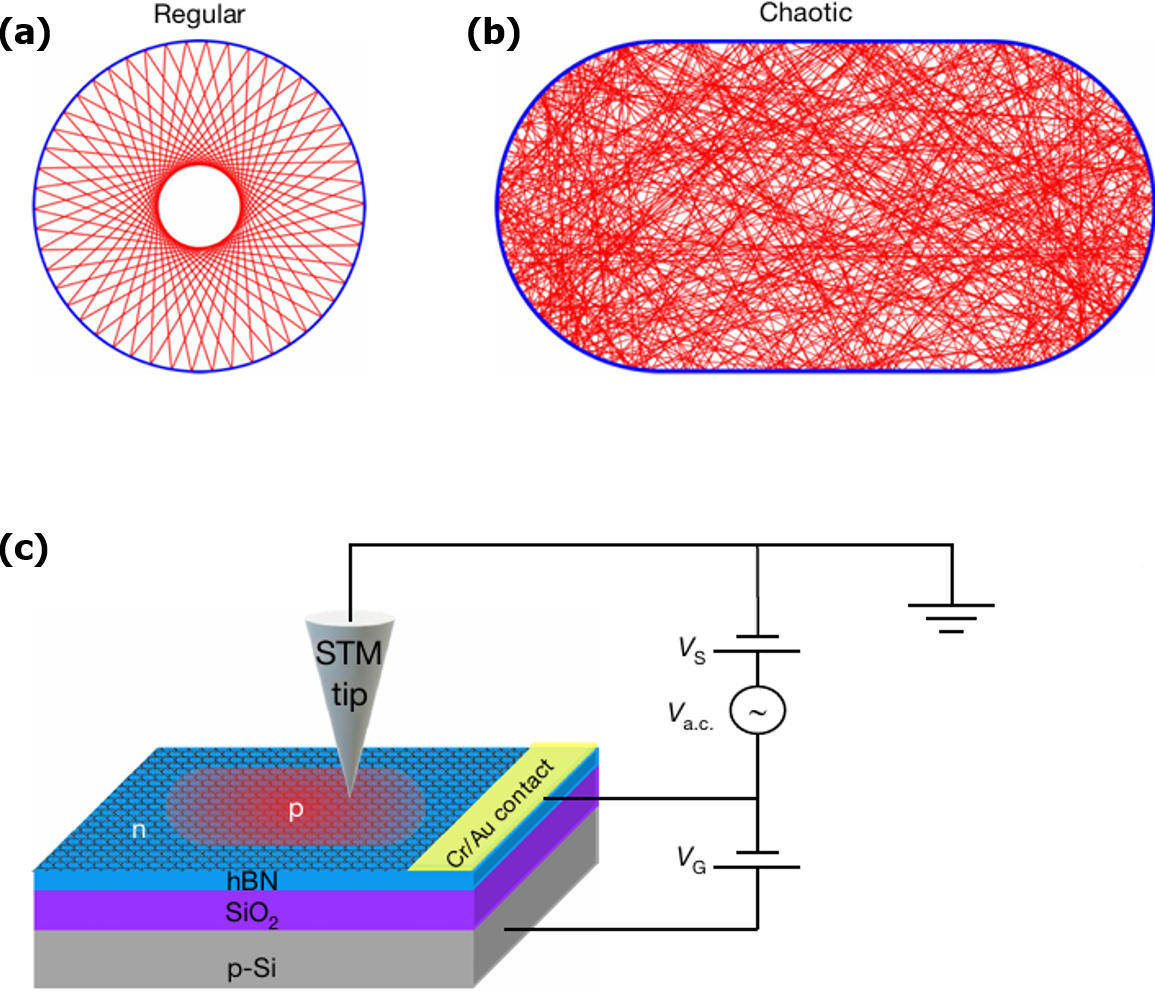}
	\caption{a,b, Classical trajectories inside a circular (a) and stadium-shaped (b) billiard. (c) A stadium-shaped graphene QDs defined by a p-n junction. Adapted from Ref.~\cite{ge2024directscar}.}
	\label{scar}
\end{figure}

Rashba and Dresselhaus SOCs can generate nontrivial spin textures in confined QDs \cite{luo2024spin}. In the classical limit, the combined effect of these SOCs can be mapped onto a nonlinear dynamical system \cite{larson2013chaos,Khomitsky2013,kirichenko2020chaotic}. When the strengths of the two SOCs differ, the corresponding canonical equations may exhibit chaotic dynamics. In anisotropic QDs hosting both Rashba and Dresselhaus SOCs, this classical chaos has been shown to manifest in the quantum regime through the emergence of scar states in the excited spectrum \cite{zhang2024controllable}. Notably, these scar states can be tuned by controlling the Rashba SOC via external gate voltages.

The study of chaotic dynamics in 2D material QDs can be formalized through the framework of Dirac billiards, which represent the relativistic analogues of classical Schrödinger billiards used in conventional semiconductor QDs\cite{ponomarenko2008chaotic,ge2024directscar,bocarsly2025coulombscar}. Dirac billiards are strongly influenced by the linear dispersion and the chirality of Dirac fermions, which modify boundary scattering and give rise to unconventional spectral features. The classical dynamics and chaotic classical dynamics schematized in Fig. \ref{scar}. The density of states and spectral correlations can be understood semiclassically in terms of contributions from classical periodic orbits, with interference effects modulated by the Berry phase acquired upon reflection. Recent theoretical and experimental studies in graphene and TMDs QDs show that these relativistic effects produce deviations from standard random matrix universality, with spectral statistics reflecting the combined influence of SOC, valley degrees of freedom, and geometric phases\cite{keski2019quantum,ge2024directscar}.

In relativistic QDs, these scarred states persist even in the presence of Klein tunneling, which allows partial leakage through the confinement barriers\cite{huang2009relativistic,wang2020quantum}. The robustness of these relativistic quantum scars arises from constructive interference of the spinor components along the classical orbits, producing enhanced local density of states and partial preservation of quantum coherence. Recent experiments in graphene based QDs and photonic analogues of Dirac billiards have observed signatures consistent with such scars, including anomalous transport features and localized wavefunction intensity along predicted trajectories, highlighting the interplay between relativistic dispersion, confinement geometry, and topological phase effects in shaping quantum chaos at the nanoscale.

\subsection{Hybrid Systems in 2D Material Quantum Dots}

\begin{figure}[htp]
	\centering
	\includegraphics[width=\linewidth]{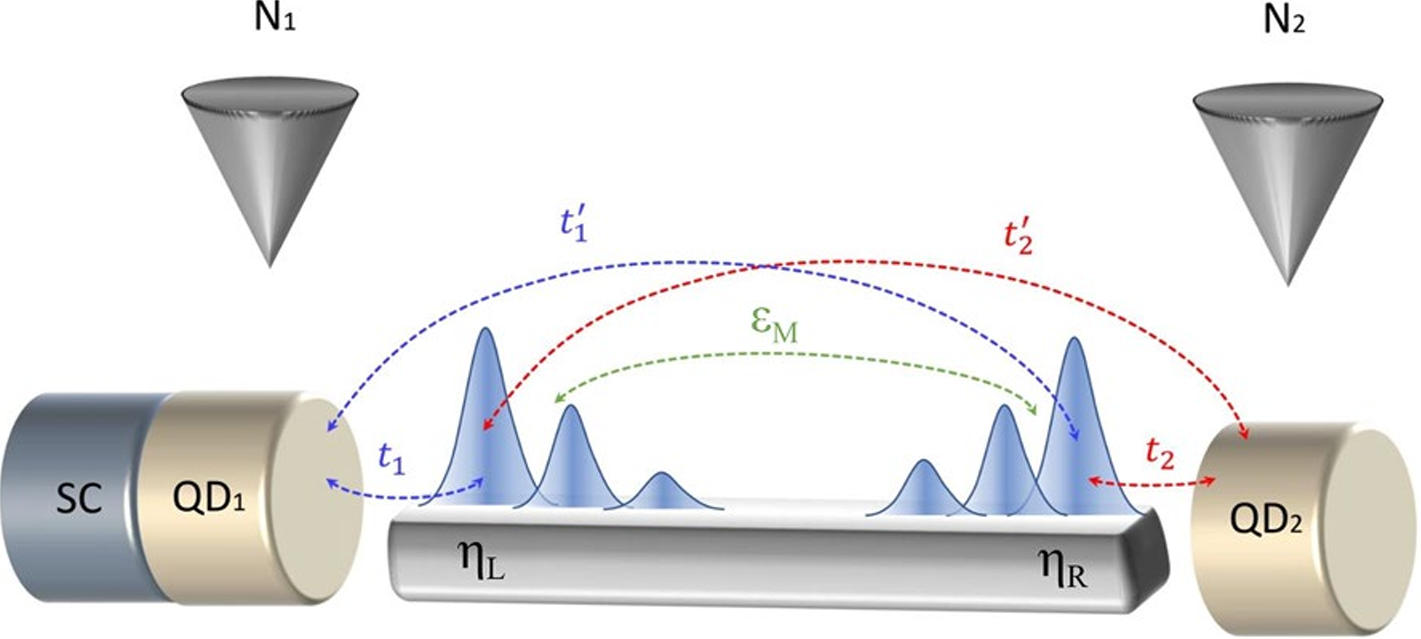}
	\caption{Schematic diagram of the hybrid system consisting of two QDs coupled through a Majorana nanowire. The Majorana bound states $\eta_L$ and $\eta_R$ appear at the ends of the nanowire with an overlap energy $\varepsilon_M$. 
	The QDs are tunnel-coupled to the Majorana modes with amplitudes $t_1, t_2, t'_1, t'_2$. 
	The left dot is connected to a superconducting lead, while the right dot is coupled to a normal lead. 
	Adapted from Ref.~\cite{gorski2024nonlocal}.}
	\label{hybird}
\end{figure}

2D materials QDs provide a versatile platform for hybrid quantum systems, combining strong electronic confinement, SOC, and the ability to couple to photonic or superconducting environments\cite{de2025rydberg,su2017andreev,liu2015strong}. The discrete energy spectrum of these QDs can be further tuned by external magnetic fields, electric gates, or mechanical strain, offering precise control over single-particle and correlated electronic states. Recent works have demonstrated that 2D materials QDs can be integrated with optical cavities, plasmonic resonators, or superconducting contacts, enabling coherent manipulation and spectroscopic access to these confined states\cite{de2025rydberg,gorski2024nonlocal}. The hybrid structure with side-attached QDs was showed in Fig. \ref{hybird}, giving rise to the emergence of molecular states due to nonlocal 
correlations.

Excitons confined in 2D materials QDs can strongly interact with cavity, forming exciton-polaritons\cite{li2021experimental}. At the microscopic level, hybrid light-matter states in quantum-dot cavity systems can be described by the Hamiltonian
\begin{equation}
	\hat H = \hat H_m + \hat H_c + \hat H_{cm}.
\end{equation}
The matter Hamiltonian describing confined electrons and holes is
\begin{equation}
	\begin{aligned}
		\hat H =&
		\sum_{i} \varepsilon^e_i \hat c_i^\dagger \hat c_i
		+
		\sum_{j} \varepsilon^h_j \hat d_j^\dagger \hat d_j \\
		&+
		\hat H_{ee}
		+
		\hat H_{hh}
		+
		\hat H_{eh}.
	\end{aligned}
\end{equation}

\begin{equation}
	\hat H_{ee}
	=
	\frac{1}{2}
	\sum_{ijkl}
	V^{ee}_{ijkl}
	\hat c_i^\dagger
	\hat c_j^\dagger
	\hat c_l
	\hat c_k ,
\end{equation}

\begin{equation}
	\hat H_{hh}
	=
	\frac{1}{2}
	\sum_{ijkl}
	V^{hh}_{ijkl}
	\hat d_i^\dagger
	\hat d_j^\dagger
	\hat d_l
	\hat d_k ,
\end{equation}

\begin{equation}
	\hat H_{eh}
	=
	-
	\sum_{ijkl}
	V^{eh}_{ijkl}
	\hat c_i^\dagger
	\hat d_j^\dagger
	\hat d_l
	\hat c_k .
\end{equation}
The cavity photon mode is described by
\begin{equation}
	\hat H_c = \omega_c \hat a^\dagger \hat a,
\end{equation}
and the light-matter interaction reads
\begin{equation}
	\hat H_{cm} =
	\sum_{ij}
	\left(
	g_{ij}\hat a \hat c_i^\dagger \hat d_j^\dagger
	+
	g_{ij}^*\hat a^\dagger \hat d_j \hat c_i
	\right).
\end{equation}
Here, $i,j,k,l$ are the indices of quantum state in QDs. The operators $\hat c_i$ and $\hat d_j$ annihilate an electron and a hole in the corresponding quantum states with single-particle energies $\varepsilon^e_i$ and $\varepsilon^h_j$, respectively. 
The quantities $V^{\sigma\sigma'}_{ijkl}$ denote Coulomb matrix elements, while $\hat a^\dagger$ ($\hat a$) creates (annihilates) a cavity photon of frequency $\omega_c$. 
The coupling constant $g_{ij}$ characterizes the strength of the cavity-induced interband transition. Interlayer excitons in van der Waals heterostructures, when localized in QDs, exhibit extended lifetimes and emission energies tunable by vertical electric fields. Some studies have demonstrated cavity enhanced emission, electrical control of linewidths, and coherent manipulation of individual excitons, highlighting the possibility of engineering valley-polarized polaritons and exploring correlated excitonic phenomena in hybrid photonic QDs systems\cite{dufferwiel2018valley}.

Coupling QDs based on 2D materials to superconducting reservoirs via the proximity effect can induce hybrid electronic states, while analogous tunable quantum platforms are realized in bosonic gases trapped in optical lattices. Andreev reflection produces discrete Andreev bound states (ABS) in the QDs, whose energies depend on the superconducting phase difference, SOC, and gate tunable chemical potential. In double QD configurations, ABS on neighboring dots hybridize into bonding and anti-bonding Andreev molecules, and Cooper pair splitting can generate spatially separated entangled electrons\cite{su2017andreev}. Extending this concept to linear or 2D arrays of 2D materials QDs allows the engineering of coupled ABS lattices, where the hybridization of states across multiple dots can produce tunable energy bands, collective excitations, and topologically nontrivial modes. Under appropriate SOC and magnetic field conditions, these extended QD arrays can access topological regimes in which the lowest energy ABS evolve into Majorana zero modes, providing a scalable and controllable platform for studying non-Abelian quasiparticles and implementing networked quantum circuits\cite{deng2016majorana,sarma2015majorana}.

The combination of optical and superconducting hybridization with strong Coulomb interactions enables 2D materials QDs to serve as flexible building blocks for quantum information. This tunability allows implementation of controlled qubit gates, long-range entanglement distribution, and simulation of strongly interacting quantum systems, positioning 2D materials QDs as promising platforms for integrated quantum photonics and topologically protected operations\cite{estrada2024correlation}. 
The synergy of optical and superconducting addressability with strong Coulomb interactions in 2D material QDs establishes them as flexible building blocks for quantum information science. Their broad tunability supports a wide range of applications, from controlled quantum gates and long-range entanglement to the simulation of strongly correlated systems, positioning these platforms at the forefront of integrated quantum photonics and topological quantum engineering\cite{estrada2024correlation}.

\section{Emergent Correlated Phenomena in 2D moir\'e Quantum Materials}
\label{correlated}

\begin{figure}[htbp]
	\centering
	\includegraphics[width=\linewidth]{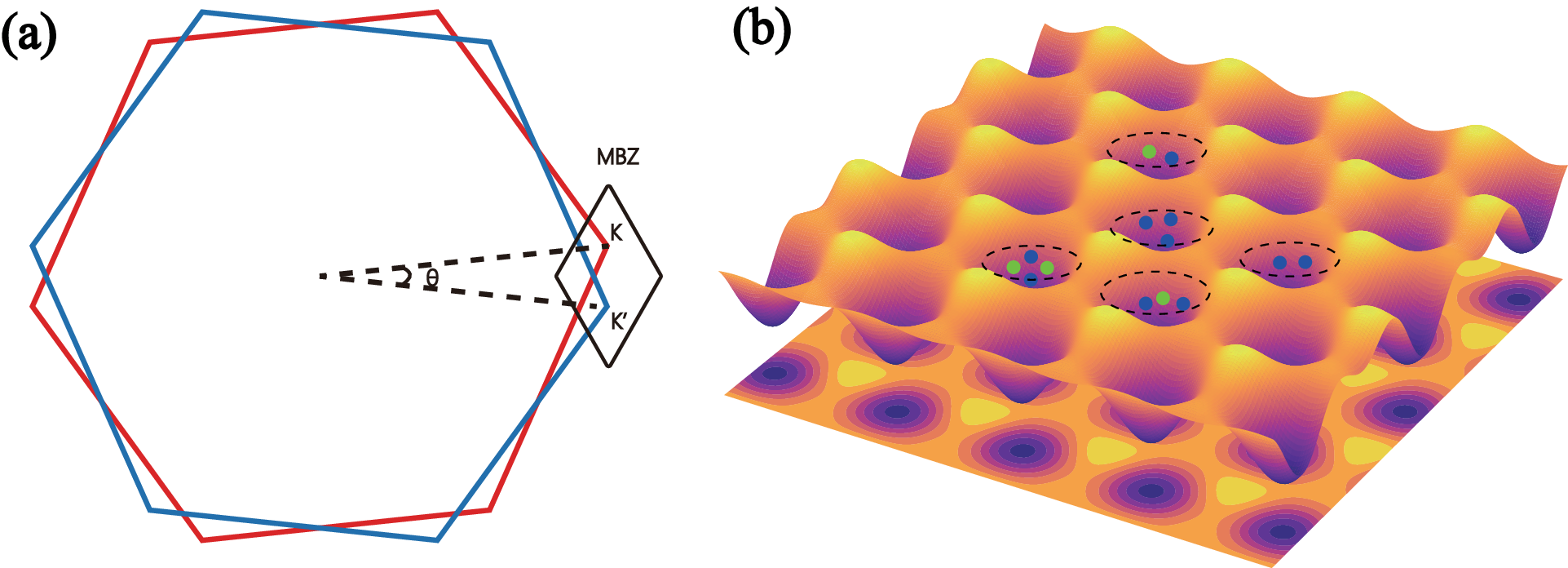}
	\caption{ (a) Single-layer BZs of graphene or TMDs (blue and red curves),
		together with the moir\'e BZ at twist angle $\theta$.
		Here, $ K $ and $K^\prime$ denote the $ K $ and $K^\prime$ points of the moir\'e Brillouin zone.
		(b) The moir\'e pattern induces a periodic modulation of the electrostatic potential,
		leading to an array of moir\'e potential minima.
		Blue and green dots represent electrons and holes, respectively, which can be trapped
		in the moir\'e potential.
	}
	\label{mp}
\end{figure}

The emergence of 2D quantum materials, particularly graphene and TMDs, has established a fertile framework for investigating correlated electron physics in the atomic thickness\cite{tapashl2010review,WehlingDirac,XuGraphene,andrei2021marvels}. Owing to the reduced dimensionality and the associated suppression of dielectric screening, Coulomb interactions in these systems are significantly enhanced compared with their bulk counterparts. As a result, their electronic properties are frequently governed by strong many-body effects rather than single-particle band descriptions\cite{regan2020mott}. The ability to assemble atomically thin layers into van der Waals heterostructures provides unprecedented control over lattice symmetry, interlayer coupling, and electrostatic environments\cite{geim2013van}. In particular, moir\'e patterns originating from lattice mismatch between stacked layers was showed in Fig. \ref{mp}, which offer an effective and highly tunable means to engineer the electronic, magnetic, and optical properties of 2D materials\cite{regan2020mott,andrei2021marvels,angeli2021gamma,2023fuliang,luo2023artificial,zeng2022strong,yannouleas2023quantum,li2024wigner,yannouleas2024crystal}. This unique confluence of intrinsically strong electronic correlations with an extensive suite of in situ tuning knobs defines a new paradigm for quantum materials science. It establishes a versatile platform for discovering and controlling a broad spectrum of emergent phenomena, ranging from correlation driven insulators and superconductors to exotic quasiparticle complexes and symmetry broken orders, thereby fundamentally reshaping our understanding of collective quantum behavior\cite{qi2025competition,gorski2024nonlocal,estrada2024correlation,gorski2024nonlocal,fu2024bilayer,Aoki02102025}.

\subsection{Moir\'e Superconductivity}

\begin{figure*}[htp]
	\centering
	\includegraphics[width=\linewidth]{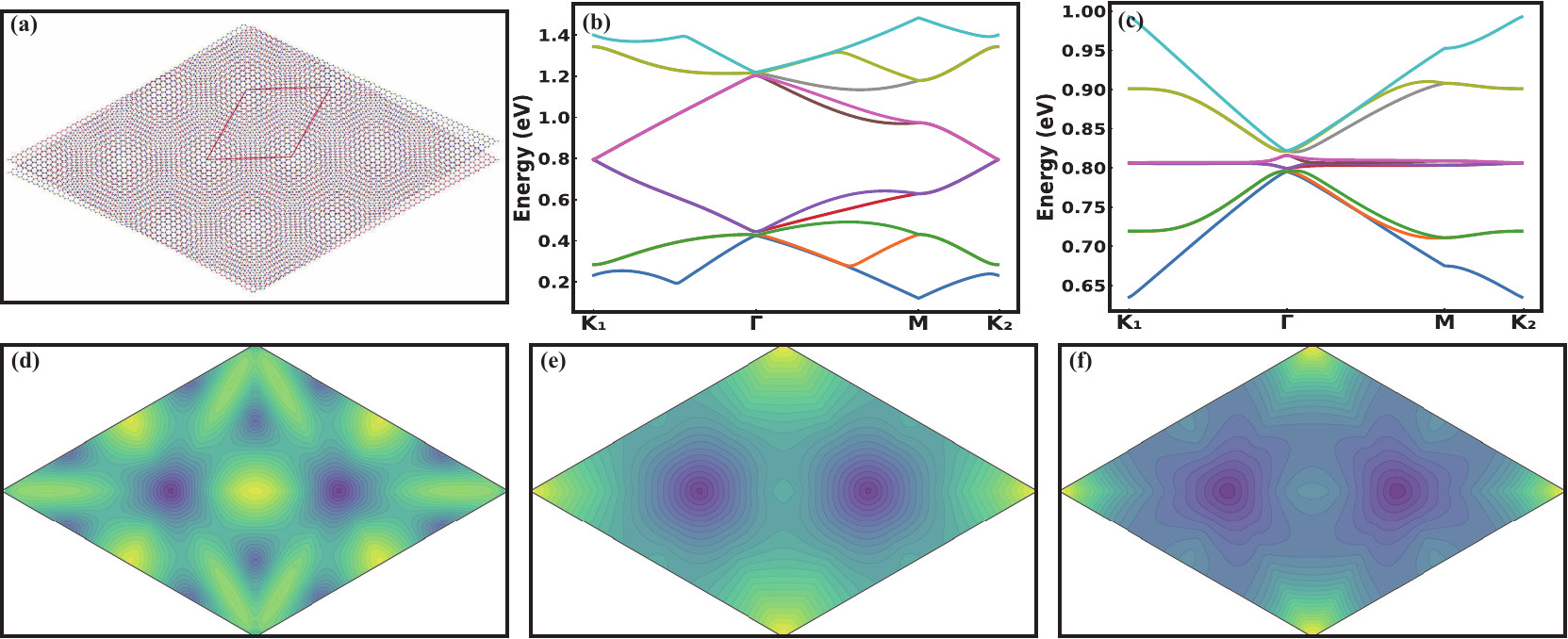}
	\caption{(a) Real-space image of TBG with a twist angle of $\theta = 9.43^\circ$; the red rhombus denotes a superlattice region.
		(b,c) Band structures of TBG for rotation angles of $2.65^\circ$ and $1.35^\circ$, respectively.
		(d)--(f) Energy contour maps of the highest valence band in the moir\'e Brillouin zones of TBG for three different twist angles.
		From left to right, the twist angles are $1.25^\circ$, $6.01^\circ$, and $2.13^\circ$.
		Dark (bright) colors represent high (low) energies.
		Notably, the number of van Hove singularities, corresponding to energy saddle points in the contour maps, changes with the twist angle.
	}
	\label{twist}
\end{figure*}

The emergence of superconductivity in twisted van der Waals heterostructures is fundamentally linked to the formation of low-energy moir\'e flat bands as showed in Fig. \ref{twist}(a)-(c). 
Crucially, the electronic bandwidth (\(W\)) in these flat bands is drastically reduced, whereas the Coulomb repulsion (\(U\)) remains substantial. This establishes a strongly correlated regime defined by the hierarchy \(U / W \gg 1\)~\cite{Trambly2026reveiw,lisi2021observation}. 

Experiments have revealed correlated insulating states at integer fillings of the moiré superlattice, often accompanied by spontaneous symmetry breaking in spin, valley, or orbital sectors. These phases are widely interpreted as manifestations of strong correlations in the flat-band regime and are frequently adjacent to superconducting phases in the phase diagram.

The mechanism of superconductivity in moiré materials remains under active debate. Within the $D_6$ symmetry of the moiré lattice, candidate pairing states include unconventional orders such as $d$-wave or chiral $p_x + i p_y$ superconductivity. Recent tunneling spectroscopy and transport measurements show evidence inconsistent with simple $s$-wave pairing, instead supporting unconventional superconducting states \cite{choi2019electronic,oh2021evidence,ko2023atomic,Aoki02102025,park2026experimental}. However, other works have proposed that electron-phonon coupling could generate conventional BCS-type pairing in these systems \cite{Wu2018PRL,Peltonen2018PRB,Lian2019PRL,zhuprb2025,zhuprb2026}. Additionally, correlation-driven pairing mechanisms, such as the Kohn-Luttinger effect \cite{KL2019}, where repulsive interactions give rise to effective attractive channels due to van Hove singularities \cite{Chen2025FiniteMomentumSC}, have also been discussed.

Superconductivity in twisted multilayer graphene has been reported to persist under in-plane magnetic fields exceeding the Pauli limit, suggesting possible spin-triplet pairing or protection by spin-valley locking \cite{cao2021pauli}. These findings highlight moiré superlattices as a unique platform where strong correlations, topology, and unconventional superconductivity intertwine.

\subsection{Chern Insulators} 

The experimental realization of fractional Chern insulators (FCIs) in zero magnetic field represents one of the most significant advances in topological matter. Unlike the conventional fractional quantum Hall effect (FQHE), which requires strong external magnetic fields to create Landau levels, these FCIs states emerge from interaction driven symmetry breaking in topological moir\'e flat bands with nonzero Berry curvature\cite{LIU2024515}. The band topology is encoded by the Chern number
\begin{equation}
	C = \frac{1}{2\pi}\int_{\mathrm{BZ}} d^2k\, \Omega(\mathbf{k}),
\end{equation}
where the Berry curvature $\Omega(\mathbf{k})$ is given by
\begin{equation}
	\Omega(\mathbf{k}) = i\!\left(
	\left\langle \partial_{k_x} u_{\mathbf{k}} \middle| \partial_{k_y} u_{\mathbf{k}} \right\rangle
	-
	\left\langle \partial_{k_y} u_{\mathbf{k}} \middle| \partial_{k_x} u_{\mathbf{k}} \right\rangle
	\right),
\end{equation}
with $|u_{\mathbf{k}}\rangle$ the cell-periodic Bloch state of the relevant moir\'e band. Materials such as twisted \ce{MoTe2} homobilayers and pentalayer graphene aligned with \ce{hBN} exhibit moir\'e bands with Chern number $C=1$\cite{chen2025robust,aronson2025displacement}. If a band with nonzero Chern number is fully filled, its nontrivial topology leads to a quantized Hall conductance, just as in the integer quantum Hall effect (IQHE). In this sense, the transport properties of an integer Chern insulator (ICI) mimic those of IQHE systems, but without requiring an external magnetic field. The lattice realization of IQHE is thus naturally achieved through the intrinsic Berry curvature of the band.

When the band is flat enough and the band only partially filled, strong electron-electron interactions drive the formation of incompressible quantum liquid states at fractional fillings, which are lattice analog to the FQHE states. Projecting the Coulomb interaction onto the isolated Chern band yields the effective many-body Hamiltonian
\begin{equation}
	H_{\mathrm{int}}
	=
	\frac{1}{2}
	\sum_{\mathbf{q},\mathbf{k},\mathbf{k}'}
	V(\mathbf{q})\,
	\lambda(\mathbf{k},\mathbf{q})
	\lambda(\mathbf{k}',-\mathbf{q})
	\,
	c^\dagger_{\mathbf{k}+\mathbf{q}}
	c^\dagger_{\mathbf{k}'-\mathbf{q}}
	c_{\mathbf{k}'}
	c_{\mathbf{k}},
\end{equation}
where the form factor 
\begin{equation}
	\lambda(\mathbf{k},\mathbf{q})
	=
	\langle u_{\mathbf{k}+\mathbf{q}} | u_{\mathbf{k}} \rangle
\end{equation}
encodes the nontrivial band geometry and Berry curvature\cite{FQH2011,xie2021fractional}. Recent experiments have observed quantized Hall resistance with vanishing longitudinal resistance at fractional fillings, establishing dissipationless FCI phases in zero magnetic field\cite{park2026fci}.

The primary experimental verification has relied on local electronic compressibility measurements using scanning single electron transistors\cite{lee2026local}. These techniques monitor the local chemical potential $\mu$ as a function of carrier density $n$\cite{tomarken2019electronic}. A key thermodynamic signature of a fractional Chern insulator is the emergence of a thermodynamic gap, characterized by a pronounced peak or discontinuity in $\mathrm{d}\mu/\mathrm{d}n$ at fractional filling factors in zero magnetic field. These gaps are robust against small perturbations but vanish when the band topology is trivialized by displacement fields. While this thermodynamic evidence confirms the existence of incompressible states, it does not directly establish their topological nature.

\begin{figure}[ht]
	\centering
	\includegraphics[width=\linewidth]{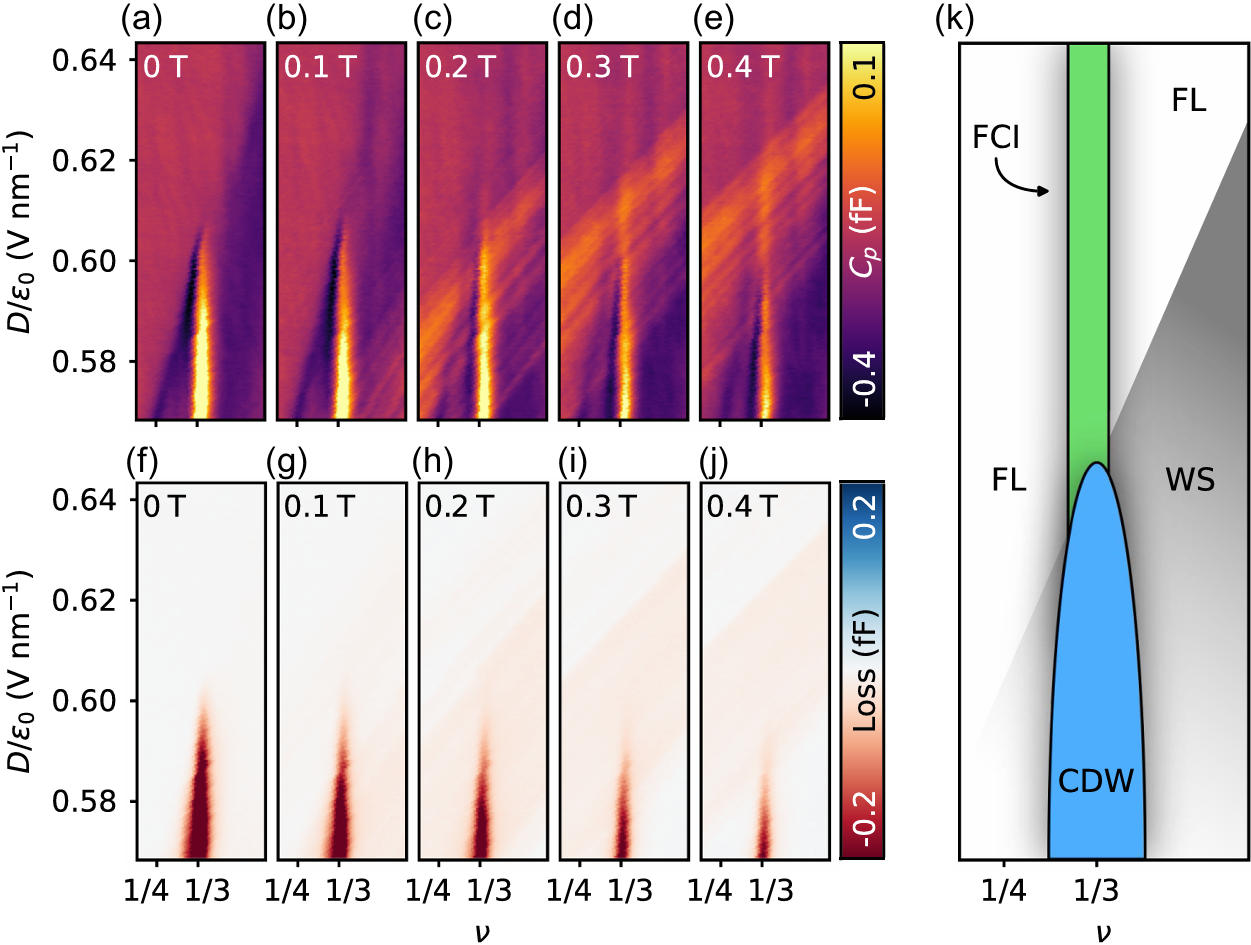}
	\caption{Compressibility maps near filling factor $\nu=1/3$ as a function of displacement field $D$ at low magnetic field. 
	(a)-(e) In-phase component of the impedance bridge signal proportional to the quantum capacitance $C_p$, reflecting the electronic compressibility. 
	The weak vertical feature indicates the FCI states when the magnetic field exceeds 0.2 T. 
	(f)-(j) Quadrature component showing the dissipative loss associated with the device resistance. 
	The color bars denote the magnitude of the quantum capacitance $C_p$ (top) and the loss signal (bottom), extracted from the complex impedance response of the sample. (k) The schematic phase diagram indicates the FCI phase competing with Fermi liquid (FL), Wigner solid (WS) and charge density wave (CDW).
	Adapted from Ref.~\cite{aronson2025displacement}. }
	\label{phase}
\end{figure}

Macroscopic transport measurements provide definitive evidence for FCIs in high-quality twisted \ce{MoTe2}, where quantized Hall conductance and vanishing longitudinal resistivity have been observed \cite{kang2024evidence,xu2023observation}. This quantization persists up to several kelvin, confirming chiral edge states and fractionalized bulk excitations, with multiple fractional states in the same device reflecting a lattice analogue of quantum Hall physics.
These states exhibit finite orbital magnetization associated with Berry curvature, giving rise to hysteretic ferromagnetic responses confined to narrow fractional filling windows, supporting spontaneous topological order \cite{cai2023signatures,redekop2024direct}.

Phase diagrams tuned by displacement field and carrier density reveal competition between FCIs and charge-ordered phases, such as charge density waves or Wigner crystals \cite{aronson2025displacement,sharma2024topological}. The topological stability at $\nu = 1/3$ is sensitive to the uniformity of Berry curvature in the moiré Brillouin zone. Efforts are underway to realize non-Abelian states, including the Moore-Read state at $\nu = 1/2$, with potential applications in fault-tolerant topological quantum computing \cite{chen2025robust}.

In the fractional quantum Hall regime, a localized charged exciton (trion) can interact with the fractionally charged quasiparticles of the Laughlin state and form a bound state. Such a process modifies the exciton recombination energy, leading to a shift in the photoluminescence frequency that reflects the fractional excitation spectrum \cite{OpticalProb1,OpticalProb2}. The system can be modeled as an exciton localized in a QD that couples to quasiparticles of the Laughlin state \cite{APALKOV2002289}. 
In a similar spirit, spectroscopic evidence for fractional charges in FCIs has recently been obtained through the formation of anyon-trion bound state in twisted bilayer \ce{MoTe2}. Photoluminescence measurements reveal bound states in which conventional trions couple to fractional anyons at $\nu=-2/3$ and $-3/5$. The observed binding-energy ratios agree with theoretical expectations for $e/3$ and $e/5$ anyons, providing the first optical confirmation of anyonic excitations in FCIs \cite{li2026signatures}. These optical approaches provide an alternative route to probe fractional excitations beyond conventional transport measurements. However, the direct observation of fractional statistics of anyons \cite{halperin1984,arovas1984} in FCIs remains challenging.

\subsection{Excitons}

\begin{figure}[htb]
	\centering
	\includegraphics[width=\linewidth]{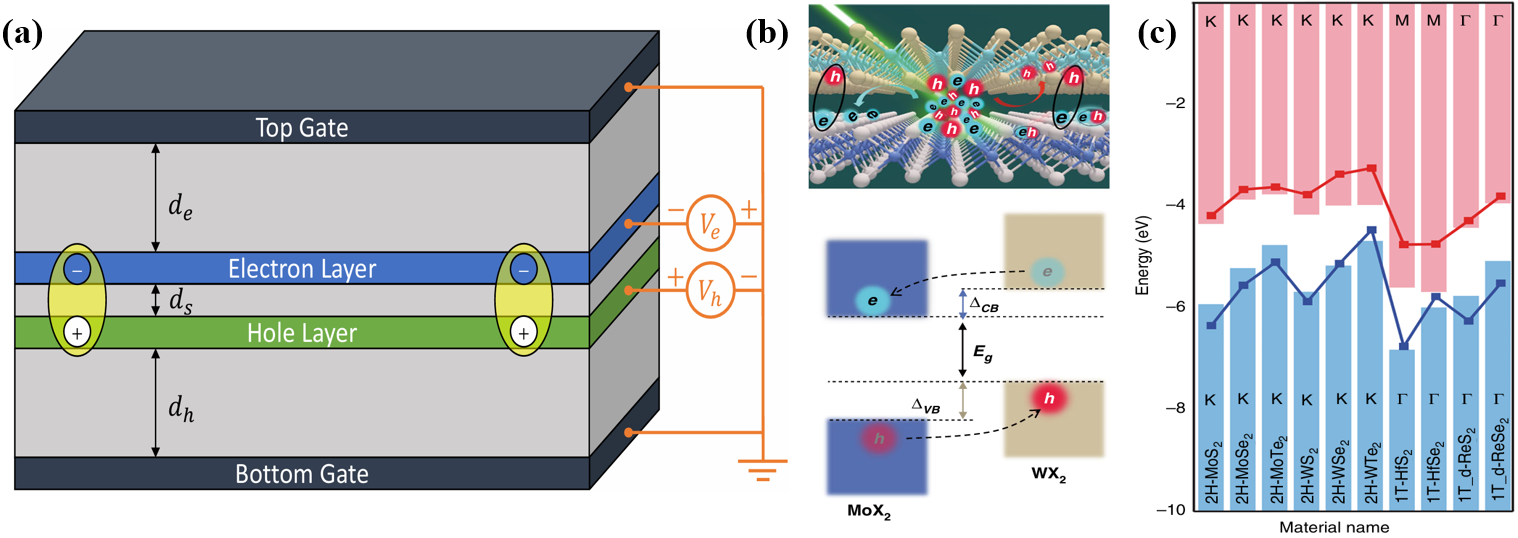}
	\caption{(a) Dual-gated electron-hole bilayer device enabling independent control of electrons and holes that form interlayer excitons.
(b) Interlayer excitons in type-II band-aligned TMD heterostructures, with electrons and holes residing in different layers.
(c) Band alignments of different stacked TMDs heterostructures at high-symmetry points of the Brillouin zone.	 }
	\label{exciton}
\end{figure}

Excitons in bilayer 2D electron systems provide a versatile platform to explore strong Coulomb correlations and collective quantum phenomena \cite{latini2017,jiang2021interlayer,lxinterlayer2025}. Spatial charge separation and reduced dielectric screening stabilize a variety of many-body states, including excitonic superfluidity, Wigner crystals, and exciton-polarons \cite{berman2016high,yannouleas2024crystal,emmanuele2020highly,li2021experimental,dufferwiel2018valley}. Advances in van der Waals assembly now allow precise fabrication of aligned bilayers and twisted moiré superlattices, where the interlayer twist angle creates a moiré potential that enhances correlations. Hybrid heterostructures combining TMDs and perovskites further expand the parameter space for excitonic engineering.

In atomically 2D materials, confinement produce bound excitons with binding energies of several hundred meV, robust at room temperature \cite{baldini2019exciton,zhang2024room}. The excitonic spectrum deviates from the simple 2D hydrogenic series due to nonlocal screening, strong SOC, and broken inversion symmetry, giving rise to valley-contrasting optical selection rules \cite{yao2012tmds}. Within an effective-mass approximation, excitons are described by
\begin{equation}
H_X = -\frac{\hbar^2}{2\mu}\nabla^2 + V(r),
\end{equation}
with electron-hole interaction in the Keldysh form
\begin{equation}
V(q) = \frac{e^2}{2\varepsilon_0 q (1 + r_0 q)},
\end{equation}
where $\mu$ is the reduced mass and $r_0$ is the characteristic screening length \cite{berman2016high,vaquero2020Rydberg}. This form softens short-range interactions, shifting low-lying levels from the hydrogenic sequence while higher states recover the 2D limit.

Strong Coulomb interactions also stabilize excitonic complexes such as trions (charged excitons) and biexcitons, which probe exciton-exciton interactions and many-body effects \cite{dostal2018direct,2024exexprx}.

Van der Waals heterostructures further enrich the excitonic landscape. Type-II heterobilayers support interlayer excitons with electrons and holes in separate layers, exhibiting long lifetimes and enhanced correlations \cite{jiang2021interlayer}, as illustrated in Fig.~\ref{exciton}. Exciton properties can be tuned via twist angle, interlayer coupling, external fields, and dielectric environment, enabling access to collective phases such as exciton condensation, superfluidity, and correlated insulators. Excitons are classified as bright or dark depending on spin configuration \cite{feng2024highly}.

Small twist angles ($\theta \sim 1^\circ$-$5^\circ$) produce a moiré superlattice that modulates exciton motion through a periodic potential, leading to quantum-dot-like confinement \cite{angeli2021gamma,2023fuliang,zeng2022strong}:
\begin{equation}
V_m(\mathbf{r}) \approx V_0 \sum_{j=1}^{3} \cos \left(\mathbf{G}_j \cdot \mathbf{r} + \phi_j\right),
\end{equation}
where $V_0 \sim 10$-$200$ meV depends on twist angle \cite{Hubbardtmd,brem2020tunable}. The moiré potential folds exciton dispersion into minibands, enhancing effective mass and suppressing kinetic energy, promoting localization at potential minima \cite{herrera2025moire,kundu2023exciton,zheng2023localization}.

Optical spectroscopy in twisted TMD homo- and heterobilayers reveals discrete photoluminescence peaks from moiré-trapped excitons, flat-band splitting, and Umklapp signatures \cite{guo2025moire}. Strain and symmetry breaking modify confinement, trion multiplicity, and valley polarization \cite{effectgraphnePR,kundu2025atomic}. Long-lived moiré excitons exhibit phonon-assisted hopping, hybridize with itinerant carriers into exciton-polarons, and form moiré polaritons with nonlinear and topological features \cite{PRRPolarons,fitzgerald2022twist}. \added{At high exciton densities, moir\'e-confined systems enter a strongly interacting many-body regime. Multiple excitons occupying the same moir\'e site experience tunneling, dipolar repulsion, thermal fluctuations, and radiative coupling, making these systems ideal candidates for bosonic Hubbard physics, excitonic Mott-like behavior, excitonic crystallization, and collective optical responses. Recent optical studies of dipole ladders and temperature-dependent excitonic Hubbard interactions show that the effective on-site interaction can be spectroscopically resolved in TMD moir\'e superlattices~\cite{park2023dipole_ladders,wang2025temperature_hubbard_u}. This many-exciton regime extends the discussion from single-exciton localization to correlated bosonic quantum matter and represents an important frontier of confinement-enhanced correlations.}

Recent advances in growth, spectroscopy, and theory have enabled observation and control of tightly bound excitons, long-lived interlayer excitons, moiré-trapped states, and hybrid exciton-polaron quasiparticles \cite{baldini2019exciton,feng2024highly,li2020dipolar,rivera2015observation,lxinterlayer2025,dufferwiel2018valley,li2021experimental,de2025rydberg,herrera2025moire,jiang2025tuning,alexeev2019resonantly}. These developments establish excitons in van der Waals heterostructures as a highly tunable platform for correlated quantum phenomena and novel optoelectronic and valleytronic functionalities.

\subsection{Ferroelectricity}

In recent years, ferroelectricity in 2D materials has attracted extensive attention due to its stability and tunability. Compared with conventional bulk ferroelectrics, atomically flat surfaces and the ultralow defect density in layered van der Waals materials can significantly suppress depolarization fields, which helps stabilize spontaneous polarization even in the monolayer limit \cite{SnTe1,Zhou2017NanoLett}. Early studies of 2D ferroelectrics mainly focused on ionic-displacement ferroelectricity, where polarization originates from relative atomic displacements within the lattice. Representative systems include layered materials such as In$_2$Se$_3$, CuInP$_2$S$_6$, SnTe, and SnSe, which established the feasibility of ferroelectricity in atomically thin crystals \cite{SnTe1,Zhou2017NanoLett,SnTe2}.

More recently, a distinct mechanism termed \textit{sliding ferroelectricity} \cite{slidingferro} has emerged in van der Waals bilayers, where polarization arises from inversion-asymmetric stacking configurations rather than internal ionic displacement. Typical examples include rhombohedral stacked 3R-MoS$_2$ \cite{3RMoS2} and twisted bilayer MoS$_2$ \cite{twistferroMoS2}, where lateral interlayer translation or moiré stacking generates switchable out-of-plane polarization. The discovery of sliding and moiré ferroelectricity has greatly expanded the landscape of 2D ferroelectrics and opened new opportunities for tunable electronic and optoelectronic functionalities. 
Recent studies further demonstrate that ferroelectric order in 2D materials can couple strongly with electronic transport, phase engineering, optoelectronic responses, and correlated electronic states in layered moir\'e systems~\cite{Wu2025ACSNano,JiaoAPR,JiaoNanoRes,Wu2023CoupledFerroelectricity}. 

Ferroelectricity is characterized by the emergence of a spontaneous macroscopic polarization in an insulating phase that lacks inversion symmetry and supports bistable states related by spatial inversion. Within Landau theory, the free energy can be expanded in terms of the polarization order parameter $P$ as
\begin{equation}
	F(P) = \alpha P^2 + \beta P^4 - EP,
\end{equation}
where $\alpha < 0$ leads to two degenerate minima at $P=\pm P_0$, corresponding to switchable polarization states under an external electric field\cite{landau2013statistical,lines2001principles}. 

Beyond this phenomenological description, polarization in crystalline solids is rigorously defined within the modern theory of polarization as a geometric Berry phase of occupied Bloch states\cite{resta1992theory,king1993theory,resta1994macroscopic}:
\begin{equation}
	\mathbf{P} = -\frac{ie}{(2\pi)^3}
	\sum_{n}^{\mathrm{occ}}
	\int_{\mathrm{BZ}} d\mathbf{k}\,
	\langle u_{n\mathbf{k}} \vert \nabla_{\mathbf{k}} \vert u_{n\mathbf{k}} \rangle .
\end{equation}
This formulation highlights that polarization is an intrinsic bulk geometric property of the electronic wave functions rather than a simple sum of localized dipole moments.

\begin{figure}[htp]
\centering
\includegraphics[width=\linewidth]{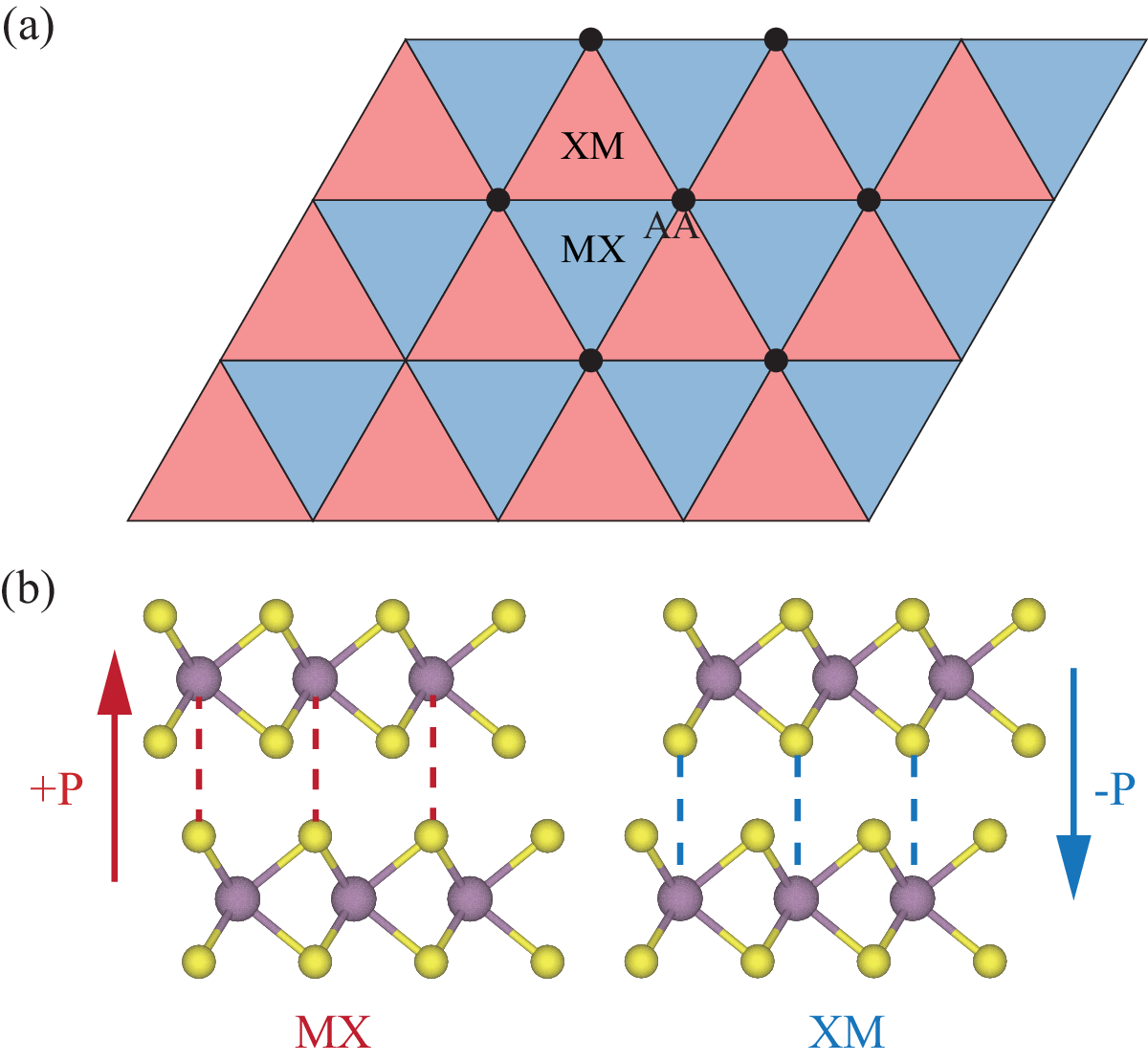}
\caption{Oppositely polarized ferroelectric domains in rhombohedral (R-phase) homobilayer TMDs.
The two inversion-related stacking configurations (\ce{MX} and \ce{XM}) generate opposite out-of-plane polarization.}
\label{domain}
\end{figure}

Ferroelectricity is unexpectedly extend to moiré systems, where a small twist between two layers creates spatially varying local stacking and alternating polarization domains, forming moiré ferroelectricity \cite{fan2025edge,yang2023atypical}. In bilayer TMDs with rhombohedral stacking, the two inversion-related configurations (\ce{MX} and \ce{XM}) generate opposite out-of-plane polarization, producing a periodic array of ferroelectric domains in Fig.~\ref{domain}.

The interlayer polarization can be expressed as a function of local stacking configuration $\mathbf{u}$ and twist angle $\theta$ \cite{ji2023general}:
\begin{equation}
P = P(\mathbf{u},\theta), \quad P(\mathbf{u},\theta) = -P(-\mathbf{u},-\theta),
\end{equation}
highlighting that inversion-related stackings correspond to opposite polarization states. The resulting periodic electrostatic potential acts as a spatially modulated superlattice, which can reconstruct electronic minibands, localize excitons, and modify interaction effects in moiré materials. The twist angle continuously tunes both the period and strength of the ferroelectric landscape \cite{fan2025edge,yang2023atypical}.

Moiré ferroelectric polarization can further couple to magnetic layers, modifying magnetic anisotropy and enabling electrically controlled switching in heterostructures \cite{heissenbuttel2021valley,liu2017wafer,xu2022coexisting}. These features establish twisted bilayer ferroelectrics as a versatile platform for engineered moiré potentials and emergent electronic states \cite{fan2025edge,yang2023atypical,wu2021sliding,ju2015topological}.

\section{Technology and Applications} 
\label{applications}

\begin{figure}[]
\centering
\includegraphics[width=0.43\textwidth]{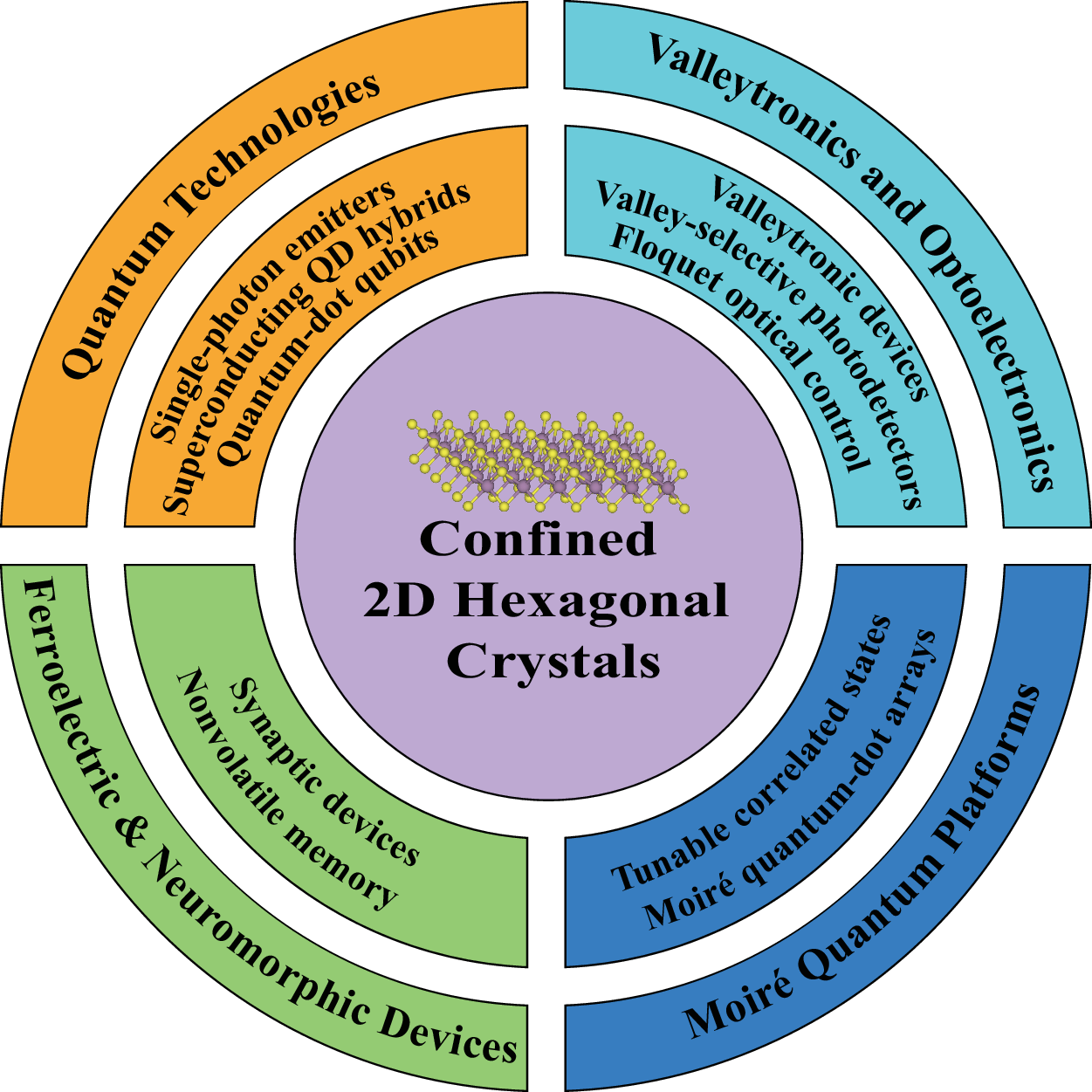}
\caption{Technology and application directions of confined two-dimensional hexagonal crystals. Graphene- and TMD-based QDs, nanoribbons, and moiré van der Waals heterostructures provide tunable platforms for quantum information and photonics, valleytronic and optoelectronic devices, moir\'e quantum platforms, and ferroelectric or neuromorphic functionalities. These device concepts exploit discrete confined levels, spin-valley coupling, excitonic responses, moir\'e potentials, and switchable polarization in atomically thin materials.}
\label{fig:2D_applications}
\end{figure}

The exotic properties of 2D materials can be harnessed across scales, from quantum dots (QDs) to wafer-scale and macroscopic architectures\cite{2Dmap}. As illustrated in Fig.~\ref{fig:2D_applications}, these atomically thin materials enable diverse technologies. Graphene and TMDs provide promising platforms for nanoelectronics. 2D QDs, with size-tunable bandgaps and edge states, support quantum information applications, including single-photon emitters and topological computing elements\cite{wang2016quantum,ren2019review,lu2025van}. Filamentary resistive switching in polycrystalline QD networks enables neuromorphic synaptic plasticity\cite{ren2019review,tian2025peculiar}, while manipulation of excitonic fine structure and valley-polarized selection rules in TMD QDs underpins valleytronic logic gates and ultrasensitive photodetectors\cite{ye2022nonvolatile}. Large-area 2D crystals further extend integrated electronics beyond conventional silicon limits\cite{das2021transistors}.

\added{TMD QDs are also promising platforms for spin--valley qubits. Spin--valley locking together with strong intrinsic SOC can suppress selected relaxation channels, making confined spin--valley states attractive for electrical and optical qubit control. Recent theoretical work on TMD QDs has clarified how SOC strength, intervalley mixing, and spin and valley $g-$factors determine the conditions for spin--valley locking in the few-electron regime~\cite{shandilya2024spin_valley_locking}. This link among relativistic band structure, confinement, and coherence distinguishes Dirac-material QDs from conventional semiconductor-dot platforms.}

\added{Time-dependent optical control offers another route for extending confinement physics in 2D Dirac materials. Early theoretical work on irradiated bilayer graphene showed that intense terahertz radiation can generate Floquet dynamical gaps, induce states inside a bias-controlled static gap, and produce valley-polarized electronic states, demonstrating that light can actively reshape the band structure and density of states in graphene-based systems \cite{Abergel_2011}. More recent observation of Floquet states and a Floquet-induced gap in graphene shows that coherent light-driven band engineering is possible even in semimetallic Dirac systems with femtosecond-scale decoherence, motivating extensions to moir\'e platforms with strong correlations~\cite{merboldt2025floquet_states,wang2026floquet_gap}. 
In confined Dirac systems, such light-driven control could provide a direct way to dynamically tune interaction effects, band topology, and correlated quantum phases.}

\section{Acknowledgments}
W.L. was supported by the Natural Science Foundation of Hunan Province (Grant No. 2025JJ50019). X.L. acknowledge Yutao Hu for helpful discussions.

%






%
%
%


\begin{thebibliography}{100}
\expandafter\ifx\csname url\endcsname\relax
  \def\url#1{\texttt{#1}}\fi
\expandafter\ifx\csname urlprefix\endcsname\relax\def\urlprefix{URL }\fi
\expandafter\ifx\csname href\endcsname\relax
  \def\href#1#2{#2} \def\path#1{#1}\fi

\bibitem{tkachov2022topological}
G.~Tkachov, Topological Quantum Materials: Concepts, Models, and Phenomena, Jenny Stanford Publishing, 2022.
\newblock \href {https://doi.org/https://doi.org/10.1201/9781003266419} {\path{doi:https://doi.org/10.1201/9781003266419}}.

\bibitem{chakraborty2002nano}
T.~Chakraborty, F.~Peeters, U.~Sivan, Nano-Physics and Bio-Electronics: A New Odyssey, Elsevier, 2002.

\bibitem{vontapashqhe40}
K.~von Klitzing, T.~Chakraborty, P.~Kim, V.~Madhavan, X.~Dai, J.~McIver, Y.~Tokura, L.~Savary, D.~Smirnova, A.~M. Rey, et~al., {40 years of the quantum Hall effect}, Nat. Rev. Phys. 2~(8) (2020) 397--401.
\newblock \href {https://doi.org/https://doi.org/10.1038/s42254-020-0209-1} {\path{doi:https://doi.org/10.1038/s42254-020-0209-1}}.

\bibitem{chakraborty2024encyclopedia}
T.~Chakraborty (Ed.), Encyclopedia of condensed matter physics (2nd Edition), Academic Press, Oxford, 2024.

\bibitem{gruber2024interplay}
C.~S. Gruber, M.~Abdel-Hafiez, Interplay of electronic orders in topological quantum materials, ACS Mater. Au 5~(1) (2024) 72--87.
\newblock \href {https://doi.org/https://doi.org/10.1021/acsmaterialsau.4c00114} {\path{doi:https://doi.org/10.1021/acsmaterialsau.4c00114}}.

\bibitem{Trambly2026reveiw}
G.~T. de~Laissardi{\`e}re, S.~Venkateswarlu, A.~Misssaoui, G.~Jema{\"\i}, K.~Chika, J.~Vahedi, O.~F. Namarvar, J.-P. Julien, A.~Honecker, L.~Magaud, et~al., {Electronic structure and transport in materials with flat bands: 2D materials and quasicrystals}, Physica E 175 (2026) 116362.
\newblock \href {https://doi.org/https://doi.org/10.1016/j.physe.2025.116362} {\path{doi:https://doi.org/10.1016/j.physe.2025.116362}}.

\bibitem{tapashl2010review}
D.~Abergel, V.~Apalkov, J.~Berashevich, K.~Ziegler, T.~Chakraborty, {Properties of graphene: a theoretical perspective}, Adv. Phys. 59~(4) (2010) 261--482.
\newblock \href {https://doi.org/https://doi.org/10.1080/00018732.2010.487978} {\path{doi:https://doi.org/10.1080/00018732.2010.487978}}.

\bibitem{WehlingDirac}
T.~Wehling, A.~Black-Schaffer, A.~Balatsky, {Dirac materials}, Adv. Phys. 63~(1) (2014) 1--76.
\newblock \href {https://doi.org/https://doi.org/10.1080/00018732.2014.927109} {\path{doi:https://doi.org/10.1080/00018732.2014.927109}}.

\bibitem{XuGraphene}
M.~Xu, T.~Liang, M.~Shi, H.~Chen, {Graphene-Like Two-Dimensional Materials}, Chem. Rev. 113~(5) (2013) 3766--3798.
\newblock \href {https://doi.org/https://doi.org/10.1021/cr300263a} {\path{doi:https://doi.org/10.1021/cr300263a}}.

\bibitem{geim2013van}
A.~K. Geim, I.~V. Grigorieva, {Van der Waals heterostructures}, Nature 499~(7459) (2013) 419--425.
\newblock \href {https://doi.org/https://doi.org/10.1038/nature12385} {\path{doi:https://doi.org/10.1038/nature12385}}.

\bibitem{jung2026hbn}
J.-H. Jung, C.-J. Kim, {Hexagonal boron nitride: interlayer with atomic scale precision for interface engineering in functional materials and devices}, npj 2D Mater. Appl. (2026).
\newblock \href {https://doi.org/https://doi.org/10.1038/s41699-026-00664-7} {\path{doi:https://doi.org/10.1038/s41699-026-00664-7}}.

\bibitem{blackph2014}
L.~Cording, J.~Liu, J.~Y. Tan, K.~Watanabe, T.~Taniguchi, A.~Avsar, B.~{\"O}zyilmaz, {Highly anisotropic spin transport in ultrathin black phosphorus}, Nat. Mater. 23~(4) (2024) 479--485.
\newblock \href {https://doi.org/https://doi.org/10.1038/s41563-023-01779-8} {\path{doi:https://doi.org/10.1038/s41563-023-01779-8}}.

\bibitem{li2025Xenes}
Z.~Li, H.~Cao, S.~Meng, {Light-induced above-room-temperature Chern insulators in group-IV Xenes}, npj Comput. Mater. 11~(1) (2025) 160.
\newblock \href {https://doi.org/https://doi.org/10.1038/s41524-025-01662-x} {\path{doi:https://doi.org/10.1038/s41524-025-01662-x}}.

\bibitem{lu2017semimetals}
H.-Z. Lu, S.-Q. Shen, {Quantum transport in topological semimetals under magnetic fields}, Front. Phys. 12~(3) (2017) 127201.
\newblock \href {https://doi.org/https://doi.org/10.1007/s11467-019-0890-7} {\path{doi:https://doi.org/10.1007/s11467-019-0890-7}}.

\bibitem{graphenedetection2024}
G.~D. Battista, K.~C. Fong, A.~Díez-Carlón, K.~Watanabe, T.~Taniguchi, D.~K. Efetov, {Infrared single-photon detection with superconducting magic-angle twisted bilayer graphene}, Sci. Adv. 10~(38) (2024) eadp3725.
\newblock \href {https://doi.org/https://doi.org/10.1126/sciadv.adp3725} {\path{doi:https://doi.org/10.1126/sciadv.adp3725}}.

\bibitem{zhouyu2025TMDs}
J.~Yi, Q.~Shuai, G.~Wu, H.~Wu, C.~Zhu, X.~Sun, Q.~Zong, Y.~Liao, X.~Zhu, X.~Fan, Z.~Zhang, Y.~Zhou, X.~Wang, D.~Li, A.~Pan, {Capacitive Amplification toward Boltzmann-Limit Switching in {WSe$_2$} Reconfigurable Field-Effect Transistors}, Adv. Funct. Mater. n/a~(n/a) (2025) e24990.
\newblock \href {https://doi.org/https://doi.org/10.1002/adfm.202524990} {\path{doi:https://doi.org/10.1002/adfm.202524990}}.

\bibitem{sunj2025TMDs}
S.~Zheng, Y.~Sun, Y.~Shen, S.~Du, H.~Chen, Y.~Jing, Y.~Yuan, F.~Yao, H.~Li, X.~Liu, Y.~Cheng, J.~Sun, {Anomalous Reconfigurable-Transport in {MoS$_2$} Transistors by Electrically-Switchable van der Waals Interfacial Dipole}, Adv. Mater. 37~(45) (2025) e02784.
\newblock \href {https://doi.org/https://doi.org/10.1002/adma.202502784} {\path{doi:https://doi.org/10.1002/adma.202502784}}.

\bibitem{2Dmap}
W.~Ren, P.~Boggild, J.~M. Redwing, K.~S. Novoselov, L.~Sun, Y.~Qi, K.~Jia, Z.~Liu, O.~Burton, J.~A. Alexander-Webber, S.~Hofmann, Y.~Cao, Y.~Long, Q.-H. Yang, D.~Li, S.~H. Choi, K.~K. Kim, Y.~H. Lee, M.~Li, Q.~Huang, Y.~Gogotsi, N.~Clark, A.~Carl, R.~Gorbachev, T.~Olsen, J.~Rosen, K.~S. Thygesen, P.~D. D.~K. Efetov, B.~S. Jessen, M.~Yankowitz, J.~Barrier, R.~K. Kumar, F.~Koppens, H.~Deng, X.~Li, S.~Dai, D.~Basov, X.~Wang, S.~Das, X.~Duan, Z.~Yu, M.~Borsch, A.~C. Ferrari, R.~Huber, M.~Kira, F.~Xia, X.~Wang, Z.-S. Wu, X.~Feng, P.~Simon, H.-M. Cheng, B.~Liu, Y.~Xie, W.~Jin, R.~R. Nair, Y.~Xu, H.-B. Zhang, V.~Pellegrini, B.~Qu, M.~Lemme, A.~Katiyar, J.-H. Ahn, I.~Aharonovich, M.~C. Hersam, S.~Roche, Q.~Hua, G.~Shen, T.-L. Ren, C.~M. Koo, N.~A. Koratkar, R.~J. Young, A.~Pollard, {The {2D} Materials Roadmap}, 2D Mater. (2025).
\newblock \href {https://doi.org/https://doi.org/10.1088/2053-1583/ae2b82} {\path{doi:https://doi.org/10.1088/2053-1583/ae2b82}}.

\bibitem{effectgraphnePR}
B.~Amorim, A.~Cortijo, F.~{de Juan}, A.~Grushin, F.~Guinea, A.~Gutiérrez-Rubio, H.~Ochoa, V.~Parente, R.~Roldán, P.~San-Jose, J.~Schiefele, M.~Sturla, M.~Vozmediano, {Novel effects of strains in graphene and other two dimensional materials}, Phys. Rep. 617 (2016) 1--54.
\newblock \href {https://doi.org/https://doi.org/10.1016/j.physrep.2015.12.006} {\path{doi:https://doi.org/10.1016/j.physrep.2015.12.006}}.

\bibitem{graphene2016pr}
A.~Rozhkov, A.~Sboychakov, A.~Rakhmanov, F.~Nori, {Electronic properties of graphene-based bilayer systems}, Phys. Rep. 648 (2016) 1--104, electronic properties of graphene-based bilayer systems.
\newblock \href {https://doi.org/https://doi.org/10.1016/j.physrep.2016.07.003} {\path{doi:https://doi.org/10.1016/j.physrep.2016.07.003}}.

\bibitem{yao2012tmds}
D.~Xiao, G.-B. Liu, W.~Feng, X.~Xu, W.~Yao, {Coupled Spin and Valley Physics in Monolayers of {MoS$_2$} and Other Group-{VI} Dichalcogenides}, Phys. Rev. Lett. 108 (2012) 196802.
\newblock \href {https://doi.org/https://doi.org/10.1103/PhysRevLett.108.196802} {\path{doi:https://doi.org/10.1103/PhysRevLett.108.196802}}.

\bibitem{2Dmater2015}
A.~Kormányos, G.~Burkard, M.~Gmitra, J.~Fabian, V.~Zólyomi, N.~D. Drummond, V.~Fal’ko, {k·p theory for two-dimensional transition metal dichalcogenide semiconductors}, 2D Mater. 2~(2) (2015) 022001.
\newblock \href {https://doi.org/https://doi.org/10.1088/2053-1583/2/2/022001} {\path{doi:https://doi.org/10.1088/2053-1583/2/2/022001}}.

\bibitem{2009electronicproperties}
A.~H. Castro~Neto, F.~Guinea, N.~M.~R. Peres, K.~S. Novoselov, A.~K. Geim, {The electronic properties of graphene}, Rev. Mod. Phys. 81 (2009) 109--162.
\newblock \href {https://doi.org/https://doi.org/10.1103/RevModPhys.81.109} {\path{doi:https://doi.org/10.1103/RevModPhys.81.109}}.

\bibitem{2010electronicproperties}
D.~Xiao, M.-C. Chang, Q.~Niu, {Berry phase effects on electronic properties}, Rev. Mod. Phys. 82 (2010) 1959--2007.
\newblock \href {https://doi.org/https://doi.org/10.1103/RevModPhys.82.1959} {\path{doi:https://doi.org/10.1103/RevModPhys.82.1959}}.

\bibitem{2008opticalgra}
A.~B. Kuzmenko, E.~van Heumen, F.~Carbone, D.~van~der Marel, {Universal Optical Conductance of Graphite}, Phys. Rev. Lett. 100 (2008) 117401.
\newblock \href {https://doi.org/https://doi.org/10.1103/PhysRevLett.100.117401} {\path{doi:https://doi.org/10.1103/PhysRevLett.100.117401}}.

\bibitem{2023highgraphene}
S.~Iwakiri, J.~Miller, F.~Lang, J.~Prettenthaler, T.~Taniguchi, K.~Watanabe, S.~S. Lee, P.~Becker, D.~G\"unther, T.~Ihn, K.~Ensslin, {High-mobility transport in isotopically enriched $^{12}\mathrm{C}$ and $^{13}\mathrm{C}$ exfoliated graphene}, Phys. Rev. Res. 5 (2023) 043212.
\newblock \href {https://doi.org/https://doi.org/10.1103/PhysRevResearch.5.043212} {\path{doi:https://doi.org/10.1103/PhysRevResearch.5.043212}}.

\bibitem{zhao2024ultrahigh}
J.~Zhao, P.~Ji, Y.~Li, R.~Li, K.~Zhang, H.~Tian, K.~Yu, B.~Bian, L.~Hao, X.~Xiao, et~al., {Ultrahigh-mobility semiconducting epitaxial graphene on silicon carbide}, Nature 625~(7993) (2024) 60--65.
\newblock \href {https://doi.org/https://doi.org/10.1038/s41586-023-06811-0} {\path{doi:https://doi.org/10.1038/s41586-023-06811-0}}.

\bibitem{yaolight2015}
G.-B. Liu, D.~Xiao, Y.~Yao, X.~Xu, W.~Yao, {Electronic structures and theoretical modelling of two-dimensional group-VIB transition metal dichalcogenides}, Chem. Soc. Rev. 44 (2015) 2643--2663.
\newblock \href {https://doi.org/https://doi.org/10.1039/C4CS00301B} {\path{doi:https://doi.org/10.1039/C4CS00301B}}.

\bibitem{mak2018light}
K.~F. Mak, D.~Xiao, J.~Shan, {Light--valley interactions in 2D semiconductors}, Nat. Photonics 12~(8) (2018) 451--460.
\newblock \href {https://doi.org/https://doi.org/10.1038/s41566-018-0204-6} {\path{doi:https://doi.org/10.1038/s41566-018-0204-6}}.

\bibitem{2013tw}
A.~Korm\'anyos, V.~Z\'olyomi, N.~D. Drummond, P.~Rakyta, G.~Burkard, V.~I. Fal'ko, {Monolayer {MoS$_2$}: Trigonal warping, the $\ensuremath{\Gamma}$ valley, and spin-orbit coupling effects}, Phys. Rev. B 88 (2013) 045416.
\newblock \href {https://doi.org/https://doi.org/10.1103/PhysRevB.88.045416} {\path{doi:https://doi.org/10.1103/PhysRevB.88.045416}}.

\bibitem{2020tw}
S.~S.~R. Bladwell, {Valley separation via trigonal warping}, Phys. Rev. B 101 (2020) 045404.
\newblock \href {https://doi.org/https://doi.org/10.1103/PhysRevB.101.045404} {\path{doi:https://doi.org/10.1103/PhysRevB.101.045404}}.

\bibitem{2018darkexciton}
E.~Malic, M.~Selig, M.~Feierabend, S.~Brem, D.~Christiansen, F.~Wendler, A.~Knorr, G.~Bergh\"auser, {Dark excitons in transition metal dichalcogenides}, Phys. Rev. Mater. 2 (2018) 014002.
\newblock \href {https://doi.org/https://doi.org/10.1103/PhysRevMaterials.2.014002} {\path{doi:https://doi.org/10.1103/PhysRevMaterials.2.014002}}.

\bibitem{xu2014spin}
X.~Xu, W.~Yao, D.~Xiao, T.~F. Heinz, {Spin and pseudospins in layered transition metal dichalcogenides}, Nat. Phys. 10~(5) (2014) 343--350.
\newblock \href {https://doi.org/https://doi.org/10.1038/nphys2942} {\path{doi:https://doi.org/10.1038/nphys2942}}.

\bibitem{wu2021sliding}
M.~Wu, J.~Li, {Sliding ferroelectricity in \text{2D} van der \text{W}aals materials: Related physics and future opportunities}, Proc. Natl. Acad. Sci. U.S.A. 118~(50) (2021) e2115703118.
\newblock \href {https://doi.org/https://doi.org/10.1073/pnas.2115703118} {\path{doi:https://doi.org/10.1073/pnas.2115703118}}.

\bibitem{ding2019defectqd}
X.~Ding, F.~Peng, J.~Zhou, W.~Gong, G.~Slaven, K.~P. Loh, C.~T. Lim, D.~T. Leong, {Defect engineered bioactive transition metals dichalcogenides quantum dots}, Nat. Commun. 10~(1) (2019) 41.
\newblock \href {https://doi.org/https://doi.org/10.1038/s41467-018-07835-1} {\path{doi:https://doi.org/10.1038/s41467-018-07835-1}}.

\bibitem{buqd2011}
R.~Liu, D.~Wu, X.~Feng, K.~M{\"u}llen, {Bottom-Up Fabrication of Photoluminescent Graphene Quantum Dots with Uniform Morphology}, J. Am. Chem. Soc. 133~(39) (2011) 15221--15223.
\newblock \href {https://doi.org/https://doi.org/10.1021/ja204953k} {\path{doi:https://doi.org/10.1021/ja204953k}}.

\bibitem{reviewqd2021}
K.~James~Singh, T.~Ahmed, P.~Gautam, A.~S. Sadhu, D.-H. Lien, S.-C. Chen, Y.-L. Chueh, H.-C. Kuo, {Recent Advances in Two-Dimensional Quantum Dots and Their Applications}, Nanomaterials 11~(6) (2021).
\newblock \href {https://doi.org/https://doi.org/10.3390/nano11061549} {\path{doi:https://doi.org/10.3390/nano11061549}}.

\bibitem{tapsh1999129}
T.~Chakraborty, {Quantum Dots}, North-Holland, Amsterdam, 1999.

\bibitem{tapashqdee}
P.~Maksym, T.~Chakraborty, {Quantum dots in a magnetic field: Role of electron-electron interactions}, Phys. Rev. Lett. 65~(1) (1990) 108.
\newblock \href {https://doi.org/https://doi.org/10.1103/PhysRevLett.65.108} {\path{doi:https://doi.org/10.1103/PhysRevLett.65.108}}.

\bibitem{brey2006graphene_nanoribbons}
L.~Brey, H.~A. Fertig, Electronic states of graphene nanoribbons studied with the {Dirac} equation, Phys. Rev. B 73 (2006) 235411.
\newblock \href {https://doi.org/https://doi.org/10.1103/PhysRevB.73.235411} {\path{doi:https://doi.org/10.1103/PhysRevB.73.235411}}.

\bibitem{chen2007fock}
H.-Y. Chen, V.~Apalkov, T.~Chakraborty, {Fock-Darwin states of Dirac electrons in graphene-based artificial atoms}, Phys. Rev. Lett. 98~(18) (2007) 186803.
\newblock \href {https://doi.org/https://doi.org/10.1103/PhysRevLett.98.186803} {\path{doi:https://doi.org/10.1103/PhysRevLett.98.186803}}.

\bibitem{kim2024confinement}
G.~Kim, B.~Huet, C.~E. Stevens, K.~Jo, J.-Y. Tsai, S.~Bachu, M.~Leger, S.~Song, M.~Rahaman, K.~Y. Ma, et~al., {Confinement of excited states in two-dimensional, in-plane, quantum heterostructures}, Nat. Commun. 15~(1) (2024) 6361.
\newblock \href {https://doi.org/https://doi.org/10.1038/s41467-024-50653-x} {\path{doi:https://doi.org/10.1038/s41467-024-50653-x}}.

\bibitem{2014TMDsqdqubit}
A.~Korm\'anyos, V.~Z\'olyomi, N.~D. Drummond, G.~Burkard, {Spin-Orbit Coupling, Quantum Dots, and Qubits in Monolayer Transition Metal Dichalcogenides}, Phys. Rev. X 4 (2014) 011034.
\newblock \href {https://doi.org/https://doi.org/10.1103/PhysRevX.4.011034} {\path{doi:https://doi.org/10.1103/PhysRevX.4.011034}}.

\bibitem{2021TMDsqdqubit}
A.~Alt\ifmmode \imath \else \i \fi{}nta\ifmmode~\mbox{\c{s}}\else \c{s}\fi{}, M.~Bieniek, A.~Dusko, M.~Korkusi\ifmmode~\acute{n}\else \'{n}\fi{}ski, J.~Paw\l{}owski, P.~Hawrylak, {Spin-valley qubits in gated quantum dots in a single layer of transition metal dichalcogenides}, Phys. Rev. B 104 (2021) 195412.
\newblock \href {https://doi.org/https://doi.org/10.1103/PhysRevB.104.195412} {\path{doi:https://doi.org/10.1103/PhysRevB.104.195412}}.

\bibitem{2021TMDsqdqubitsBrooks}
M.~Brooks, G.~Burkard, {Electric dipole spin resonance of two-dimensional semiconductor spin qubits}, Phys. Rev. B 101 (2020) 035204.
\newblock \href {https://doi.org/https://doi.org/10.1103/PhysRevB.101.035204} {\path{doi:https://doi.org/10.1103/PhysRevB.101.035204}}.

\bibitem{2009Tapashchirality}
D.~S.~L. Abergel, P.~Pietil\"ainen, T.~Chakraborty, {Electronic compressibility of graphene: The case of vanishing electron correlations and the role of chirality}, Phys. Rev. B 80 (2009) 081408.
\newblock \href {https://doi.org/https://doi.org/10.1103/PhysRevB.80.081408} {\path{doi:https://doi.org/10.1103/PhysRevB.80.081408}}.

\bibitem{2009TapashBilayerGraphene}
D.~S.~L. Abergel, T.~Chakraborty, {Long-Range Coulomb Interaction in Bilayer Graphene}, Phys. Rev. Lett. 102 (2009) 056807.
\newblock \href {https://doi.org/https://doi.org/10.1103/PhysRevLett.102.056807} {\path{doi:https://doi.org/10.1103/PhysRevLett.102.056807}}.

\bibitem{tapashmoire}
J.~Berashevich, T.~Chakraborty, Interlayer repulsion and decoupling effects in stacked turbostratic graphene flakes, Phys. Rev. B 84 (2011) 033403.
\newblock \href {https://doi.org/https://doi.org/10.1103/PhysRevB.84.033403} {\path{doi:https://doi.org/10.1103/PhysRevB.84.033403}}.

\bibitem{berashevich2011nature}
J.~Berashevich, T.~Chakraborty, On the nature of interlayer interactions in a system of two graphene fragments, J. Phys. Chem. C 115~(50) (2011) 24666--24673.
\newblock \href {https://doi.org/https://doi.org/10.1021/jp2095032} {\path{doi:https://doi.org/10.1021/jp2095032}}.

\bibitem{CHAKRABORTY2013123}
T.~Chakraborty, V.~M. Apalkov, Traits and characteristics of interacting dirac fermions in monolayer and bilayer graphene, Solid State Commun. 175-176 (2013) 123--131.
\newblock \href {https://doi.org/https://doi.org/10.1016/j.ssc.2013.04.002} {\path{doi:https://doi.org/10.1016/j.ssc.2013.04.002}}.

\bibitem{2014TapashHofstadterSystem}
V.~M. Apalkov, T.~Chakraborty, {Gap Structure of the Hofstadter System of Interacting Dirac Fermions in Graphene}, Phys. Rev. Lett. 112 (2014) 176401.
\newblock \href {https://doi.org/https://doi.org/10.1103/PhysRevLett.112.176401} {\path{doi:https://doi.org/10.1103/PhysRevLett.112.176401}}.

\bibitem{2012Tapashtrilayergraphene}
V.~M. Apalkov, T.~Chakraborty, {Electrically tunable charge and spin transitions in Landau levels of interacting Dirac fermions in trilayer graphene}, Phys. Rev. B 86 (2012) 035401.
\newblock \href {https://doi.org/https://doi.org/10.1103/PhysRevB.86.035401} {\path{doi:https://doi.org/10.1103/PhysRevB.86.035401}}.

\bibitem{regan2020mott}
E.~C. Regan, D.~Wang, C.~Jin, M.~I. Bakti~Utama, B.~Gao, X.~Wei, S.~Zhao, W.~Zhao, Z.~Zhang, K.~Yumigeta, et~al., {Mott and generalized Wigner crystal states in WSe2/WS2 moir{\'e} superlattices}, Nature 579~(7799) (2020) 359--363.
\newblock \href {https://doi.org/https://doi.org/10.1038/s41586-020-2092-4} {\path{doi:https://doi.org/10.1038/s41586-020-2092-4}}.

\bibitem{andrei2021marvels}
E.~Y. Andrei, D.~K. Efetov, P.~Jarillo-Herrero, A.~H. MacDonald, K.~F. Mak, T.~Senthil, E.~Tutuc, A.~Yazdani, A.~F. Young, {The marvels of moir{\'e} materials}, Nat. Rev. Mater. 6~(3) (2021) 201--206.
\newblock \href {https://doi.org/https://doi.org/10.1038/s41578-021-00284-1} {\path{doi:https://doi.org/10.1038/s41578-021-00284-1}}.

\bibitem{ghiotto2021quantum}
A.~Ghiotto, E.-M. Shih, G.~S. Pereira, D.~A. Rhodes, B.~Kim, J.~Zang, A.~J. Millis, K.~Watanabe, T.~Taniguchi, J.~C. Hone, et~al., {Quantum criticality in twisted transition metal dichalcogenides}, Nature 597~(7876) (2021) 345--349.
\newblock \href {https://doi.org/ttps://doi.org/10.1038/s41586-021-03815-6} {\path{doi:ttps://doi.org/10.1038/s41586-021-03815-6}}.

\bibitem{sheng2024terahertz}
S.~Sheng, M.~Abdo, S.~Rolf-Pissarczyk, K.~Lichtenberg, S.~Baumann, J.~A. Burgess, L.~Malavolti, S.~Loth, {Terahertz spectroscopy of collective charge density wave dynamics at the atomic scale}, Nat. Phys. 20~(10) (2024) 1603--1608.
\newblock \href {https://doi.org/https://doi.org/10.1038/s41567-024-02552-7} {\path{doi:https://doi.org/10.1038/s41567-024-02552-7}}.

\bibitem{2008berry}
P.~Carmier, D.~Ullmo, {Berry phase in graphene: Semiclassical perspective}, Phys. Rev. B 77 (2008) 245413.
\newblock \href {https://doi.org/https://link.aps.org/doi/10.1103/PhysRevB.77.245413} {\path{doi:https://link.aps.org/doi/10.1103/PhysRevB.77.245413}}.

\bibitem{2019berry}
R.~Battilomo, N.~Scopigno, C.~Ortix, {Berry Curvature Dipole in Strained Graphene: A Fermi Surface Warping Effect}, Phys. Rev. Lett. 123 (2019) 196403.
\newblock \href {https://doi.org/https://link.aps.org/doi/10.1103/PhysRevLett.123.196403} {\path{doi:https://link.aps.org/doi/10.1103/PhysRevLett.123.196403}}.

\bibitem{2020berry}
Y.~Zhang, Y.~Su, L.~He, {Local Berry Phase Signatures of Bilayer Graphene in Intervalley Quantum Interference}, Phys. Rev. Lett. 125 (2020) 116804.
\newblock \href {https://doi.org/https://link.aps.org/doi/10.1103/PhysRevLett.125.116804} {\path{doi:https://link.aps.org/doi/10.1103/PhysRevLett.125.116804}}.

\bibitem{2011soczhu}
Z.~Y. Zhu, Y.~C. Cheng, U.~Schwingenschl\"ogl, {Giant spin-orbit-induced spin splitting in two-dimensional transition-metal dichalcogenide semiconductors}, Phys. Rev. B 84 (2011) 153402.
\newblock \href {https://doi.org/https://doi.org/10.1103/PhysRevB.84.153402} {\path{doi:https://doi.org/10.1103/PhysRevB.84.153402}}.

\bibitem{Qi2014}
F.~Qi, G.~Jin, \href{https://doi.org/10.1063/1.4874676}{Mechanically and optically controlled graphene valley filter}, Journal of Applied Physics 115~(17) (2014) 173701.
\newblock \href {https://doi.org/10.1063/1.4874676} {\path{doi:10.1063/1.4874676}}.
\newline\urlprefix\url{https://doi.org/10.1063/1.4874676}

\bibitem{wang2017valley}
Z.~Wang, J.~Shan, K.~F. Mak, {Valley-and spin-polarized Landau levels in monolayer {WSe$_2$}}, Nat. Nanotechnol. 12~(2) (2017) 144--149.
\newblock \href {https://doi.org/https://doi.org/10.1038/nnano.2016.213} {\path{doi:https://doi.org/10.1038/nnano.2016.213}}.

\bibitem{valleyqhe}
K.~F. Mak, K.~L. McGill, J.~Park, P.~L. McEuen, {The valley Hall effect in {MoS$_2$} transistors}, Science 344~(6191) (2014) 1489--1492.
\newblock \href {https://doi.org/https://doi.org/10.1126/science.1250140} {\path{doi:https://doi.org/10.1126/science.1250140}}.

\bibitem{cappelluti2013tight}
E.~Cappelluti, R.~Rold{\'a}n, J.~Silva-Guill{\'e}n, P.~Ordej{\'o}n, F.~Guinea, {Tight-binding model and direct-gap/indirect-gap transition in single-layer and multilayer MoS$_2$}, Phys. Rev. B 88~(7) (2013) 075409.
\newblock \href {https://doi.org/https://doi.org/10.1103/PhysRevB.88.075409} {\path{doi:https://doi.org/10.1103/PhysRevB.88.075409}}.

\bibitem{2018WeylRevModPhys}
N.~P. Armitage, E.~J. Mele, A.~Vishwanath, {Weyl and Dirac semimetals in three-dimensional solids}, Rev. Mod. Phys. 90 (2018) 015001.
\newblock \href {https://doi.org/https://link.aps.org/doi/10.1103/RevModPhys.90.015001} {\path{doi:https://link.aps.org/doi/10.1103/RevModPhys.90.015001}}.

\bibitem{2025Vadym}
V.~Apalkov, W.~Luo, T.~Chakraborty, {Selective enhancement of Coulomb interactions in planar Weyl fermions}, Phys. Rev. B 111 (2025) 195409.
\newblock \href {https://doi.org/https://link.aps.org/doi/10.1103/PhysRevB.111.195409} {\path{doi:https://link.aps.org/doi/10.1103/PhysRevB.111.195409}}.

\bibitem{2014Weylhall}
A.~A. Burkov, {Anomalous Hall Effect in Weyl Metals}, Phys. Rev. Lett. 113 (2014) 187202.
\newblock \href {https://doi.org/https://link.aps.org/doi/10.1103/PhysRevLett.113.187202} {\path{doi:https://link.aps.org/doi/10.1103/PhysRevLett.113.187202}}.

\bibitem{lu2024realization}
Q.~Lu, P.~S. Reddy, H.~Jeon, A.~R. Mazza, M.~Brahlek, W.~Wu, S.~A. Yang, J.~Cook, C.~Conner, X.~Zhang, et~al., Realization of a two-dimensional weyl semimetal and topological fermi strings, Nat. Commun. 15~(1) (2024) 6001.
\newblock \href {https://doi.org/https://doi.org/10.1038/s41467-024-50329-6} {\path{doi:https://doi.org/10.1038/s41467-024-50329-6}}.

\bibitem{chakraborty2012fractional}
T.~Chakraborty, P.~Pietil{\"a}inen, The fractional quantum Hall effect: properties of an incompressible quantum fluid, Vol.~85, Springer Science \& Business Media, 2012.
\newblock \href {https://doi.org/https://doi.org/10.1007/978-3-642-97101-3} {\path{doi:https://doi.org/10.1007/978-3-642-97101-3}}.

\bibitem{latini2017}
S.~Latini, K.~T. Winther, T.~Olsen, K.~S. Thygesen, {Interlayer Excitons and Band Alignment in {MoS$_2$/hBN/WSe$_2$} van der Waals Heterostructures}, Nano Lett. 17~(2) (2017) 938--945.
\newblock \href {https://doi.org/https://doi.org/10.1021/acs.nanolett.6b04275} {\path{doi:https://doi.org/10.1021/acs.nanolett.6b04275}}.

\bibitem{emmanuele2020highly}
R.~Emmanuele, M.~Sich, O.~Kyriienko, V.~Shahnazaryan, F.~Withers, A.~Catanzaro, P.~Walker, F.~Benimetskiy, M.~Skolnick, A.~Tartakovskii, et~al., {Highly nonlinear trion-polaritons in a monolayer semiconductor}, Nat. Commun. 11~(1) (2020) 3589.
\newblock \href {https://doi.org/https://doi.org/10.1038/s41467-020-17340-z} {\path{doi:https://doi.org/10.1038/s41467-020-17340-z}}.

\bibitem{lisi2021observation}
S.~Lisi, X.~Lu, T.~Benschop, T.~A. de~Jong, P.~Stepanov, J.~R. Duran, F.~Margot, I.~Cucchi, E.~Cappelli, A.~Hunter, et~al., {Observation of flat bands in twisted bilayer graphene}, Nat. Phys. 17~(2) (2021) 189--193.
\newblock \href {https://doi.org/https://doi.org/10.1038/s41567-020-01041-x} {\path{doi:https://doi.org/10.1038/s41567-020-01041-x}}.

\bibitem{riley2014direct}
J.~M. Riley, F.~Mazzola, M.~Dendzik, M.~Michiardi, T.~Takayama, L.~Bawden, C.~Graner{\o}d, M.~Leandersson, T.~Balasubramanian, M.~Hoesch, et~al., {Direct observation of spin-polarized bulk bands in an inversion-symmetric semiconductor}, Nat. Phys. 10~(11) (2014) 835--839.
\newblock \href {https://doi.org/https://doi.org/10.1038/nphys3105} {\path{doi:https://doi.org/10.1038/nphys3105}}.

\bibitem{lxinterlayer2025}
X.~Liu, Z.~Tao, W.~Luo, T.~Chakraborty, {Interlayer excitons in double-layer transition metal dichalcogenide quantum dots}, Phys. Rev. B 111 (2025) 085424.
\newblock \href {https://doi.org/https://link.aps.org/doi/10.1103/PhysRevB.111.085424} {\path{doi:https://link.aps.org/doi/10.1103/PhysRevB.111.085424}}.

\bibitem{Vadym2024ultrafast}
A.~Mitra, A.~J. Zafar, S.~J. Hosseini, V.~Apalkov, {Ultrafast high harmonic generation in transition metal dichalcogenide quantum dots}, Phys. Rev. B 109~(15) (2024) 155425.
\newblock \href {https://doi.org/https://doi.org/10.1103/PhysRevB.109.155425} {\path{doi:https://doi.org/10.1103/PhysRevB.109.155425}}.

\bibitem{berrymassless1987}
M.~V. Berry, R.~Mondragon, {Neutrino billiards: time-reversal symmetry-breaking without magnetic fields}, Proc. R. Soc. Lond. A 412~(1842) (1987) 53--74.
\newblock \href {https://doi.org/https://doi.org/10.1098/rspa.1987.0080} {\path{doi:https://doi.org/10.1098/rspa.1987.0080}}.

\bibitem{liu2014intervalley}
G.-B. Liu, H.~Pang, Y.~Yao, W.~Yao, {Intervalley coupling by quantum dot confinement potentials in monolayer transition metal dichalcogenides}, New J. Phys. 16~(10) (2014) 105011.
\newblock \href {https://doi.org/https://doi.org/10.1088/1367-2630/16/10/105011} {\path{doi:https://doi.org/10.1088/1367-2630/16/10/105011}}.

\bibitem{grujic2011electronic}
M.~Gruji{\'c}, M.~Zarenia, A.~Chaves, M.~Tadi{\'c}, G.~Farias, F.~Peeters, Electronic and optical properties of a circular graphene quantum dot in a magnetic field: Influence of the boundary conditions, Phys. Rev. B 84~(20) (2011) 205441.
\newblock \href {https://doi.org/https://doi.org/10.1103/PhysRevB.84.205441} {\path{doi:https://doi.org/10.1103/PhysRevB.84.205441}}.

\bibitem{paananen2011finite}
T.~Paananen, R.~Egger, Finite-size version of the excitonic instability in graphene quantum dots, Phys. Rev. B 84~(15) (2011) 155456.
\newblock \href {https://doi.org/https://doi.org/10.1103/PhysRevB.84.155456} {\path{doi:https://doi.org/10.1103/PhysRevB.84.155456}}.

\bibitem{paananen2011signatures}
T.~Paananen, R.~Egger, H.~Siedentop, {Signatures of Wigner molecule formation in interacting Dirac fermion quantum dots}, Phys. Rev. B 83~(8) (2011) 085409.
\newblock \href {https://doi.org/https://doi.org/10.1103/PhysRevB.83.085409} {\path{doi:https://doi.org/10.1103/PhysRevB.83.085409}}.

\bibitem{raca2017excitonic}
V.~Raca, M.~Milovanovi{\'c}, Excitonic physics in a dirac quantum dot, Phys. Rev. B 96~(19) (2017) 195434.
\newblock \href {https://doi.org/https://doi.org/10.1103/PhysRevB.96.195434} {\path{doi:https://doi.org/10.1103/PhysRevB.96.195434}}.

\bibitem{grapheneqds}
A.~Pena, Electron trapping in twisted light driven graphene quantum dots, Phys. Rev. B 105 (2022) 045405.
\newblock \href {https://doi.org/https://doi.org/10.1103/PhysRevB.105.045405} {\path{doi:https://doi.org/10.1103/PhysRevB.105.045405}}.

\bibitem{qu2017tunable}
F.~Qu, A.~Dias, J.~Fu, L.~Villegas-Lelovsky, D.~L. Azevedo, Tunable spin and valley dependent magneto-optical absorption in molybdenum disulfide quantum dots, Sci. Rep. 7~(1) (2017) 41044.
\newblock \href {https://doi.org/https://doi.org/10.1038/srep41044} {\path{doi:https://doi.org/10.1038/srep41044}}.

\bibitem{2017trilayerqd}
M.~Mirzakhani, M.~Zarenia, P.~Vasilopoulos, F.~Peeters, {Electrostatically confined trilayer graphene quantum dots}, Phys. Rev. B 95~(15) (2017) 155434.
\newblock \href {https://doi.org/https://doi.org/10.1103/PhysRevB.95.155434} {\path{doi:https://doi.org/10.1103/PhysRevB.95.155434}}.

\bibitem{hecker2023coherent}
K.~Hecker, L.~Banszerus, A.~Sch{\"a}pers, S.~M{\"o}ller, A.~Peters, E.~Icking, K.~Watanabe, T.~Taniguchi, C.~Volk, C.~Stampfer, {Coherent charge oscillations in a bilayer graphene double quantum dot}, Nat. Commun. 14~(1) (2023) 7911.
\newblock \href {https://doi.org/https://doi.org/10.1038/s41467-023-43541-3} {\path{doi:https://doi.org/10.1038/s41467-023-43541-3}}.

\bibitem{mirzakhani2020circular}
M.~Mirzakhani, F.~Peeters, M.~Zarenia, {Circular quantum dots in twisted bilayer graphene}, Phys. Rev. B 101~(7) (2020) 075413.
\newblock \href {https://doi.org/https://doi.org/10.1103/PhysRevB.101.075413} {\path{doi:https://doi.org/10.1103/PhysRevB.101.075413}}.

\bibitem{tiutiunnyk2019opto}
A.~Tiutiunnyk, C.~Duque, F.~Caro-Lopera, M.~Mora-Ramos, J.~Correa, {Opto-electronic properties of twisted bilayer graphene quantum dots}, Physica E 112 (2019) 36--48.
\newblock \href {https://doi.org/https://doi.org/10.1016/j.physe.2019.03.028} {\path{doi:https://doi.org/10.1016/j.physe.2019.03.028}}.

\bibitem{tapash2013ee}
M.~Zarenia, B.~Partoens, T.~Chakraborty, F.~M. Peeters, {Electron-electron interactions in bilayer graphene quantum dots}, Phys. Rev. B 88 (2013) 245432.
\newblock \href {https://doi.org/https://doi.org/10.1103/PhysRevB.88.245432} {\path{doi:https://doi.org/10.1103/PhysRevB.88.245432}}.

\bibitem{2016trilayerPeeters}
M.~Mirzakhani, M.~Zarenia, S.~A. Ketabi, D.~R. da~Costa, F.~M. Peeters, Energy levels of hybrid monolayer-bilayer graphene quantum dots, Phys. Rev. B 93 (2016) 165410.
\newblock \href {https://doi.org/https://doi.org/10.1103/PhysRevB.93.165410} {\path{doi:https://doi.org/10.1103/PhysRevB.93.165410}}.

\bibitem{wang2025spin}
L.~Wang, G.~Burkard, Spin relaxation in a single-electron bilayer graphene quantum dot, Phys. Rev. Res. 7~(4) (2025) 043061.
\newblock \href {https://doi.org/https://doi.org/10.1103/j6cp-6m5j} {\path{doi:https://doi.org/10.1103/j6cp-6m5j}}.

\bibitem{Tapash2006}
P.~Pietil\"ainen, T.~Chakraborty, Energy levels and magneto-optical transitions in parabolic quantum dots with spin-orbit coupling, Phys. Rev. B 73 (2006) 155315.
\newblock \href {https://doi.org/https://doi.org/10.1103/PhysRevB.73.155315} {\path{doi:https://doi.org/10.1103/PhysRevB.73.155315}}.

\bibitem{zeng2022strong}
Y.~Zeng, A.~H. MacDonald, {Strong modulation limit of excitons and trions in moir{\'e} materials}, Phys. Rev. B 106~(3) (2022) 035115.
\newblock \href {https://doi.org/https://doi.org/10.1103/PhysRevB.106.035115} {\path{doi:https://doi.org/10.1103/PhysRevB.106.035115}}.

\bibitem{yannouleas2023quantum}
C.~Yannouleas, U.~Landman, {Quantum Wigner molecules in moir{\'e} materials}, Phys. Rev. B 108~(12) (2023) L121411.
\newblock \href {https://doi.org/https://doi.org/10.1103/PhysRevB.108.L121411} {\path{doi:https://doi.org/10.1103/PhysRevB.108.L121411}}.

\bibitem{brotons2019coulomb}
M.~Brotons-Gisbert, A.~Branny, S.~Kumar, R.~Picard, R.~Proux, M.~Gray, K.~S. Burch, K.~Watanabe, T.~Taniguchi, B.~D. Gerardot, {Coulomb blockade in an atomically thin quantum dot coupled to a tunable Fermi reservoir}, Nat. Nanotechnol. 14~(5) (2019) 442--446.
\newblock \href {https://doi.org/https://doi.org/10.1038/s41565-019-0402-5} {\path{doi:https://doi.org/10.1038/s41565-019-0402-5}}.

\bibitem{Halonen_1996}
V.~Halonen, P.~Pietiläinen, T.~Chakraborty, \href{https://doi.org/10.1209/epl/i1996-00350-5}{Optical-absorption spectra of quantum dots and rings with a repulsive scattering centre}, Europhysics Letters 33~(5) (1996) 377.
\newblock \href {https://doi.org/10.1209/epl/i1996-00350-5} {\path{doi:10.1209/epl/i1996-00350-5}}.
\newline\urlprefix\url{https://doi.org/10.1209/epl/i1996-00350-5}

\bibitem{dias2016robust}
A.~Dias, J.~Fu, L.~Villegas-Lelovsky, F.~Qu, {Robust effective Zeeman energy in monolayer MoS2 quantum dots}, J. Phys.: Condens. Matter 28~(37) (2016) 375803.
\newblock \href {https://doi.org/https://doi.org/10.1088/0953-8984/28/37/375803} {\path{doi:https://doi.org/10.1088/0953-8984/28/37/375803}}.

\bibitem{mitra2024ultrafast_valley}
A.~Mitra, A.~J. Zafar, V.~Apalkov, Ultrafast field-driven valley polarization of transition metal dichalcogenide quantum dots, J. Phys.: Condens. Matter 36~(20) (2024) 205302.
\newblock \href {https://doi.org/https://doi.org/10.1088/1361-648X/ad271a} {\path{doi:https://doi.org/10.1088/1361-648X/ad271a}}.

\bibitem{mitra2024ultrafast_hhg}
A.~Mitra, A.~J. Zafar, S.~J. Hosseini, V.~Apalkov, Ultrafast high harmonic generation in transition metal dichalcogenide quantum dots, Phys. Rev. B 109~(15) (2024) 155425.
\newblock \href {https://doi.org/https://doi.org/10.1103/PhysRevB.109.155425} {\path{doi:https://doi.org/10.1103/PhysRevB.109.155425}}.

\bibitem{avetisyan2012strong}
S.~Avetisyan, P.~Pietil{\"a}inen, T.~Chakraborty, {Strong enhancement of Rashba spin-orbit coupling with increasing anisotropy in the Fock-Darwin states of a quantum dot}, Phys. Rev. B 85~(15) (2012) 153301.
\newblock \href {https://doi.org/https://doi.org/10.1103/PhysRevB.85.153301} {\path{doi:https://doi.org/10.1103/PhysRevB.85.153301}}.

\bibitem{avetisyan2013superintense}
S.~Avetisyan, P.~Pietil{\"a}inen, T.~Chakraborty, Superintense highly anisotropic optical transitions in anisotropic quantum dots, Phys. Rev. B 88~(20) (2013) 205310.
\newblock \href {https://doi.org/https://doi.org/10.1103/PhysRevB.88.205310} {\path{doi:https://doi.org/10.1103/PhysRevB.88.205310}}.

\bibitem{chakraborty2018controllable}
T.~Chakraborty, A.~Manaselyan, M.~Barseghyan, D.~Laroze, Controllable continuous evolution of electronic states in a single quantum ring, Phys. Rev. B 97~(4) (2018) 041304.
\newblock \href {https://doi.org/https://doi.org/10.1103/PhysRevB.97.041304} {\path{doi:https://doi.org/10.1103/PhysRevB.97.041304}}.

\bibitem{angeli2021gamma}
M.~Angeli, A.~H. MacDonald, {$\Gamma$ valley transition metal dichalcogenide moir{\'e} bands}, Proc. Natl. Acad. Sci. U.S.A. 118~(10) (2021) e2021826118.
\newblock \href {https://doi.org/https://doi.org/10.1073/pnas.2021826118} {\path{doi:https://doi.org/10.1073/pnas.2021826118}}.

\bibitem{2023fuliang}
A.~P. Reddy, T.~Devakul, L.~Fu, {Artificial Atoms, Wigner Molecules, and an Emergent Kagome Lattice in Semiconductor Moir\'e Superlattices}, Phys. Rev. Lett. 131 (2023) 246501.
\newblock \href {https://doi.org/https://doi.org/10.1103/PhysRevLett.131.246501} {\path{doi:https://doi.org/10.1103/PhysRevLett.131.246501}}.

\bibitem{reviewChakraborty1992}
T.~Chakraborty, Physics of the artificial atoms: Quantum dots in a magnetic field, Comments on Condensed Matter Physics 16 (1992) 35--68.

\bibitem{luo2023artificial}
D.~Luo, A.~P. Reddy, T.~Devakul, L.~Fu, {Artificial intelligence for artificial materials: moir$\backslash$'e atom}, arXiv:2303.08162 (2023).
\newblock \href {https://doi.org/https://doi.org/10.48550/arXiv.2303.08162} {\path{doi:https://doi.org/10.48550/arXiv.2303.08162}}.

\bibitem{sanchez2025chiral}
J.~Sanchez-Lopez, Z.-X. Lin, D.~Luo, et~al., {Chiral Cavity Control of the Interlayer Exciton Energy Spectrum}, arXiv:2511.05751 (2025).
\newblock \href {https://doi.org/https://doi.org/10.48550/arXiv.2511.05751} {\path{doi:https://doi.org/10.48550/arXiv.2511.05751}}.

\bibitem{luo2025solving}
D.~Luo, T.~Zaklama, L.~Fu, {Solving fractional electron states in twisted MoTe$_2$ with deep neural network}, arXiv:2503.13585 (2025).
\newblock \href {https://doi.org/https://doi.org/10.48550/arXiv.2503.13585} {\path{doi:https://doi.org/10.48550/arXiv.2503.13585}}.

\bibitem{park2023dipole_ladders}
H.~Park, J.~Zhu, X.~Wang, Y.~Wang, W.~Holtzmann, T.~Taniguchi, K.~Watanabe, J.~Yan, L.~Fu, T.~Cao, D.~Xiao, D.~R. Gamelin, H.~Yu, W.~Yao, X.~Xu, Dipole ladders with large {Hubbard} interaction in a moir{\'e} exciton lattice, Nat. Phys. 19 (2023) 1286--1292.
\newblock \href {https://doi.org/https://doi.org/10.1038/s41567-023-02077-5} {\path{doi:https://doi.org/10.1038/s41567-023-02077-5}}.

\bibitem{pasek2023magnetic_moire_qd_arrays}
W.~Pasek, M.~Kupczynski, P.~Potasz, Magnetic properties of moir{\'e} quantum dot arrays, Phys. Rev. B 108 (2023) 165152.
\newblock \href {https://doi.org/https://doi.org/10.1103/PhysRevB.108.165152} {\path{doi:https://doi.org/10.1103/PhysRevB.108.165152}}.

\bibitem{nakada1996edge}
K.~Nakada, M.~Fujita, G.~Dresselhaus, M.~S. Dresselhaus, Edge state in graphene ribbons: Nanometer size effect and edge shape dependence, Phys. Rev. B 54 (1996) 17954--17961.
\newblock \href {https://doi.org/https://doi.org/10.1103/PhysRevB.54.17954} {\path{doi:https://doi.org/10.1103/PhysRevB.54.17954}}.

\bibitem{yao2009edge_states}
W.~Yao, S.~A. Yang, Q.~Niu, {Edge States in Graphene: From Gapped Flat-Band to Gapless Chiral Modes}, Phys. Rev. Lett. 102 (2009) 096801.
\newblock \href {https://doi.org/https://doi.org/10.1103/PhysRevLett.102.096801} {\path{doi:https://doi.org/10.1103/PhysRevLett.102.096801}}.

\bibitem{martin2008topological_confinement}
I.~Martin, Y.~M. Blanter, A.~F. Morpurgo, {Topological Confinement in Bilayer Graphene}, Phys. Rev. Lett. 100 (2008) 036804.
\newblock \href {https://doi.org/https://doi.org/10.1103/PhysRevLett.100.036804} {\path{doi:https://doi.org/10.1103/PhysRevLett.100.036804}}.

\bibitem{Brito2022}
F.~M.~O. Brito, L.~Li, J.~a. M. V.~P. Lopes, E.~V. Castro, \href{https://link.aps.org/doi/10.1103/PhysRevB.105.195130}{Edge magnetism in transition metal dichalcogenide nanoribbons: Mean field theory and determinant quantum monte carlo}, Phys. Rev. B 105 (2022) 195130.
\newblock \href {https://doi.org/10.1103/PhysRevB.105.195130} {\path{doi:10.1103/PhysRevB.105.195130}}.
\newline\urlprefix\url{https://link.aps.org/doi/10.1103/PhysRevB.105.195130}

\bibitem{jiang2025}
Y.~Jiang, M.~Dan, S.~Fan, G.~Hu, F.~Huang, W.-Y. Wang, Q.~Chen, \href{https://doi.org/10.1038/s42005-025-02110-4}{Geometry engineering of tunneling transistors in transition metal dichalcogenide nanoribbon heterojunctions}, Communications Physics 8~(1) (2025) 189.
\newblock \href {https://doi.org/10.1038/s42005-025-02110-4} {\path{doi:10.1038/s42005-025-02110-4}}.
\newline\urlprefix\url{https://doi.org/10.1038/s42005-025-02110-4}

\bibitem{ren2024interfacial_qds}
Y.-N. Ren, M.-H. Zhang, X.-F. Zhou, Q.~Zheng, H.-Y. Ren, L.~He, In situ creation and tailoring of interfacial quantum dots in graphene/transition metal dichalcogenide heterostructures, Phys. Rev. B 110 (2024) 125416.
\newblock \href {https://doi.org/https://doi.org/10.1103/PhysRevB.110.125416} {\path{doi:https://doi.org/10.1103/PhysRevB.110.125416}}.

\bibitem{keren2020quantum}
I.~Keren, T.~Dvir, A.~Zalic, A.~Iluz, D.~LeBoeuf, K.~Watanabe, T.~Taniguchi, H.~Steinberg, {Quantum-dot assisted spectroscopy of degeneracy-lifted Landau levels in graphene}, Nat. Commun. 11~(1) (2020) 3408.
\newblock \href {https://doi.org/https://doi.org/10.1038/s41467-020-17225-1} {\path{doi:https://doi.org/10.1038/s41467-020-17225-1}}.

\bibitem{mao2025orbital}
Y.~Mao, H.-Y. Ren, X.-F. Zhou, H.~Sheng, Y.-H. Xiao, Y.-C. Zhuang, Y.-N. Ren, L.~He, Q.-F. Sun, {Orbital hybridization in graphene-based artificial atoms}, Nature (2025) 1--6\href {https://doi.org/https://doi.org/10.1038/s41586-025-08620-z} {\path{doi:https://doi.org/10.1038/s41586-025-08620-z}}.

\bibitem{zhou2025recent}
X.~Zhou, L.~He, {Recent progresses on artificial graphene quantum dots}, Fundam. Res. (2025).
\newblock \href {https://doi.org/https://doi.org/10.1016/j.fmre.2025.03.020} {\path{doi:https://doi.org/10.1016/j.fmre.2025.03.020}}.

\bibitem{jiang2021interlayer}
Y.~Jiang, S.~Chen, W.~Zheng, B.~Zheng, A.~Pan, {Interlayer exciton formation, relaxation, and transport in TMD van der Waals heterostructures}, Light Sci. Appl. 10~(1) (2021) 72.
\newblock \href {https://doi.org/https://doi.org/10.1038/s41377-021-00500-1} {\path{doi:https://doi.org/10.1038/s41377-021-00500-1}}.

\bibitem{robert2020measurement}
C.~Robert, B.~Han, P.~Kapuscinski, A.~Delhomme, C.~Faugeras, T.~Amand, M.~R. Molas, M.~Bartos, K.~Watanabe, T.~Taniguchi, et~al., {Measurement of the spin-forbidden dark excitons in \ce{MoS2} and \ce{MoSe2} monolayers}, Nat. Commun. 11~(1) (2020) 4037.
\newblock \href {https://doi.org/https://doi.org/10.1038/s41467-020-17608-4} {\path{doi:https://doi.org/10.1038/s41467-020-17608-4}}.

\bibitem{kapuscinski2021rydberg}
P.~Kapu{\'s}ci{\'n}ski, A.~Delhomme, D.~Vaclavkova, A.~O. Slobodeniuk, M.~Grzeszczyk, M.~Bartos, K.~Watanabe, T.~Taniguchi, C.~Faugeras, M.~Potemski, {Rydberg series of dark excitons and the conduction band spin-orbit splitting in monolayer {WSe$_2$}}, Commun. Phys. 4~(1) (2021) 186.
\newblock \href {https://doi.org/https://doi.org/10.1038/s42005-021-00692-3} {\path{doi:https://doi.org/10.1038/s42005-021-00692-3}}.

\bibitem{rivera2015observation}
P.~Rivera, J.~R. Schaibley, A.~M. Jones, J.~S. Ross, S.~Wu, G.~Aivazian, P.~Klement, K.~Seyler, G.~Clark, N.~J. Ghimire, et~al., {Observation of long-lived interlayer excitons in monolayer \ce{MoSe2}--\ce{WSe2} heterostructures}, Nat. Commun. 6~(1) (2015) 6242.
\newblock \href {https://doi.org/https://doi.org/10.1038/ncomms7242} {\path{doi:https://doi.org/10.1038/ncomms7242}}.

\bibitem{li2020dipolar}
W.~Li, X.~Lu, S.~Dubey, L.~Devenica, A.~Srivastava, {Dipolar interactions between localized interlayer excitons in van der Waals heterostructures}, Nat. Mater. 19~(6) (2020) 624--629.
\newblock \href {https://doi.org/https://doi.org/10.1038/s41563-020-0661-4} {\path{doi:https://doi.org/10.1038/s41563-020-0661-4}}.

\bibitem{feng2024highly}
S.~Feng, A.~J. Campbell, M.~Brotons-Gisbert, D.~Andres-Penares, H.~Baek, T.~Taniguchi, K.~Watanabe, B.~Urbaszek, I.~C. Gerber, B.~D. Gerardot, {Highly tunable ground and excited state excitonic dipoles in multilayer {2H-MoSe$_2$}}, Nat. Commun. 15~(1) (2024) 4377.
\newblock \href {https://doi.org/https://doi.org/10.1038/s41467-024-48476-x} {\path{doi:https://doi.org/10.1038/s41467-024-48476-x}}.

\bibitem{li2024wigner}
H.~Li, Z.~Xiang, A.~P. Reddy, T.~Devakul, R.~Sailus, R.~Banerjee, T.~Taniguchi, K.~Watanabe, S.~Tongay, A.~Zettl, et~al., {Wigner molecular crystals from multielectron moir{\'e} artificial atoms}, Science 385~(6704) (2024) 86--91.
\newblock \href {https://doi.org/https://doi.org/10.1126/science.adk1348} {\path{doi:https://doi.org/10.1126/science.adk1348}}.

\bibitem{goldberg2024electronic}
A.~Goldberg, C.~Yannouleas, U.~Landman, {Electronic Wigner-molecule polymeric chains in elongated silicon quantum dots and finite-length quantum wires}, Phys. Rev. Appl. 21~(6) (2024) 064063.
\newblock \href {https://doi.org/https://doi.org/10.1103/PhysRevApplied.21.064063} {\path{doi:https://doi.org/10.1103/PhysRevApplied.21.064063}}.

\bibitem{yannouleas2024crystal}
C.~Yannouleas, U.~Landman, {Crystal-field effects in the formation of Wigner-molecule supercrystals in moir{\'e} transition metal dichalcogenide superlattices}, Phys. Rev. Lett. 133~(24) (2024) 246502.
\newblock \href {https://doi.org/https://doi.org/10.1103/PhysRevLett.133.246502} {\path{doi:https://doi.org/10.1103/PhysRevLett.133.246502}}.

\bibitem{luo2019unique}
W.~Luo, A.~Naseri, J.~Sirker, T.~Chakraborty, {Unique Spin Vortices and Topological Charges in Quantum Dots with Spin-orbit Couplings}, Sci. Rep. 9~(1) (2019) 672.
\newblock \href {https://doi.org/https://doi.org/10.1038/s41598-018-35837-y} {\path{doi:https://doi.org/10.1038/s41598-018-35837-y}}.

\bibitem{luo2019tuning}
W.~Luo, T.~Chakraborty, {Tuning the topological features of quantum-dot hydrogen and helium by a magnetic field}, Phys. Rev. B 100~(8) (2019) 085309.
\newblock \href {https://doi.org/https://doi.org/10.1103/PhysRevB.100.085309} {\path{doi:https://doi.org/10.1103/PhysRevB.100.085309}}.

\bibitem{liu2024unconventional}
Y.~Liu, J.~Li, P.~Liu, Q.~Liu, {Unconventional spin textures emerging from a universal symmetry theory of spin-momentum locking}, npj Quantum Materials 9~(1) (2024) 69.
\newblock \href {https://doi.org/https://doi.org/10.1038/s41535-024-00682-y} {\path{doi:https://doi.org/10.1038/s41535-024-00682-y}}.

\bibitem{xia2023universal}
J.~Xia, X.~Zhang, X.~Liu, Y.~Zhou, M.~Ezawa, {Universal quantum computation based on nanoscale skyrmion helicity qubits in frustrated magnets}, Phys. Rev. Lett. 130~(10) (2023) 106701.
\newblock \href {https://doi.org/https://doi.org/10.1103/PhysRevLett.130.106701} {\path{doi:https://doi.org/10.1103/PhysRevLett.130.106701}}.

\bibitem{yang2021non}
L.-P. Yang, Z.~Jacob, {Non-classical photonic spin texture of quantum structured light}, Commun. Phys. 4~(1) (2021) 221.
\newblock \href {https://doi.org/https://doi.org/10.1038/s42005-021-00726-w} {\path{doi:https://doi.org/10.1038/s42005-021-00726-w}}.

\bibitem{wan2023optical}
C.~Wan, A.~Chong, Q.~Zhan, {Optical spatiotemporal vortices}, eLight 3~(1) (2023) 11.
\newblock \href {https://doi.org/https://doi.org/10.1186/s43593-023-00042-6} {\path{doi:https://doi.org/10.1186/s43593-023-00042-6}}.

\bibitem{bliokh2015spin}
K.~Y. Bliokh, F.~J. Rodr{\'\i}guez-Fortu{\~n}o, F.~Nori, A.~V. Zayats, {Spin--orbit interactions of light}, Nat. Photonics 9~(12) (2015) 796--808.
\newblock \href {https://doi.org/https://doi.org/10.1038/nphoton.2015.201} {\path{doi:https://doi.org/10.1038/nphoton.2015.201}}.

\bibitem{ge2024directscar}
Z.~Ge, A.~M. Graf, J.~Keski-Rahkonen, S.~Slizovskiy, P.~Polizogopoulos, T.~Taniguchi, K.~Watanabe, R.~Van~Haren, D.~Lederman, V.~I. Fal’ko, et~al., {Direct visualization of relativistic quantum scars in graphene quantum dots}, Nature 635~(8040) (2024) 841--846.
\newblock \href {https://doi.org/https://doi.org/10.1038/s41586-024-08190-6} {\path{doi:https://doi.org/10.1038/s41586-024-08190-6}}.

\bibitem{luo2024spin}
W.~Luo, S.~Peng, T.~Chakraborty, Spin textures in quantum dots and quantum rings, in: T.~Chakraborty (Ed.), Encyclopedia of Condensed Matter Physics (2nd Edition), second edition Edition, Academic Press, Oxford, 2024, pp. 400--414.
\newblock \href {https://doi.org/https://doi.org/10.1016/B978-0-323-90800-9.00046-9} {\path{doi:https://doi.org/10.1016/B978-0-323-90800-9.00046-9}}.

\bibitem{larson2013chaos}
J.~Larson, B.~M. Anderson, A.~Altland, Chaos-driven dynamics in spin-orbit-coupled atomic gases, Phys. Rev. A 87~(1) (2013) 013624.
\newblock \href {https://doi.org/https://doi.org/10.1103/PhysRevA.87.013624} {\path{doi:https://doi.org/10.1103/PhysRevA.87.013624}}.

\bibitem{Khomitsky2013}
D.~V. Khomitsky, A.~I. Malyshev, E.~Y. Sherman, M.~Di~Ventra, Spin chaos manifestation in a driven quantum billiard with spin-orbit coupling, Phys. Rev. B 88 (2013) 195407.
\newblock \href {https://doi.org/https://doi.org/10.1103/PhysRevB.88.195407} {\path{doi:https://doi.org/10.1103/PhysRevB.88.195407}}.

\bibitem{kirichenko2020chaotic}
E.~V. Kirichenko, V.~A. Stephanovich, E.~Y. Sherman, Chaotic cyclotron and hall trajectories due to spin-orbit coupling, Ann. Phys. 532~(7) (2020) 2000012.
\newblock \href {https://doi.org/https://doi.org/10.1002/andp.202000012} {\path{doi:https://doi.org/10.1002/andp.202000012}}.

\bibitem{zhang2024controllable}
L.~Zhang, Y.~Hu, Z.~Yao, X.~Liu, W.~Luo, K.~Sun, T.~Chakraborty, {Controllable quantum scars induced by spin--orbit couplings in quantum dots}, Discover Nano 19~(1) (2024) 72.
\newblock \href {https://doi.org/https://doi.org/10.1186/s11671-024-04015-7} {\path{doi:https://doi.org/10.1186/s11671-024-04015-7}}.

\bibitem{ponomarenko2008chaotic}
L.~A. Ponomarenko, F.~Schedin, M.~I. Katsnelson, R.~Yang, E.~W. Hill, K.~S. Novoselov, A.~K. Geim, {Chaotic Dirac billiard in graphene quantum dots}, Science 320~(5874) (2008) 356--358.
\newblock \href {https://doi.org/https://doi.org/10.1126/science.1154663} {\path{doi:https://doi.org/10.1126/science.1154663}}.

\bibitem{bocarsly2025coulombscar}
M.~Bocarsly, I.~Roy, V.~Bhardwaj, M.~Uzan, P.~Ledwith, G.~Shavit, N.~Banu, Y.~Zhou, Y.~Myasoedov, K.~Watanabe, et~al., {Coulomb interactions and migrating Dirac cones imaged by local quantum oscillations in twisted graphene}, Nat. Phys. (2025) 1--9.

\bibitem{keski2019quantum}
J.~Keski-Rahkonen, A.~Ruhanen, E.~Heller, E.~R{\"a}s{\"a}nen, {Quantum lissajous scars}, Phys. Rev. Lett. 123~(21) (2019) 214101.
\newblock \href {https://doi.org/https://doi.org/10.1103/PhysRevLett.123.214101} {\path{doi:https://doi.org/10.1103/PhysRevLett.123.214101}}.

\bibitem{huang2009relativistic}
L.~Huang, Y.-C. Lai, D.~K. Ferry, S.~M. Goodnick, R.~Akis, {Relativistic quantum scars}, Phys. Rev. Lett. 103~(5) (2009) 054101.
\newblock \href {https://doi.org/https://doi.org/10.1103/PhysRevLett.103.054101} {\path{doi:https://doi.org/10.1103/PhysRevLett.103.054101}}.

\bibitem{wang2020quantum}
X.~Wang, G.-D. Wei, {Quantum Scars in Microwave Dielectric Photonic Graphene Billiards}, Chinese Physics Letters 37~(1) (2020) 014201.
\newblock \href {https://doi.org/https://doi.org/10.1088/0256-307X/37/1/014201} {\path{doi:https://doi.org/10.1088/0256-307X/37/1/014201}}.

\bibitem{de2025rydberg}
D.~de~la Fuente~Pico, J.~Levinsen, E.~Laird, M.~M. Parish, F.~M. Marchetti, {Rydberg excitons and polaritons in monolayer transition metal dichalcogenides in a magnetic field}, Phys. Rev. B 111~(3) (2025) 035432.
\newblock \href {https://doi.org/https://doi.org/10.1103/PhysRevB.111.035432} {\path{doi:https://doi.org/10.1103/PhysRevB.111.035432}}.

\bibitem{su2017andreev}
Z.~Su, A.~B. Tacla, M.~Hocevar, D.~Car, S.~R. Plissard, E.~P. Bakkers, A.~J. Daley, D.~Pekker, S.~M. Frolov, {Andreev molecules in semiconductor nanowire double quantum dots}, Nat. Commun. 8~(1) (2017) 585.
\newblock \href {https://doi.org/https://doi.org/10.1038/s41467-017-00665-7} {\path{doi:https://doi.org/10.1038/s41467-017-00665-7}}.

\bibitem{liu2015strong}
X.~Liu, T.~Galfsky, Z.~Sun, F.~Xia, E.-c. Lin, Y.-H. Lee, S.~K{\'e}na-Cohen, V.~M. Menon, {Strong light--matter coupling in two-dimensional atomic crystals}, Nat. Photonics 9~(1) (2015) 30--34.
\newblock \href {https://doi.org/https://doi.org/10.1038/nphoton.2014.304} {\path{doi:https://doi.org/10.1038/nphoton.2014.304}}.

\bibitem{gorski2024nonlocal}
G.~G{\'o}rski, K.~P. W{\'o}jcik, J.~Bara{\'n}ski, I.~Weymann, T.~Doma{\'n}ski, {Nonlocal correlations transmitted between quantum dots via short topological superconductor}, Sci. Rep. 14~(1) (2024) 13848.
\newblock \href {https://doi.org/https://doi.org/10.1038/s41598-024-64578-4} {\path{doi:https://doi.org/10.1038/s41598-024-64578-4}}.

\bibitem{li2021experimental}
M.~Li, I.~Sinev, F.~Benimetskiy, T.~Ivanova, E.~Khestanova, S.~Kiriushechkina, A.~Vakulenko, S.~Guddala, M.~Skolnick, V.~M. Menon, et~al., {Experimental observation of topological Z2 exciton-polaritons in transition metal dichalcogenide monolayers}, Nat. Commun. 12~(1) (2021) 4425.
\newblock \href {https://doi.org/https://doi.org/10.1038/s41467-021-24728-y} {\path{doi:https://doi.org/10.1038/s41467-021-24728-y}}.

\bibitem{dufferwiel2018valley}
S.~Dufferwiel, T.~Lyons, D.~Solnyshkov, A.~Trichet, A.~Catanzaro, F.~Withers, G.~Malpuech, J.~Smith, K.~Novoselov, M.~Skolnick, et~al., {Valley coherent exciton-polaritons in a monolayer semiconductor}, Nat. Commun. 9~(1) (2018) 4797.
\newblock \href {https://doi.org/https://doi.org/10.1038/s41467-018-07249-z} {\path{doi:https://doi.org/10.1038/s41467-018-07249-z}}.

\bibitem{deng2016majorana}
M.~Deng, S.~Vaitiek{\.e}nas, E.~B. Hansen, J.~Danon, M.~Leijnse, K.~Flensberg, J.~Nyg{\aa}rd, P.~Krogstrup, C.~M. Marcus, {Majorana bound state in a coupled quantum-dot hybrid-nanowire system}, Science 354~(6319) (2016) 1557--1562.
\newblock \href {https://doi.org/https://doi.org/10.1126/science.aaf3961} {\path{doi:https://doi.org/10.1126/science.aaf3961}}.

\bibitem{sarma2015majorana}
S.~D. Sarma, M.~Freedman, C.~Nayak, {Majorana zero modes and topological quantum computation}, npj Quantum Inf. 1~(1) (2015) 1--13.
\newblock \href {https://doi.org/https://doi.org/10.1038/npjqi.2015.1} {\path{doi:https://doi.org/10.1038/npjqi.2015.1}}.

\bibitem{estrada2024correlation}
J.~C. Estrada~Salda{\~n}a, A.~Vekris, L.~Pave{\v{s}}i{\v{c}}, R.~{\v{Z}}itko, K.~Grove-Rasmussen, J.~Nyg{\aa}rd, {Correlation between two distant quasiparticles in separate superconducting islands mediated by a single spin}, Nat. Commun. 15~(1) (2024) 3465.
\newblock \href {https://doi.org/https://doi.org/10.1038/s41467-024-47694-7} {\path{doi:https://doi.org/10.1038/s41467-024-47694-7}}.

\bibitem{qi2025competition}
R.~Qi, Q.~Li, Z.~Zhang, J.~Nie, B.~Zou, Z.~Cui, H.~Kim, C.~Sanborn, S.~Chen, J.~Xie, et~al., {Competition between excitonic insulators and quantum Hall states in correlated electron--hole bilayers}, Nat. Mater. (2025) 1--7\href {https://doi.org/https://doi.org/10.1038/s41563-025-02316-5} {\path{doi:https://doi.org/10.1038/s41563-025-02316-5}}.

\bibitem{fu2024bilayer}
D.~D. Dai, L.~Fu, {Strong-coupling phases of trions and excitons in electron-hole bilayers at commensurate densities}, Phys. Rev. Lett. 132~(19) (2024) 196202.
\newblock \href {https://doi.org/https://doi.org/10.1103/PhysRevLett.132.196202} {\path{doi:https://doi.org/10.1103/PhysRevLett.132.196202}}.

\bibitem{Aoki02102025}
H.~Aoki, \href{https://doi.org/10.1080/00107514.2025.2550105}{Flat bands in condensed-matter systems – perspective for magnetism and superconductivity}, Contemporary Physics 66~(1-4) (2025) 1--38.
\newblock \href {https://doi.org/10.1080/00107514.2025.2550105} {\path{doi:10.1080/00107514.2025.2550105}}.
\newline\urlprefix\url{https://doi.org/10.1080/00107514.2025.2550105}

\bibitem{choi2019electronic}
Y.~Choi, J.~Kemmer, Y.~Peng, A.~Thomson, H.~Arora, R.~Polski, Y.~Zhang, H.~Ren, J.~Alicea, G.~Refael, et~al., {Electronic correlations in twisted bilayer graphene near the magic angle}, Nat. Phys. 15~(11) (2019) 1174--1180.
\newblock \href {https://doi.org/https://doi.org/10.1038/s41567-019-0606-5} {\path{doi:https://doi.org/10.1038/s41567-019-0606-5}}.

\bibitem{oh2021evidence}
M.~Oh, K.~P. Nuckolls, D.~Wong, R.~L. Lee, X.~Liu, K.~Watanabe, T.~Taniguchi, A.~Yazdani, {Evidence for unconventional superconductivity in twisted bilayer graphene}, Nature 600~(7888) (2021) 240--245.
\newblock \href {https://doi.org/https://doi.org/10.1038/s41586-021-04121-x} {\path{doi:https://doi.org/10.1038/s41586-021-04121-x}}.

\bibitem{ko2023atomic}
W.~Ko, S.~Y. Song, J.~Yan, J.~L. Lado, P.~Maksymovych, {Atomic-Scale Andreev Probe of Unconventional Superconductivity}, Nano Lett. 23~(17) (2023) 8310--8318.
\newblock \href {https://doi.org/https://doi.org/10.3390/nano15100737} {\path{doi:https://doi.org/10.3390/nano15100737}}.

\bibitem{park2026experimental}
J.~M. Park, S.~Sun, K.~Watanabe, T.~Taniguchi, P.~Jarillo-Herrero, {Experimental evidence for nodal superconducting gap in moir{\'e} graphene}, Science 391~(6780) (2026) 79--83.
\newblock \href {https://doi.org/https://doi.org/10.1126/science.adv8376} {\path{doi:https://doi.org/10.1126/science.adv8376}}.

\bibitem{Wu2018PRL}
F.~Wu, A.~H. MacDonald, I.~Martin, {Theory of Phonon-Mediated Superconductivity in Twisted Bilayer Graphene}, Phys. Rev. Lett. 121 (2018) 257001.
\newblock \href {https://doi.org/https://doi.org/10.1103/PhysRevLett.121.257001} {\path{doi:https://doi.org/10.1103/PhysRevLett.121.257001}}.

\bibitem{Peltonen2018PRB}
T.~J. Peltonen, R.~Ojajärvi, T.~T. Heikkilä, Mean-field theory for superconductivity in twisted bilayer graphene, Phys. Rev. B 98 (2018) 220504.
\newblock \href {https://doi.org/https://doi.org/10.1103/PhysRevB.98.220504} {\path{doi:https://doi.org/10.1103/PhysRevB.98.220504}}.

\bibitem{Lian2019PRL}
B.~Lian, Z.~Wang, B.~A. Bernevig, {Twisted Bilayer Graphene: A Phonon-Driven Superconductor}, Phys. Rev. Lett. 122 (2019) 257002.
\newblock \href {https://doi.org/https://doi.org/10.1103/PhysRevLett.122.257002} {\path{doi:https://doi.org/10.1103/PhysRevLett.122.257002}}.

\bibitem{zhuprb2025}
J.~Zhu, Y.-Z. Chou, M.~Xie, S.~Das~Sarma, Superconductivity in twisted transition metal dichalcogenide homobilayers, Phys. Rev. B 111 (2025) L060501.
\newblock \href {https://doi.org/https://doi.org/10.1103/PhysRevB.111.L060501} {\path{doi:https://doi.org/10.1103/PhysRevB.111.L060501}}.

\bibitem{zhuprb2026}
Z.~Zhu, T.~P. Devereaux, Microscopic theory for electron-phonon coupling in twisted bilayer graphene, Phys. Rev. B 113 (2026) 035446.
\newblock \href {https://doi.org/https://doi.org/10.1103/tpww-cq4k} {\path{doi:https://doi.org/10.1103/tpww-cq4k}}.

\bibitem{KL2019}
J.~Gonz\'alez, T.~Stauber, {Kohn-Luttinger Superconductivity in Twisted Bilayer Graphene}, Phys. Rev. Lett. 122 (2019) 026801.
\newblock \href {https://doi.org/https://doi.org/10.1103/PhysRevLett.122.026801} {\path{doi:https://doi.org/10.1103/PhysRevLett.122.026801}}.

\bibitem{Chen2025FiniteMomentumSC}
Y.~Chen, C.~Xu, Y.~Zhang, C.~Schrade, {Finite-momentum superconductivity from chiral bands in twisted MoTe$_2$}, Nat. Commun. (2025).
\newblock \href {https://doi.org/https://doi.org/10.1038/s41467-025-67836-9} {\path{doi:https://doi.org/10.1038/s41467-025-67836-9}}.

\bibitem{cao2021pauli}
Y.~Cao, J.~M. Park, K.~Watanabe, T.~Taniguchi, P.~Jarillo-Herrero, {Pauli-limit violation and re-entrant superconductivity in moir{\'e} graphene}, Nature 595~(7868) (2021) 526--531.
\newblock \href {https://doi.org/https://doi.org/10.1038/s41586-021-03685-y} {\path{doi:https://doi.org/10.1038/s41586-021-03685-y}}.

\bibitem{LIU2024515}
Z.~Liu, E.~J. Bergholtz, Recent developments in fractional chern insulators, in: T.~Chakraborty (Ed.), Encyclopedia of Condensed Matter Physics (2nd Edition), Academic Press, Oxford, 2024, pp. 515--538.
\newblock \href {https://doi.org/https://doi.org/10.1016/B978-0-323-90800-9.00136-0} {\path{doi:https://doi.org/10.1016/B978-0-323-90800-9.00136-0}}.

\bibitem{chen2025robust}
F.~Chen, W.-W. Luo, W.~Zhu, D.~Sheng, {Robust non-Abelian even-denominator fractional Chern insulator in twisted bilayer {MoTe$_2$}}, Nat. Commun. 16~(1) (2025) 2115.
\newblock \href {https://doi.org/https://doi.org/10.1038/s41467-025-57326-3} {\path{doi:https://doi.org/10.1038/s41467-025-57326-3}}.

\bibitem{aronson2025displacement}
S.~H. Aronson, T.~Han, Z.~Lu, Y.~Yao, J.~P. Butler, K.~Watanabe, T.~Taniguchi, L.~Ju, R.~C. Ashoori, {Displacement field-controlled fractional Chern insulators and charge density waves in a graphene/{hBN} moir{\'e} superlattice}, Phys. Rev. X 15~(3) (2025) 031026.
\newblock \href {https://doi.org/https://doi.org/10.1103/75gl-jzl6} {\path{doi:https://doi.org/10.1103/75gl-jzl6}}.

\bibitem{FQH2011}
T.~Neupert, L.~Santos, C.~Chamon, C.~Mudry, Fractional quantum hall states at zero magnetic field, Phys. Rev. Lett. 106 (2011) 236804.
\newblock \href {https://doi.org/https://doi.org/10.1103/PhysRevLett.106.236804} {\path{doi:https://doi.org/10.1103/PhysRevLett.106.236804}}.

\bibitem{xie2021fractional}
Y.~Xie, A.~T. Pierce, J.~M. Park, D.~E. Parker, E.~Khalaf, P.~Ledwith, Y.~Cao, S.~H. Lee, S.~Chen, P.~R. Forrester, et~al., Fractional chern insulators in magic-angle twisted bilayer graphene, Nature 600~(7889) (2021) 439--443.
\newblock \href {https://doi.org/https://doi.org/10.1038/s41586-021-04002-3} {\path{doi:https://doi.org/10.1038/s41586-021-04002-3}}.

\bibitem{park2026fci}
H.~Park, W.~Li, C.~Hu, C.~Beach, M.~Gon{\c{c}}alves, J.~F. Mendez-Valderrama, J.~Herzog-Arbeitman, T.~Taniguchi, K.~Watanabe, D.~Cobden, et~al., Observation of dissipationless fractional chern insulator, Nat. Phys. (2026) 1--7\href {https://doi.org/https://doi.org/10.1038/s41567-025-03167-2} {\path{doi:https://doi.org/10.1038/s41567-025-03167-2}}.

\bibitem{lee2026local}
N.~Lee, H.~Park, S.~Jung, B.~Jang, S.~Lee, J.~Jang, {Local thermodynamic DOS measurement and twist-angle mapping in graphene--{hBN} superlattices}, Appl. Phys. Lett. 128~(1) (2026).
\newblock \href {https://doi.org/https://doi.org/10.1063/5.0309548} {\path{doi:https://doi.org/10.1063/5.0309548}}.

\bibitem{tomarken2019electronic}
S.~L. Tomarken, Y.~Cao, A.~Demir, K.~Watanabe, T.~Taniguchi, P.~Jarillo-Herrero, R.~Ashoori, {Electronic compressibility of magic-angle graphene superlattices}, Phys. Rev. Lett. 123~(4) (2019) 046601.
\newblock \href {https://doi.org/https://doi.org/10.1103/PhysRevLett.123.046601} {\path{doi:https://doi.org/10.1103/PhysRevLett.123.046601}}.

\bibitem{kang2024evidence}
K.~Kang, B.~Shen, Y.~Qiu, Y.~Zeng, Z.~Xia, K.~Watanabe, T.~Taniguchi, J.~Shan, K.~F. Mak, {Evidence of the fractional quantum spin Hall effect in moir{\'e} {MoTe$_2$}}, Nature 628~(8008) (2024) 522--526.
\newblock \href {https://doi.org/https://doi.org/10.1038/s41586-024-07214-5} {\path{doi:https://doi.org/10.1038/s41586-024-07214-5}}.

\bibitem{xu2023observation}
F.~Xu, Z.~Sun, T.~Jia, C.~Liu, C.~Xu, C.~Li, Y.~Gu, K.~Watanabe, T.~Taniguchi, B.~Tong, et~al., {Observation of integer and fractional quantum anomalous Hall effects in twisted bilayer {MoTe$_2$}}, Phys. Rev. X 13~(3) (2023) 031037.
\newblock \href {https://doi.org/https://doi.org/10.1103/PhysRevX.13.031037} {\path{doi:https://doi.org/10.1103/PhysRevX.13.031037}}.

\bibitem{cai2023signatures}
J.~Cai, E.~Anderson, C.~Wang, X.~Zhang, X.~Liu, W.~Holtzmann, Y.~Zhang, F.~Fan, T.~Taniguchi, K.~Watanabe, et~al., {ignatures of fractional quantum anomalous Hall states in twisted {MoTe$_2$}}, Nature 622~(7981) (2023) 63--68.
\newblock \href {https://doi.org/https://doi.org/10.1038/s41586-023-06289-w} {\path{doi:https://doi.org/10.1038/s41586-023-06289-w}}.

\bibitem{redekop2024direct}
E.~Redekop, C.~Zhang, H.~Park, J.~Cai, E.~Anderson, O.~Sheekey, T.~Arp, G.~Babikyan, S.~Salters, K.~Watanabe, et~al., {Direct magnetic imaging of fractional Chern insulators in twisted {MoTe$_2$}}, Nature 635~(8039) (2024) 584--589.
\newblock \href {https://doi.org/https://doi.org/10.1038/s41586-024-08153-x} {\path{doi:https://doi.org/10.1038/s41586-024-08153-x}}.

\bibitem{sharma2024topological}
P.~Sharma, Y.~Peng, D.~Sheng, {Topological quantum phase transitions driven by a displacement field in twisted {MoTe$_2$} bilayers}, Phys. Rev. B 110~(12) (2024) 125142.
\newblock \href {https://doi.org/https://doi.org/10.1103/PhysRevB.110.125142} {\path{doi:https://doi.org/10.1103/PhysRevB.110.125142}}.

\bibitem{OpticalProb1}
C.~Sch\"uller, K.-B. Broocks, P.~Schr\"oter, C.~Heyn, D.~Heitmann, M.~Bichler, W.~Wegscheider, T.~Chakraborty, V.~M. Apalkov, Optical probing of a fractionally charged quasihole in an incompressible liquid, Phys. Rev. Lett. 91 (2003) 116403.
\newblock \href {https://doi.org/https://doi.org/10.1103/PhysRevLett.91.116403} {\path{doi:https://doi.org/10.1103/PhysRevLett.91.116403}}.

\bibitem{OpticalProb2}
C.~Sch\"uller, K.-B. Broocks, P.~Schröter, C.~Heyn, D.~Heitmann, M.~Bichler, W.~Wegscheider, T.~Chakraborty, V.~Apalkov, How to probe a fractionally charged quasihole?, Physica E: Low-dimensional Systems and Nanostructures 22~(1) (2004) 131--134.
\newblock \href {https://doi.org/https://doi.org/10.1016/j.physe.2003.11.233} {\path{doi:https://doi.org/10.1016/j.physe.2003.11.233}}.

\bibitem{APALKOV2002289}
V.~Apalkov, T.~Chakraborty, Interaction of a quantum dot with an incompressible two-dimensional electron gas, Physica E: Low-dimensional Systems and Nanostructures 14~(3) (2002) 289--293.
\newblock \href {https://doi.org/https://doi.org/10.1016/S1386-9477(01)00266-1} {\path{doi:https://doi.org/10.1016/S1386-9477(01)00266-1}}.

\bibitem{li2026signatures}
W.~Li, C.~Wang~Beach, C.~Hu, T.~Taniguchi, K.~Watanabe, J.-H. Chu, A.~Imamo{\u{g}}lu, T.~Cao, D.~Xiao, X.~Xu, Signatures of fractional charges via anyon--trions in twisted \ce{MoTe2}, Nature (2026) 1--6\href {https://doi.org/https://doi.org/10.1038/s41586-026-10101-w} {\path{doi:https://doi.org/10.1038/s41586-026-10101-w}}.

\bibitem{halperin1984}
B.~I. Halperin, Statistics of quasiparticles and the hierarchy of fractional quantized hall states, Phys. Rev. Lett. 52 (1984) 1583.
\newblock \href {https://doi.org/https://doi.org/10.1103/PhysRevLett.52.1583} {\path{doi:https://doi.org/10.1103/PhysRevLett.52.1583}}.

\bibitem{arovas1984}
D.~Arovas, J.~R. Schrieffer, F.~Wilczek, Sfractional statistics and the quantum hall effect, Phys. Rev. Lett. 53 (1984) 722.
\newblock \href {https://doi.org/https://doi.org/10.1103/PhysRevLett.53.722} {\path{doi:https://doi.org/10.1103/PhysRevLett.53.722}}.

\bibitem{berman2016high}
O.~L. Berman, R.~Y. Kezerashvili, {High-temperature superfluidity of the two-component Bose gas in a transition metal dichalcogenide bilayer}, Phys. Rev. B 93~(24) (2016) 245410.
\newblock \href {https://doi.org/https://doi.org/10.1103/PhysRevB.93.245410} {\path{doi:https://doi.org/10.1103/PhysRevB.93.245410}}.

\bibitem{baldini2019exciton}
E.~Baldini, A.~Dominguez, T.~Palmieri, O.~Cannelli, A.~Rubio, P.~Ruello, M.~Chergui, {Exciton control in a room temperature bulk semiconductor with coherent strain pulses}, Sci. Adv. 5~(11) (2019) eaax2937.
\newblock \href {https://doi.org/https://doi.org/10.1126/sciadv.aax2937} {\path{doi:https://doi.org/10.1126/sciadv.aax2937}}.

\bibitem{zhang2024room}
L.~Zhang, M.~Ge, B.~Zhao, K.~Xu, W.~Xie, Z.~Zou, W.~Li, J.~Zhao, T.~Wang, W.~Du, {Room-Temperature Exciton Polaritons in a Monolayer Molecular Crystal}, Nano Lett. 24~(50) (2024) 16072--16080.
\newblock \href {https://doi.org/https://doi.org/10.1021/acs.nanolett.4c04562} {\path{doi:https://doi.org/10.1021/acs.nanolett.4c04562}}.

\bibitem{vaquero2020Rydberg}
D.~Vaquero, V.~Cleric{\`o}, J.~Salvador-S{\'a}nchez, A.~Mart{\'\i}n-Ramos, E.~D{\'\i}az, F.~Dom{\'\i}nguez-Adame, Y.~M. Meziani, E.~Diez, J.~Quereda, Excitons, trions and rydberg states in monolayer \ce{MoS2} revealed by low-temperature photocurrent spectroscopy, Commun. Phys. 3~(1) (2020) 194.
\newblock \href {https://doi.org/https://doi.org/10.1038/s42005-020-00460-9} {\path{doi:https://doi.org/10.1038/s42005-020-00460-9}}.

\bibitem{dostal2018direct}
J.~Dost{\'a}l, F.~Fennel, F.~Koch, S.~Herbst, F.~W{\"u}rthner, T.~Brixner, {Direct observation of exciton--exciton interactions}, Nat. Commun. 9~(1) (2018) 2466.
\newblock \href {https://doi.org/https://doi.org/10.1038/s41467-018-04884-4} {\path{doi:https://doi.org/10.1038/s41467-018-04884-4}}.

\bibitem{2024exexprx}
A.~Steinhoff, E.~Wietek, M.~Florian, T.~Schulz, T.~Taniguchi, K.~Watanabe, S.~Zhao, A.~H\"ogele, F.~Jahnke, A.~Chernikov, {Exciton-Exciton Interactions in Van der Waals Heterobilayers}, Phys. Rev. X 14 (2024) 031025.
\newblock \href {https://doi.org/https://doi.org/10.1103/PhysRevX.14.031025} {\path{doi:https://doi.org/10.1103/PhysRevX.14.031025}}.

\bibitem{Hubbardtmd}
F.~Wu, T.~Lovorn, E.~Tutuc, A.~H. MacDonald, Hubbard model physics in transition metal dichalcogenide moir\'e bands, Phys. Rev. Lett. 121 (2018) 026402.
\newblock \href {https://doi.org/https://doi.org/10.1103/PhysRevLett.121.026402} {\path{doi:https://doi.org/10.1103/PhysRevLett.121.026402}}.

\bibitem{brem2020tunable}
S.~Brem, C.~Linderalv, P.~Erhart, E.~Malic, Tunable phases of moir{\'e} excitons in van der waals heterostructures, Nano Lett. 20~(12) (2020) 8534--8540.
\newblock \href {https://doi.org/https://doi.org/10.1021/acs.nanolett.0c03019} {\path{doi:https://doi.org/10.1021/acs.nanolett.0c03019}}.

\bibitem{herrera2025moire}
S.~A. Herrera-Gonz{\'a}lez, H.~A. Lara-Garc{\'\i}a, G.~Pirruccio, D.~A. Ruiz-Tijerina, A.~Camacho-Guardian, Moir{\'e} excitons and exciton--polaritons: a review, J. Phys.: Condens. Matter. 37~(48) (2025) 483002.
\newblock \href {https://doi.org/https://doi.org/10.1088/1361-648X/ae20e5} {\path{doi:https://doi.org/10.1088/1361-648X/ae20e5}}.

\bibitem{kundu2023exciton}
S.~Kundu, T.~Amit, H.~Krishnamurthy, M.~Jain, S.~Refaely-Abramson, Exciton fine structure in twisted transition metal dichalcogenide heterostructures, npj Comput Mater. 9~(1) (2023) 186.
\newblock \href {https://doi.org/https://doi.org/10.1038/s41524-023-01145-x} {\path{doi:https://doi.org/10.1038/s41524-023-01145-x}}.

\bibitem{zheng2023localization}
H.~Zheng, B.~Wu, S.~Li, J.~Ding, J.~He, Z.~Liu, C.-T. Wang, J.-T. Wang, A.~Pan, Y.~Liu, Localization-enhanced moir{\'e} exciton in twisted transition metal dichalcogenide heterotrilayer superlattices, Light Sci Appl. 12~(1) (2023) 117.
\newblock \href {https://doi.org/https://doi.org/10.1038/s41377-023-01171-w} {\path{doi:https://doi.org/10.1038/s41377-023-01171-w}}.

\bibitem{guo2025moire}
J.~Guo, Z.~H. Withers, Z.~Li, B.~Hou, A.~Adler, J.~Ding, V.~Chang~Lee, R.~K. Kawakami, G.~Sch{\"o}nhense, A.~Kunin, et~al., {Moir{\'e}-controllable exciton localization and dynamics through spatially-modulated inter-and intralayer excitons in a {MoSe$_2$/WS$_2$} heterobilayer}, Nat. Commun. (2025).
\newblock \href {https://doi.org/https://doi.org/10.1038/s41467-025-66127-7} {\path{doi:https://doi.org/10.1038/s41467-025-66127-7}}.

\bibitem{kundu2025atomic}
S.~Kundu, I.~Maity, R.~Bajaj, H.~Krishnamurthy, M.~Jain, {Atomic relaxation and flat bands in strain-engineered transition metal dichalcogenide bilayer moir{\'e} systems}, Phys. Rev. B 112~(15) (2025) 155412.
\newblock \href {https://doi.org/https://doi.org/10.1103/75fh-xf7g} {\path{doi:https://doi.org/10.1103/75fh-xf7g}}.

\bibitem{PRRPolarons}
A.~Julku, S.~Ding, G.~M. Bruun, Exciton interacting with a moir\'e lattice: Polarons, strings, and optical probing of spin correlations, Phys. Rev. Res. 6 (2024) 033119.
\newblock \href {https://doi.org/https://doi.org/10.1103/PhysRevResearch.6.033119} {\path{doi:https://doi.org/10.1103/PhysRevResearch.6.033119}}.

\bibitem{fitzgerald2022twist}
J.~M. Fitzgerald, J.~J. Thompson, E.~Malic, Twist angle tuning of moir{\'e} exciton polaritons in van der waals heterostructures, Nano Lett. 22~(11) (2022) 4468--4474.
\newblock \href {https://doi.org/https://doi.org/10.1021/acs.nanolett.2c01175} {\path{doi:https://doi.org/10.1021/acs.nanolett.2c01175}}.

\bibitem{wang2025temperature_hubbard_u}
Z.~Wang, H.~Xu, S.~Chen, Y.~Wang, R.~Han, Z.~Sun, X.~Zhang, S.~Huang, W.~Gao, H.~Liu, D.~Liu, Temperature dependence of excitonic {Hubbard} {$U$} interaction in {WS$_2$/WSe$_2$} moir{\'e} superlattices, Nanoscale 17 (2025) 24263--24271.
\newblock \href {https://doi.org/https://doi.org/10.1039/D5NR02149A} {\path{doi:https://doi.org/10.1039/D5NR02149A}}.

\bibitem{jiang2025tuning}
Y.~Jiang, Y.~She, X.~Cheng, Q.~Tan, J.~Yang, Y.~Zhao, P.~Liu, M.~Wu, X.~Dai, Z.~Wang, et~al., Tuning valley polarization of moir{\'e} trapped biexcitons by fine structure occupation in \ce{WS2}/\ce{WSe2} heterostructures, Nat. Commun. (2025).
\newblock \href {https://doi.org/https://doi.org/10.1038/s41467-025-67846-7} {\path{doi:https://doi.org/10.1038/s41467-025-67846-7}}.

\bibitem{alexeev2019resonantly}
E.~M. Alexeev, D.~A. Ruiz-Tijerina, M.~Danovich, M.~J. Hamer, D.~J. Terry, P.~K. Nayak, S.~Ahn, S.~Pak, J.~Lee, J.~I. Sohn, et~al., Resonantly hybridized excitons in moir{\'e} superlattices in van der \text{W}aals heterostructures, Nature 567~(7746) (2019) 81--86.
\newblock \href {https://doi.org/https://doi.org/10.1038/s41586-019-0986-9} {\path{doi:https://doi.org/10.1038/s41586-019-0986-9}}.

\bibitem{SnTe1}
K.~Chang, J.~Liu, H.~Lin, N.~Wang, K.~Zhao, A.~Zhang, F.~Jin, Y.~Zhong, X.~Hu, W.~Duan, Q.~Zhang, L.~Fu, Q.-K. Xue, X.~Chen, , S.-H. Ji, Discovery of robust in-plane ferroelectricity in atomic-thick snte, Science 353 (2016) 274--278.
\newblock \href {https://doi.org/https://doi.org/10.1126/science.aad8609} {\path{doi:https://doi.org/10.1126/science.aad8609}}.

\bibitem{Zhou2017NanoLett}
Y.~Zhou, D.~Wu, Y.~Zhu, Y.~Cho, Q.~He, X.~Yang, K.~Herrera, Z.~Chu, Y.~Han, M.~C. Downer, H.~Peng, K.~Lai, Out-of-plane piezoelectricity and ferroelectricity in layered {$\alpha$}-in$_2$se$_3$ nanoflakes, Nano Lett. 7~(9) (2017) 5508--5513.
\newblock \href {https://doi.org/https://doi.org/10.1021/acs.nanolett.7b02198} {\path{doi:https://doi.org/10.1021/acs.nanolett.7b02198}}.

\bibitem{SnTe2}
K.~Chang, J.-R. Ji, Z.~Gao, A.~Huamán, R.-Q. Cao, W.-L. Wang, C.~Yue, C.~Liu, S.~B.-L. . S. S.~P. Parkin, Ferroelectrically switched valley-dependent transmission in snte-pbte-snte monolayer lateral heterostructures, Nat. Commun. 16 (2025) 11063.
\newblock \href {https://doi.org/https://doi.org/10.1038/s41467-025-66005-2} {\path{doi:https://doi.org/10.1038/s41467-025-66005-2}}.

\bibitem{slidingferro}
P.~Tang, G.~E.~W. Bauer, \href{https://link.aps.org/doi/10.1103/PhysRevLett.130.176801}{Sliding phase transition in ferroelectric van der waals bilayers}, Phys. Rev. Lett. 130 (2023) 176801.
\newblock \href {https://doi.org/10.1103/PhysRevLett.130.176801} {\path{doi:10.1103/PhysRevLett.130.176801}}.
\newline\urlprefix\url{https://link.aps.org/doi/10.1103/PhysRevLett.130.176801}

\bibitem{3RMoS2}
X.~Wang, K.~Yasuda, Y.~Zhang, S.~Liu, K.~Watanabe, T.~Taniguchi, J.~Hone, L.~Fu, P.~Jarillo-Herrero, Interfacial ferroelectricity in rhombohedral-stacked bilayer transition metal dichalcogenides, Nat. Nanotechnol. 17 (2022) 367--371.
\newblock \href {https://doi.org/https://doi.org/10.1038/s41565-021-01059-z} {\path{doi:https://doi.org/10.1038/s41565-021-01059-z}}.

\bibitem{twistferroMoS2}
T.~H. Yang, B.-W. Liang, H.-C. Hu, F.-X. Chen, S.-Z. Ho, W.-H. Chang, L.~Yang, H.-C. Lo, T.-H. Kuo, J.-H. Chen, P.-Y. Lin, K.~B. Simbulan, Z.-F. Luo, A.~C. Chang, Y.-H. Kuo, Y.-S. Ku, Y.-C. Chen, Y.-J. Huang, Y.-C. Chang, Y.-F. Chiang, T.-H. Lu, M.-H. Lee, K.-S. Li, M.~Wu, Y.-C. Chen, C.-L. Lin, Y.-W. Lan, Ferroelectric transistors based on shear-transformation-mediated rhombohedral-stacked molybdenum disulfide, Nat. Electron. 7 (2024) 29--38.
\newblock \href {https://doi.org/https://doi.org/10.1038/s41928-023-01073-0} {\path{doi:https://doi.org/10.1038/s41928-023-01073-0}}.

\bibitem{Wu2025ACSNano}
G.~Wu, F.~Yu, J.~Yi, H.~Liu, X.~Fan, C.~Li, C.~Zhu, X.~Sun, Y.~Liu, Q.~Shuai, T.~Xie, S.~Li, Y.~Zhou, D.~Li, A.~Pan, Opto-electrical decoupling of phototransistors via light-induced ferroelectric depolarization for in-sensor computing, ACS nano 19~(22) (2025) 20980--20990.
\newblock \href {https://doi.org/https://doi.org/10.1021/acsnano.5c04090} {\path{doi:https://doi.org/10.1021/acsnano.5c04090}}.

\bibitem{JiaoAPR}
C.~Jiao, S.~Pei, Z.~Zhang, C.~Li, J.~Zhu, J.~Qin, M.~Zhang, T.~Wen, Y.~Zhou, Z.~Wang, et~al., Exploration toward a new stacking-pressure phase diagram in bilayer aa-and ab-mos$_2$, Appl. Phys. Rev. 11~(3) (2024).
\newblock \href {https://doi.org/https://doi.org/10.1063/5.0202832} {\path{doi:https://doi.org/10.1063/5.0202832}}.

\bibitem{JiaoNanoRes}
J.~Qin, C.~Jiao, S.~Pei, Z.~Zhang, C.~Li, J.~Liang, Y.~Zhou, Z.~Wang, W.~Cai, J.~Xia, et~al., Completing the stacking--pressure phase diagram in bilayer mos$_2$, Nano Res. (2025).
\newblock \href {https://doi.org/https://doi.org/10.26599/NR.2025.94908129} {\path{doi:https://doi.org/10.26599/NR.2025.94908129}}.

\bibitem{Wu2023CoupledFerroelectricity}
F.~Wu, L.~Li, Q.~Xu, L.~Liu, Y.~Yuan, J.~Zhao, Z.~Huang, X.~Zan, K.~Watanabe, T.~Taniguchi, D.~Shi, L.~Xian, W.~Yang, L.~Du, G.~Zhang, \href{https://doi.org/10.1088/0256-307X/40/4/047303}{Coupled ferroelectricity and correlated states in a twisted quadrilayer mos2 moiré superlattice}, Chinese Physics Letters 40~(4) (2023) 047303.
\newblock \href {https://doi.org/10.1088/0256-307X/40/4/047303} {\path{doi:10.1088/0256-307X/40/4/047303}}.
\newline\urlprefix\url{https://doi.org/10.1088/0256-307X/40/4/047303}

\bibitem{landau2013statistical}
L.~D. Landau, E.~M. Lifshitz, Statistical physics: volume 5, Vol.~5, Elsevier, 2013.
\newblock \href {https://doi.org/https://doi.org/10.1016/c2009-0-24487-4} {\path{doi:https://doi.org/10.1016/c2009-0-24487-4}}.

\bibitem{lines2001principles}
M.~E. Lines, A.~M. Glass, Principles and applications of ferroelectrics and related materials, Oxford university press, 2001.
\newblock \href {https://doi.org/https://doi.org/10.1093/acprof:oso/9780198507789.001.0001} {\path{doi:https://doi.org/10.1093/acprof:oso/9780198507789.001.0001}}.

\bibitem{resta1992theory}
R.~Resta, Theory of the electric polarization in crystals, Ferroelectrics 136~(1) (1992) 51--55.
\newblock \href {https://doi.org/https://doi.org/10.1080/00150199208016065} {\path{doi:https://doi.org/10.1080/00150199208016065}}.

\bibitem{king1993theory}
R.~King-Smith, D.~Vanderbilt, Theory of polarization of crystalline solids, Phys. Rev. B 47~(3) (1993) 1651.
\newblock \href {https://doi.org/https://doi.org/10.1103/PhysRevB.47.1651} {\path{doi:https://doi.org/10.1103/PhysRevB.47.1651}}.

\bibitem{resta1994macroscopic}
R.~Resta, Macroscopic polarization in crystalline dielectrics: the geometric phase approach, Rev. Mod. Phys. 66~(3) (1994) 899.
\newblock \href {https://doi.org/https://doi.org/10.1103/RevModPhys.66.899} {\path{doi:https://doi.org/10.1103/RevModPhys.66.899}}.

\bibitem{fan2025edge}
W.-C. Fan, Z.~Guan, L.-Q. Wei, H.-W. Xu, W.-Y. Tong, M.~Tian, N.~Wan, C.-S. Yao, J.-D. Zheng, B.-B. Chen, et~al., {Edge polarization topology integrated with sliding ferroelectricity in Moir{\'e} system}, Nat. Commun. 16~(1) (2025) 3557.
\newblock \href {https://doi.org/https://doi.org/10.1038/s41467-025-58877-1} {\path{doi:https://doi.org/10.1038/s41467-025-58877-1}}.

\bibitem{yang2023atypical}
L.~Yang, S.~Ding, J.~Gao, M.~Wu, {Atypical sliding and moir{\'e} ferroelectricity in pure multilayer graphene}, Phys. Rev. Lett. 131~(9) (2023) 096801.
\newblock \href {https://doi.org/https://doi.org/10.1103/PhysRevLett.131.096801} {\path{doi:https://doi.org/10.1103/PhysRevLett.131.096801}}.

\bibitem{ji2023general}
J.~Ji, G.~Yu, C.~Xu, H.~Xiang, General theory for bilayer stacking ferroelectricity, Phys. Rev. Lett. 130~(14) (2023) 146801.
\newblock \href {https://doi.org/https://doi.org/10.1103/PhysRevLett.130.146801} {\path{doi:https://doi.org/10.1103/PhysRevLett.130.146801}}.

\bibitem{heissenbuttel2021valley}
M.-C. Hei{\ss}enb\"uttel, T.~Deilmann, P.~Kr\"uger, M.~Rohlfing, {Valley-dependent interlayer excitons in magnetic {WSe$_2$/CrI$_3$}}, Nano Lett. 21~(12) (2021) 5173--5178.
\newblock \href {https://doi.org/https://doi.org/10.1021/acs.nanolett.1c01232} {\path{doi:https://doi.org/10.1021/acs.nanolett.1c01232}}.

\bibitem{liu2017wafer}
S.~Liu, X.~Yuan, Y.~Zou, Y.~Sheng, C.~Huang, E.~Zhang, J.~Ling, Y.~Liu, W.~Wang, C.~Zhang, et~al., {Wafer-scale two-dimensional ferromagnetic {Fe$_3$GeTe$_2$} thin films grown by molecular beam epitaxy}, npj 2D Mater. Appl. 1~(1) (2017) 30.
\newblock \href {https://doi.org/https://doi.org/10.1038/s41699-017-0033-3} {\path{doi:https://doi.org/10.1038/s41699-017-0033-3}}.

\bibitem{xu2022coexisting}
Y.~Xu, A.~Ray, Y.-T. Shao, S.~Jiang, K.~Lee, D.~Weber, J.~E. Goldberger, K.~Watanabe, T.~Taniguchi, D.~A. Muller, et~al., {Coexisting ferromagnetic--antiferromagnetic state in twisted bilayer {CrI$_3$}}, Nat. Nanotechnol. 17~(2) (2022) 143--147.
\newblock \href {https://doi.org/https://doi.org/10.1038/s41565-021-01014-y} {\path{doi:https://doi.org/10.1038/s41565-021-01014-y}}.

\bibitem{ju2015topological}
L.~Ju, Z.~Shi, N.~Nair, Y.~Lv, C.~Jin, J.~Velasco~Jr, C.~Ojeda-Aristizabal, H.~A. Bechtel, M.~C. Martin, A.~Zettl, et~al., {Topological valley transport at bilayer graphene domain walls}, Nature 520~(7549) (2015) 650--655.
\newblock \href {https://doi.org/https://doi.org/10.1038/nature14364} {\path{doi:https://doi.org/10.1038/nature14364}}.

\bibitem{wang2016quantum}
X.~Wang, G.~Sun, N.~Li, P.~Chen, {Quantum dots derived from two-dimensional materials and their applications for catalysis and energy}, Chem. Soc. Rev. 45~(8) (2016) 2239--2262.
\newblock \href {https://doi.org/https://doi.org/10.1039/C5CS00811E} {\path{doi:https://doi.org/10.1039/C5CS00811E}}.

\bibitem{ren2019review}
S.~Ren, Q.~Tan, J.~Zhang, {Review on the quantum emitters in two-dimensional materials}, J. Semicond. 40~(7) (2019) 071903.
\newblock \href {https://doi.org/https://doi.org/10.1088/1674-4926/40/7/071903} {\path{doi:https://doi.org/10.1088/1674-4926/40/7/071903}}.

\bibitem{lu2025van}
R.~Lu, C.~Lu, Y.~Li, H.~Song, K.~Gou, X.~Yuan, J.~Jiang, {Van der Waals Epitaxy of {CsPbI$_3$/MoS$_2$} Heterojunction Phototransistors for Neuromorphic Computing}, J. Phys. Chem. Lett. 16~(37) (2025) 9830--9838.
\newblock \href {https://doi.org/https://doi.org/10.1021/acs.jpclett.5c02548} {\path{doi:https://doi.org/10.1021/acs.jpclett.5c02548}}.

\bibitem{tian2025peculiar}
Z.-k. Tian, Z.-y. Luo, J.-j. Guo, J.-m. Ding, Y.-z. Nie, Q.-l. Xia, Y.~Zhou, G.-h. Guo, {Peculiar spin Hall magnetoresistance in polycrystalline {WTe$_2$/Ni$_{80}$Fe$_{20}$} heterostructures}, Appl. Phys. Lett. 126~(6) (2025).
\newblock \href {https://doi.org/https://doi.org/10.1063/5.0229028} {\path{doi:https://doi.org/10.1063/5.0229028}}.

\bibitem{ye2022nonvolatile}
T.~Ye, Y.~Li, J.~Li, H.~Shen, J.~Ren, C.-Z. Ning, D.~Li, {Nonvolatile electrical switching of optical and valleytronic properties of interlayer excitons}, Light Sci. Appl. 11~(1) (2022) 23.
\newblock \href {https://doi.org/https://doi.org/10.1038/s41377-022-00718-7} {\path{doi:https://doi.org/10.1038/s41377-022-00718-7}}.

\bibitem{das2021transistors}
S.~Das, A.~Sebastian, E.~Pop, C.~J. McClellan, A.~D. Franklin, T.~Grasser, T.~Knobloch, Y.~Illarionov, A.~V. Penumatcha, J.~Appenzeller, et~al., {Transistors based on two-dimensional materials for future integrated circuits}, Nat. Electron. 4~(11) (2021) 786--799.
\newblock \href {https://doi.org/https://doi.org/10.1038/s41928-021-00670-1} {\path{doi:https://doi.org/10.1038/s41928-021-00670-1}}.

\bibitem{shandilya2024spin_valley_locking}
A.~Shandilya, S.~Kapila, R.~Krishnan, B.~Weber, B.~Muralidharan, Unifying recent experiments on spin-valley locking in {TMDC} quantum dots (2024).
\newblock \href {http://arxiv.org/abs/2410.21814} {\path{arXiv:2410.21814}}, \href {https://doi.org/https://doi.org/10.48550/arXiv.2410.21814} {\path{doi:https://doi.org/10.48550/arXiv.2410.21814}}.

\bibitem{Abergel_2011}
D.~S.~L. Abergel, T.~Chakraborty, \href{https://doi.org/10.1088/0957-4484/22/1/015203}{Irradiated bilayer graphene}, Nanotechnology 22~(1) (2010) 015203.
\newblock \href {https://doi.org/10.1088/0957-4484/22/1/015203} {\path{doi:10.1088/0957-4484/22/1/015203}}.
\newline\urlprefix\url{https://doi.org/10.1088/0957-4484/22/1/015203}

\bibitem{merboldt2025floquet_states}
M.~Merboldt, M.~Sch{\"u}ler, D.~Schmitt, J.~P. Bange, W.~Bennecke, K.~Gadge, K.~Pierz, H.~W. Schumacher, D.~Momeni, D.~Steil, S.~R. Manmana, M.~A. Sentef, M.~Reutzel, S.~Mathias, Observation of {Floquet} states in graphene, Nat. Phys. 21 (2025) 1093--1099.
\newblock \href {https://doi.org/https://doi.org/10.1038/s41567-025-02889-7} {\path{doi:https://doi.org/10.1038/s41567-025-02889-7}}.

\bibitem{wang2026floquet_gap}
F.~Wang, X.~Cai, X.~Tang, J.~Lu, W.~Chen, T.~Sheng, R.~Feng, H.~Zhong, H.~Zhang, P.~Yu, S.~Zhou, Observation of {Floquet}-induced gap in graphene, Nat. Mater. (2026).
\newblock \href {https://doi.org/https://doi.org/10.1038/s41563-026-02549-y} {\path{doi:https://doi.org/10.1038/s41563-026-02549-y}}.

\end{thebibliography}

\end{document}